\documentclass[preprint,5p,times, twocolumn]{elsarticle}
\usepackage{hyperref}
\usepackage{graphicx}
\usepackage{epstopdf}
\usepackage{makecell}
\usepackage{amsmath}
\usepackage{float}
\usepackage{adjustbox}
\usepackage{rotating}
\usepackage{booktabs}

\usepackage{titlesec}

\titlespacing\subsection{0pt}{12pt plus 4pt minus 2pt}{0pt plus 2pt minus 2pt}
\titlespacing\subsubsection{0pt}{12pt plus 4pt minus 2pt}{0pt plus 2pt minus 2pt}

\usepackage{appendix}

\usepackage{amssymb}
\newtheorem{definition}{Definition}[section]

\journal{Journal Name}

\begin{document}

\begin{frontmatter}



\title{Towards a Formal Modelling, Analysis, and Verification of a Clone Node Attack Detection Scheme in the Internet of Things}

%\title{A Clone Node Attack Detection Scheme in the Internet of Things: Formal Modeling, Analysis, and Verification}


\author[]{Khizar Hameed \corref{mycorrespondingauthor}}
\ead{hameed.khizar@utas.edu.au}
\author[]{Saurabh Garg}
\ead{saurabh.garg@utas.edu.au}
\author[]{Muhammad Bilal Amin}
\ead{bilal.amin@utas.edu.au}
\author[]{Byeong Kang}
\ead{byeong.kang@utas.edu.au}
%\author[2]{Abid Khan}
%\ead{abk15@aber.ac.uk}

\address[]{Discipline of ICT, School of Technology, Environments, and Design,  University of Tasmania, Australia}
%\address[2]{Department of Computer Science, Aberystwyth University, Wales, United Kingdom}

\cortext[mycorrespondingauthor]{Corresponding author}

\begin{abstract}
%An Internet of Things (IoT) enables individuals to monitor and control their integrated environment through a network of interconnected IoT devices, each of which acts as a sensor and actuator, and actively participates in sensing, processing, storing, and sharing information. However,

A substantial component of the Internet of Things (IoT) network is made up of unmonitored IoT devices that are generally deployed in hostile situations where attackers attempt to capture and compromise them to gain control of the entire network. One such illustration of this malevolent behaviour on the part of an adversary is the cloning of IoT devices. In a clone node attack, an attacker attempted to physically capture the devices to gather sensitive information to conduct various insider attacks. Several solutions for detecting clone node attacks on IoT networks have been presented in the viewpoints above. These solutions are focused on specific system designs, processes, and feature sets and act as a high-level abstraction of underlying system architectures based on a few performance requirements. However, critical features like formal analysis, modelling, and verification are frequently overlooked in existing proposed solutions aimed at verifying the correctness and robustness of systems in order to ensure that no problematic scenarios or anomalies exist. This paper presents a formal analysis, modelling, and verification of our existing proposed clone node attack detection scheme in IoT. Firstly, we modelled the architectural components of the proposed scheme using High-Level Petri Nets (HLPNs) and then mapped them using their specified functionalities. Secondly, we defined and analysed the behavioural properties of the proposed scheme using Z specification language. Furthermore, we used the Satisfiability Modulo Theories Library (SMT-Lib) and the Z3 Solver to validate and demonstrate the overall functionality of the proposed scheme. Finally, in addition to modelling and analysis, this work employs Coloured Petri Nets (CPNs), which combine Petri Nets with a high-level programming language, making them more suitable for large-scale system modelling. To perform the simulations in CPN, we used both timed and untimed models, where timed models are used to evaluate performance, and untimed models are used to validate logical validity. 

\end{abstract}

\begin{keyword}

Clone Node \sep Internet of things \sep Formal modelling \sep Analysis \sep Verification \sep High level Petri Nets \sep Coloured Petri Nets


\end{keyword}

\end{frontmatter}



\section{Introduction}

An emerging and promising network paradigm known as the Internet of Things (IoT) involves numerous devices connecting people and objects to the internet to accomplish various functions \cite{al2015internet}. These devices are heterogeneous in nature and are placed in a variety of environments to collect data and information that is then sent to some controlling authorities, e.g. clouds, for analysis or decision-making enhancement \cite{diaz2016state}. Some of the advanced IoT-powered smart home features include thermostats, doorbells, security alarms, and similar automated devices, which enable people to remotely control and manage their homes and notify them if something suspicious happens in their absence. However, several risks have pervaded IoT devices, ranging from security threats to privacy concerns, owing to the numerous constraints imposed by their functional capabilities (processing, storage, and power) and the various design features \cite{alaa2017review}. For example, IoT devices are frequently non-tempered resistant and versatile, making them easy targets for attackers \cite{yang2017survey, frustaci2017evaluating}.

There are numerous types of attacks on IoT devices, and one type is a clone-node attack, also known as a device replication attack \cite{parno2005distributed}. In a clone-node attack, the attacker can extract the personal device ID, public and private keys from the IoT network and use this information to control the physical device(s). In order to leverage this vulnerability, an attacker must first locate the physical device, collect the secret credentials, and then use these credentials to manipulate the device's function before reinstalling it in the network \cite{numan2020systematic}. Clone-node attacks are more likely to occur if the IoT devices are not updated with the latest security software and their security certificates are outdated. Additionally, the clone node attack can be used in tandem with other malevolent attack vectors on the IoT network, including a selective forwarding attack, wormhole attack, and blackhole attack \cite{raza2013svelte}.

In recent years, the surge in popularity of IoT-based specific applications has intensified the interest in designing and developing security solutions to protect IoT devices and their data. Existing studies have been proven that a clone node attack detection mechanism effectively protects IoT devices from adversaries, as it enables systems to detect when the attacker initiates abnormal activity, which results in duplicate nodes being established in the network \cite{numan2020systematic}. One of the effective methods to mitigate the risk of a clone-node attack on an IoT network is that every device uses a data forwarding protocol to send authentication messages containing all of the information of their neighbours to a base station. The base station receives the message and validates it by utilising a key exclusively assigned to that device \cite{xing2008real}. Some other methods for finding the cloned node attacks include finding network witnesses \cite{parno2005distributed}, measuring the signal strength of devices \cite{xing2008real}, randomly picking secret keys \cite{brooks2007}, calculating trust values \cite{choi2007set, rikli2016lightweight}, managing distributed trust \cite{zhou2016improved}, localization algorithms \cite{yu2013localized} and multi-dimensional scaling \cite{lee2018mdsclone}. A detailed analysis of existing schemes based on wireless sensor networks, IoT, and mobile ad-hoc networks for detecting clone node attacks is summarised here \cite{hameed2021contextaware}.

Existing clone node attack detection approaches, on the other hand, are strictly limited to the system's architectures, including its functions and feature sets. Additionally, another common trend observed in existing clone node attack detection schemes is that they provide a high level of abstraction for system architectures and evaluate schemes solely on well-known performance parameters such as detection rate, detection time, and various system overhead complexities such as computation, communication, and storage, without providing any modelling and verification of their works to verify the correctness and robustness. In this work, we conducted formal modelling, analysis and verification of our existing proposed clone node attack detection scheme \cite{hameed2021contextaware} to provide the fine-grained abstraction level and to ensure that our proposed scheme does not contain any problematic scenarios or anomalies. Our proposed clone node attack detection scheme uses semantic information about IoT devices, known as context information, sensed from the deployed environment to locate them securely. Furthermore, our proposed scheme designed the location proof mechanism by combining location proofs and batch verification of the extended elliptic curve digital signature technique (ECDSA*) to accelerate the verification process at selected trusted nodes. Section \ref{overview} goes into detail regarding our existing proposed scheme and its working mechanism.

Formal verification is a systematic process that evaluates a proposed system or approach by defining rational arguments and determining whether or not the proposed system satisfies the provided design, engineering, and implementation requirements \cite{hameed2021formally}. We follow three fundamental steps of the formal verification process in this work: modelling, analysis, and verification. First, we use HLPNs to illustrate our proposed system model and information about the various components and internal links to get a detailed understanding of the design. Then, HLPN is utilised to model the systems and present mathematics to understand and analyse system behaviour and structural properties.  Moreover, the model is used to help analyse the interconnection of the components and processes, reveal the flow of information and processing of that information in-depth, and answer how the flow of information and processing occurs. Next, we used the PIPE+ tool to estimate the HLPNs' estimated incidence markers and confidence interval values during the result analysis phase. Afterwards, Z-language is applied to the systematic and behavioural characteristics of the system to model, describe and analyse. Additionally, we evaluated the models in three-fold: (i) we applied the automated model checking technique using the Satisfiability Modulo Theories Library (SMT-Lib) and Z3 solver to perform automated verification of the models. First, the models are translated into SMT and the specified properties, then SMT is used to validate the Petri Net models, (ii) The Z3 solver is used to examine the model to see if it is adhering to the specifications, (iii) Finally, we extend this work by modelling and analysis of proposed scheme using CPNs.

To the best of our knowledge, no existing work has been conducted to formally model, analyse, and verify the detection scheme for clone node attacks on IoT networks. This work aims to demonstrate the correctness of our proposed clone node attack detection scheme, which will enable researchers to comprehend the design, modelling, analysis, and verification processes required to analyse the efficient functioning of any system architecture and its underlying set of features.

The following constitute the primary contributions to this paper:

\begin{itemize}
    \item Model our proposed scheme with HLPNs and analyse the results with the help of incidence marking and confidence intervals.
    \item Analyse the proposed HLPNs using the Z specification language.
    \item Provide formal verification of the proposed HLPNs and their defined specifications and properties using SMT-Lib and Z3 solver. 
    \item Model our proposed scheme with CPNs gives the user the ability to design models as a hierarchy of modules supporting timed and untimed simulations.
\end{itemize}

The organisation of this paper is structured as follows: Section \ref{overview} provides an overview of our proposed clone node attack detection scheme. The preliminaries used in this paper are discussed in section \ref{preliminaries}. Section \ref{modelling} presents the modelling and analysis of the proposed scheme using high-level Petri nets and Z specification language, respectively. Formal verification of the proposed system along with its properties and results is illustrated in the section \ref{verification}. Section \ref{coloured} provides the modelling and analysis of our proposed scheme using CPNs. Finally, section \ref{conclusion} includes the conclusion to our work.

\section{An Overview of Our Existing Proposed Clone Node Attack Detection Scheme} \label{overview}

This section provides an overview of our existing proposed clone node attack detection scheme \cite{hameed2021contextaware}, including the background description, methodology with the network components and proposed algorithms, to assist readers in understanding the proposed scheme's procedure and helping with the system modelling, analysis and verification processes.

The proposed scheme leveraged context-aware systems in IoT networks to detect clone node attacks using context information from deployed IoT devices. Context-aware systems are IoT systems that keep track of surrounding objects and provide timely feedback to users. \textit{Context Information} is the semantic information that enables users to comprehend the networking environment and locate network entities based on their relationships with the environment. The proposed scheme also integrates location-based services (LBS) and context-aware systems to track network nodes. The LBS can enable numerous services in IoT-based applications, such as tracking and monitoring patients, tracking vehicles on the road, and determining an individual's actual position. LPSs are being implemented in IoT-based applications to create and share digitally signed context data to prove the user's location at any given time. The proposed scheme used digital signatures to verify the authenticity of devices or messages during communication to ensure IoT device and data security. Since the scalability of IoT networks is crucial in most applications, verifying individual device signatures is not recommended. For this reason, the proposed scheme makes use of batch verification, which verifies digital signatures in batches. \textit{Batch verification} is a concept that involves verifying multiple signatures simultaneously in order to minimise the time taken to validate each signature submitted by thousands of sensor nodes in large-scale IoT networks.

There are two primary components to the proposed scheme: an enhanced ECDSA* and a location-proof system (LPS), as explained in sections \ref{section2.1} and \ref{section2.2} respectively. These two components help us design an HLPN-based model, which is subsequently verified and analysed.

\subsection{An ECDSA* Technique}\label{section2.1}

As stated above, the proposed scheme used batch verification rather than individual signature validation to reduce verification time. In batch verification, the signer interacts with the verifier to generate \textit{t} signatures, and the verifier validates them all at once. An ECDSA is a widely used digital signature technique in the IoT, as it provides the same level of security as public-key cryptography but requires smaller key sizes. An ECDSA* is a variation of ECDSA that provides 40\% improved verification efficiency without sacrificing security. 

The ECDSA* algorithms used in the proposed scheme are explained in the following sections.

\subsubsection{Key Generation}
An ECDSA* Key Generation algorithm generates a pair of public and private keys for use in the signing and verification processes. The algorithm accepts as inputs standard domain parameters from elliptic curve cryptography (ECC), including \textit{p}, \textit{E}, \textit{P}, \textit{n}, \textit{h}. These parameters are defined in the following order: \textit{p} denotes order of prime field, \textit{E} denotes an elliptic curve generated over the prime field, \textit{P} denotes a non-zero random base point in \textit{E}, \textit{n} denotes the ordinal value of \textit{P}, which is typically a prime integer, and \textit{h} denotes a co-factor.

\subsubsection{Signature Generation}
An ECDSA* Signature Generation procedure takes the following parameters as an input such as message \textit{m}, hash function \textit{H}, and domain parameters such as \textit{P}, and then outputs the signature (\textit{r}, \textit{s}) for each device. This process started by selecting a random integer between 1 and n-1 for the \textit{k} parameter. Following that, a coordinate value \textit{X} is then determined by multiplying the random integer \textit{k} by the random point \textit{P}. The hash function \textit{H} (in this example SHA-1) takes the message \textit{m} and outputs a digest string value, which is then transformed to an integer \textit{e}. Finally, a signature value \textit{s} is calculated by taking the inverse of \textit{k} random integers and multiplying the sum of integer \textit{e} and private key \textit{d} by \textit{r}. The final output of ECDSA* signature generation is a pair, such as  (\textit{r}, \textit{s}). 
 
\subsubsection{Signature Verification}
An ECDSA* Signature Verification mechanism verifies the signer's signatures sent with his/her public key. The verification process is entirely dependent on the signature size in terms of computational time. For instance, the lengthy signature requires additional verification time. The verification algorithm requires a signature value (\textit{r}, \textit{s}) and a public key \textit{Q} as inputs. To be more specific, the signature verification output is a binary choice (accept or reject). The signature verification process starts by determining if the signature values \textit{r} and \textit{s} are in the interval [1, \textit{n}-1]. The hash \textit{H} function then computes the hash value of the message \textit{m} for comparison. Similarly to the signature generation algorithm, the hash value is converted into an integer \textit{e}. Further, by taking the modulus of the inverse value of the signature, an integer value \textit{w} is generated. Two coordinates, $u_{1}$ and $u_{2}$, are determined by multiplying the integers \textit{e} and \textit{r} by the value \textit{w}, respectively. An \textit{X} value is generated by combining the multiplications of \textit{P} and \textit{Q} by the calculated coordinates ($u_{1}$, $u_{2}$) from the previous step. If X $=$ $\mathcal{O}$, the signature will be rejected; otherwise, it will be accepted if and only if \textit{$\upsilon$} = \textit{r}.
\subsection{Location Proof System (LPS)}\label{section2.2}

The LPS component used context-aware modalities localisation to detect clone node attacks on IoT networks. As the name implies, context-based localisation collects contextual information about the IoT device's environment (e.g. ambient acoustic light, noise level, humidity, temperature, Wi-Fi and Bluetooth signal power) and then use proofs to determine the device's exact location. Contextual information is collected concurrently by devices and verifiers, with the device generating proofs of presence from the collected data and the verifier validating such proofs. The proposed LPS mechanism consists of the following interacting components: prover, clone node, verifier, and LBS. According to the proposed scheme, verifiers aim to inform an LBS about the existence of provers at a specific place to detect a clone node attack. A prover and a verifier collect context information from the deployed environment. To validate the proofs, the verifier IoT devices compare the context information obtained from the prover IoT devices to their context information to determine whether or not the IoT device has been compromised. The following sections illustrate the working of LPS in the proposed scheme: calculate location, generate location proof, and verify location proof. 
\subsubsection{Calculate Location}

The Calculate Location algorithm shows how to locate network devices. The proposed scheme calculates the location of each device (such as the prover and verifier) by measuring the distance between them in two-dimensional (2-D) space.
\subsubsection{Generate Location Proof}

The Generate Location Proof algorithm shows how to generate location proofs for IoT devices based on LBS. The prover and the verifier first collect contextual information about their deployed environment. A verifier first requests the prover to generate a location proof utilising sensed context information as \textit{CI}. The proof is then signed using the prover's private key $K_{Pr}$. 

\subsubsection{Verify Location Proof}

Finally, the Verify Location Proof algorithm describes the process of verifying location proofs for IoT devices claiming to be at a given location. The verifiers utilised the prover's public key $K_{Pb}$ to validate the signature $P_{sign}$. After successfully validating each prover's signature, the verifier will accept the proof with location confirmation and other credentials. If the signature is not validated, the verifier informs the LBS about the compromise of devices.

\section{Preliminaries} \label{preliminaries}

This section explains the preliminaries for various techniques, technologies, and tools that helped us achieve our goals and ensure the paper's readability. This work includes HLPNs, SMT-Lib, Z3 solver, formal verification, and CPNs as preliminaries.

\subsection{The High-Level Petri Nets}
The HLPN determines system behaviour using modular design and mathematical principles. The HLPN approach is useful for representing systems with various schemas, such as serial and parallel, synchronous or asynchronous distributed, pseudo-deterministic and stochastic \cite{jensen2012high} \cite{genrich1981system}. Each system has its analytical capabilities and limitations. For large networks-based systems, it is necessary to specify fine-grained modules and generalisation for protocol analysis. HLPNs, or Petri nets, are widely accepted as the best technique to model and construct a system for evaluation and validation. In simple terms, HLPN is an array \textit{N} which has seven tuples, for example {\textit{N}=\{\textit{P}, \textit{T}, \textit{F}, \textit{$\varphi$}, \textit{R}, \textit{L}, \textit{$M_{0}$}\}}. The definition of each tuple is as follows \cite{sibertin1985high} \cite{jensen1983high}:

\begin{itemize}

\item \textit{P} is a set of fixed places.

\item \textit{T} is a set of fixed transitions, thus, \textit{P}  and \textit{T} are two different sets denoted by \textit{P} $\cap$  \textit{T} = $\phi$.

\item \textit{F} representing a directed flow from transition to place or place to transition, so \textit{F} $\subseteq$ $( \textit{P} \times \textit{T} )$ $\cup$ $( \textit{T} \times \textit{P})$.

\item \textit{$\varphi$} denotes a mapping function that maps the places \textit{P} to defined data types such as $\varphi$ : \textit{P} $\cup$ \textit{T} $\rightarrow$ Data types.

\item \textit{R} is a set of finite rules that map \textit{T} to logical formulae, such as \textit{R}: $\rightarrow$ Formula or logical reasoning

\item \textit{L} is the label associated with each flow as \textit{F}, such that \textit{L} : \textit{F} $\rightarrow$ Label.

\item \textit{$M_{0}$} is the Petri Net's start state that initiates flow and produces tokens  \textit{M} : \textit{P} $\rightarrow$ Tokens.

\end{itemize}

The three building blocks (\textit{P}, \textit{T}, \textit{F}) define the structure of Petri nets. Additionally, three blocks (\textit{$\varphi$}, \textit{R}, \textit{L}) contain information about the Petri net metadata that is used during implementation and validation part. In HLPNs, places are represented by circles, transitions by rectangles and directed arrows can be used to connect places and transitions, and vice versa. However, no links between places or transitions are established. For each HLPN, the ``Start'' transition is used to activate the HLPN, and the ``Inputs'' places include the tokens that can input the model and travel through several places and transitions before reaching the end place in the Petri net. To create tokens and initiate the process from the input place, the R (Input) = $\exists$ x $\in$ X $|$ . x = $\theta$ rule is used.

\subsection{SMT Lib and Z3 Solver}

Satisfiability Modulo Theories (SMT) \cite{de2009satisfiability, dutertre2006yices} is a commonly used decision-making technique that can solve large decision-making problems based on first-order logic formulae and provide system satisfaction using a broad range of dynamically typed theorems. SMT-Lib is evolved from Boolean Satisfiability solvers (SAT); however, SMT-Lib relies on first-order formulae while SAT relies on propositional formulae. SMT has the advantage of supporting a wide range of theories and decision problem domains, including integers, real numbers, rationals, arrays, and even supported equality, as well as bit-vectors and uninterrupted functions. The SMT is used for model verification, inductive reasoning software verification, test generation, and simulation. SMT-Lib is included to quantify the verification facility for a large number of different solvers. In SMT-Lib, the suggested system's behaviour required abstract models and bounded model procedures executed by bounded symbolic execution. SMT-Lib is a Microsoft Research library that supports and integrates various solvers, including STP \cite{ganesh2007decision}, OpenSMT \cite{bruttomesso2010opensmt}, \cite{cimatti2013mathsat5}, Boolector \cite{brummayer2009boolector}, Beaver \cite{jha2009beaver} and Z3 \cite{de2008z3}. 

The Z3 is an automated satisfiability prover developed by Microsoft Research that uses the standard SMT-LIB built-in theories. It also supports other theories such as bit vectors, empty theory, arrays, data types, linear and nonlinear arithmetic, and quantifiers. Like other SMT solvers such as MathSatb \cite{dutertre2006yices} and CVC4 \cite{DBLP:conf/cav/BarrettCDHJKRT11}, Z3 has its own verification language.

\subsection{Formal Verification}

Formal verification is a mathematical technique for proving or disproving the correctness of a model built for an underlying system  \cite{de2008z3}. The criteria for measuring the system correctness are often specified in a property specification language as logical properties. The correctness procedure determines and verifies that the model acts as indicated in the properties. Unlike simulation and testing, formal verification thoroughly evaluates the behaviour of a model. It presents a formal proof based on a complex logical description of the system. The verification approach examines all relevant states of the system model to ensure that the properties given for each state are fulfilled. Model checker demonstrates how to achieve the failure state of the model if any state is found to be violating the observed property \cite{baier2008principles}. The most extensively used formal verification techniques are bounded model checking and theorem proving.

We first built our system as a finite-state model with specifications denoted by temporal logic properties to conduct formal verification. We then verified these models using the bounded model checking technique specified as an automated technique and analysed the results using SMT-Lib and the Z3 solver. The bounded model checking problem is shown below: For example, a model checking tool accepts inputs as a finite state model \textit{$F_{m}$} and a property \textit{p} to find and validate the property expressed as \textit{$F_{m}$}$|$ = \textit{p}.

The solver checks the expression and returns sat (satisfiable) or unsat (unsatisfiable). If the outcome is sat, the property fails (i.e. \textit{$F_{m}$}$|$ = \textit{p}) and the solver provides a counterexample showing how the model fails. If the solution results are unsatisfactory, the property is true in \textit{$F_{m}$}. To define Bounded Model Checking, we use the Kripke Structure \cite{barrett2010satisfiability}.

\begin{definition}
\textbf{(Bounded Model Checking) \cite{barrett2010satisfiability}} Given a Kripke structure, \textit{M} =  (\textit{S}, \textit{$S_{0}$}, \textit{T}, \textit{L}), where \textit{S} denotes all states, \textit{$S_{0}$} denotes initial states,\textit{T} denotes transitions, and \textit{L} denotes the labelling function. The problem of model checking is to determine the set of all states in \textit{S} that satisfy \textit{f}, i.e. \{\textit{s} $\in$ \textit{S} $|$, \textit{M}, \textit{s} $\neq$ \textit{f}\} where \textit{f} is a temporal logic formula expressing some desired specification.

\end{definition}

\subsection{Coloured Petri Nets}

Coloured Petri Nets (CP-nets) is a modelling language or graphical tool extensively used to describe and specify concurrent and distributed systems and investigate their behaviour using simulation and verification based on mathematical properties \cite{jensen1981coloured}. Petri net is a fundamental layer upon which graphical notations for specified models are constructed. These graphical notations can support control structure, concurrency, communication, and synchronisation using basic modelling primitives. For formal analysis, utilising CPNs provides a specification language with dynamic nature and formal semantics. Moreover, unlike domain-specific modelling, CPN modelling is more generic, allowing it to build a broad range of distributed and concurrent systems \cite{jensen1983high, jensen1987coloured}.

CPNs have been used in a wide range of systems, including concurrent and distributed systems, complex data networks, communication protocols, business process management, scientific workflows, embedded systems, multi-agent systems, and manufacturing processes. CP-nets can also be used with current modelling languages and methods like Unified Modelling Language (UML). UML is a functional programming language that includes primitives for defining data types, expressing data manipulation and generating parametric graphical models. Furthermore, state spaces and place invariant simulation are utilised to validate CPN models \cite{jensen2012high}.

CPNs improve the overall usability of Petri Nets and extend their applications to large-scale systems. For example, CPN models concurrent systems using Petri Net graphical components for process communication and synchronisation, while ML functional programming allows for new data types and data value modification. Furthermore, complex networks and systems can be described in the CPN language as collections of modules \cite{jensen1997brief}.

The following two types of models can be built with CPNs.

\begin{itemize}
    \item \textbf{Timed Models:} Time is a key performance indicator for concurrent systems. Timed models are used to evaluate system performance in a CPN-based simulation environment. 
    \item \textbf{Untimed Models:} Unlike timed models, untimed models are commonly used to verify a system's logical or theoretical correctness during a simulation.
\end{itemize}

The main objective of CPN is to define a modelling language for concurrent systems that can be scaled to industrial applications. In light of the preceding, several formal modelling languages, notably CPNs, have been developed to address the complexity of concurrent system designs, which can occasionally result in small, undetected errors. Furthermore, with growing interest in the correctness of concurrent systems, formal verification approaches have evolved into a critical component of developing reliable systems.

The main differences between CPNs and standard Petri nets are the following: (i) Petri Net tokens have no data attached to them, (ii) Tokens usually are black dots in Petri Nets, but in the CPN, tokens can be any colour (types), (iii) Tokens can be defined as a sequence of complex data types such as colour sets using CPN, (iv) Despite the differences in tokens between CPNs and Petri Nets models, CPNs allow a variety of data types to be added into the model,  (v) The arc inscriptions used in CPNs are expressions, not constants, (vi) The CPN ML used a variety of data types, including integers, strings, reals, Booleans, and the void unit, (vii) Custom data types like the product, subrange, enumeration, record, union, and the list can be created using CPN ML, (viii) CPNs make Petri Nets more practical by introducing a programming language with a higher degree of abstraction, increasing their scalability, and effectiveness \cite{jensen2009coloured}.

CPNs is an array \textit{N} which has nine tuples, for example {\textit{N}=\{\textit{P}, \textit{T}, \textit{A}, \textit{$\varphi$}, \textit{C}, \textit{F}, \textit{G}, \textit{L}, \textit{$M_{0}$}\}}. The definition of each tuple is as follows \cite{jensen1983high}:

\begin{itemize}

\item \textit{P} is a set comprising a fixed number of places.

\item \textit{T} is a set of transitions with a finite number, such as \textit{P}  and \textit{T} are two different sets denoted by \textit{P} $\cap$  \textit{T} = $\phi$.

\item \textit{A} representing the arcs from transition to place or place to transition.

\item \textit{$\varphi$} representing the universal colour set consisting of all colours, actions, and functions. 

\item \textit{C} denotes a colour function that maps the places \textit{P} and transitions \textit{T} to defined colour set such as \textit{C} : \textit{P} $\cup$ \textit{T} $\rightarrow$ $\varphi$.

\item \textit{F} representing a node function that maps \textit{A} into places and transitions such as \textit{A} $\subseteq$ $( \textit{P} \times \textit{T} )$ $\cup$ $( \textit{T} \times \textit{P})$.

\item \textit{G} $\in$ \textit{g} is a function that acts as a guard. Each transition \textit{t} $\in$ \textit{T} is associated with a guard expression \textit{g}. The guard expression's output should be a Boolean value: true or false.

\item \textit{L} is the label (or expression) associated with each arc as \textit{A}, such that \textit{L} : \textit{A} $\rightarrow$ Label or Expression.

\item \textit{$M_{0}$ $\in$ \textit{M}} is the CPNs start state that initiates arc and produces markers \textit{$M_{0}$ } : \textit{P} $\rightarrow$ Markers.

\end{itemize}

The places are depicted as circles or eclipses. Transitions are represented by rectangles. The acrs are depicted as directed arrows. When transitions occur, the arc expression describes the change in each state of the CPNs. Each place contains a collection of tokens. As compared to HLPNs, the CPN's tokens represent some data values tied with the place.

\section{Modelling and Analysis of Proposed Scheme Using High-Level Petri Nets} \label{modelling}

This section details the formal modelling and analysis of the proposed scheme for detecting clone node attacks. We used HLPNs to model our proposed scheme and the Z specification language to specify the rules or properties that defined the behaviour of the underlying proposed system's model. 

We divide our proposed scheme into parts to facilitate the modelling process, such as an ECDSA* technique and a LPS technique. Each part, along with its HLPN modelling, is presented separately in the sections \ref{modelling1} and \ref{modelling2}. Furthermore, in section \ref{rules}, a set of rules consisting of syntax and semantics is presented to specify the interaction behaviours of the proposed scheme's components. Finally, the analysis of the modelling results obtained using incidence marking and confidence intervals are presented and discussed in detail in the section  \ref{analysis}.

\subsection{An ECDSA* Technique} \label{modelling1}

An ECDSA* technique is divided into three HLPNs based on the algorithms used to generate keys, generate signatures, and verify signatures.

\subsubsection{ECDSA* Key Generation}

\begin{figure}[!t]
  \centerline{\includegraphics[width=9cm, height=4cm]{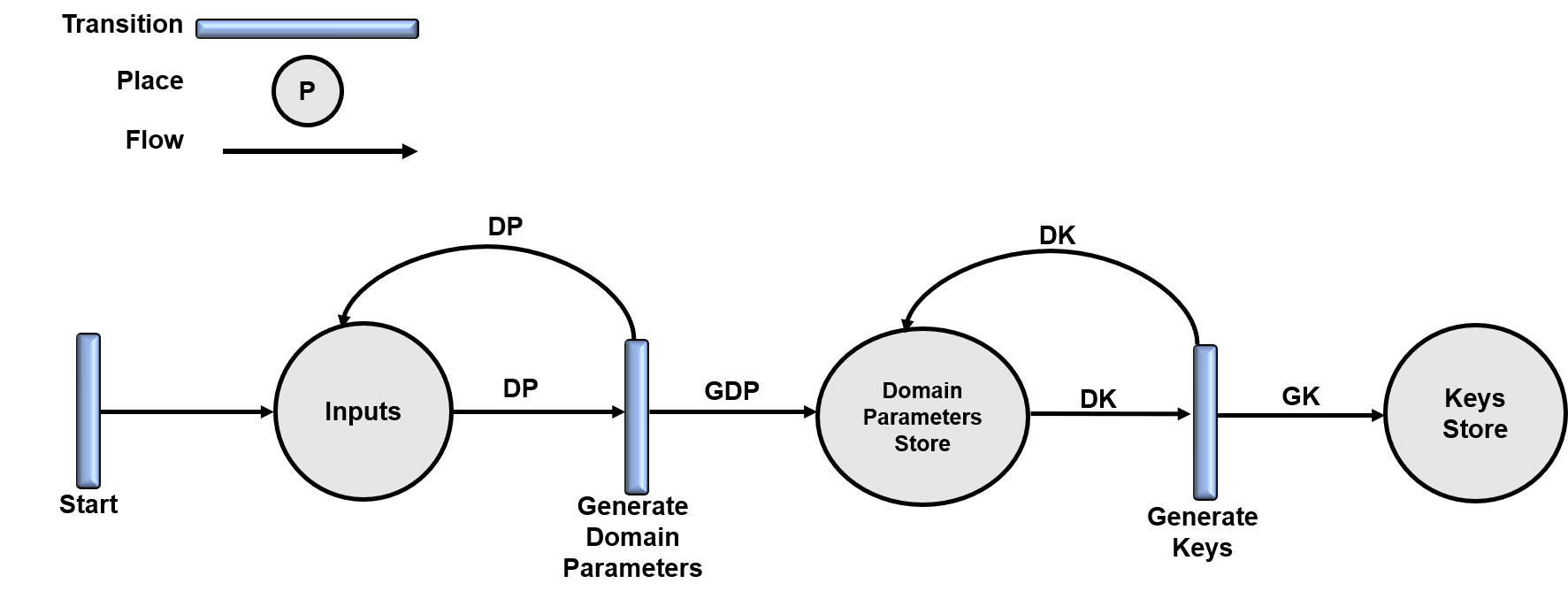}}
  \caption{An ECDSA* Key Generation - Petri Net}
  \label{petrinet1}
  \vspace{-4mm}%Put here to reduce too much white space after your table 
\end{figure}

Fig. \ref{petrinet1} illustrates the HLPN of the key generation process in the ECDSA* technique. In Petri nets, the process starts with the ``Start'' transition, the initial state of any HLPN and is responsible for creating tokens that transit through all subsequent transitions. The terms ``Inputs'', ``Domain Parameters Store'' and ``Keys Store'' refer to the places or stores used to store the preliminary data (or variables). The ``Inputs'' place contains the initial data required to generate the domain parameters. The ``Domain Parameters Store'' place has the domain parameter values used during the key generation process, and the ``Keys Store'' place contains the public and private keys. Transitions are responsible for carrying out the anticipated actions or activities in the algorithms. For example, as indicated previously, the ``Start'' transition initiates the entire process. The ``Generate Domain Parameters'' transition generates the domain parameters, and the ``Generate Keys'' transition generates the keys based on inputs from the ``Domain Parameters Store'' place.

\subsubsection{An ECDSA* Signature Generation}

\begin{figure}[ht]  
\centerline{\includegraphics[width=9cm, height=4cm]{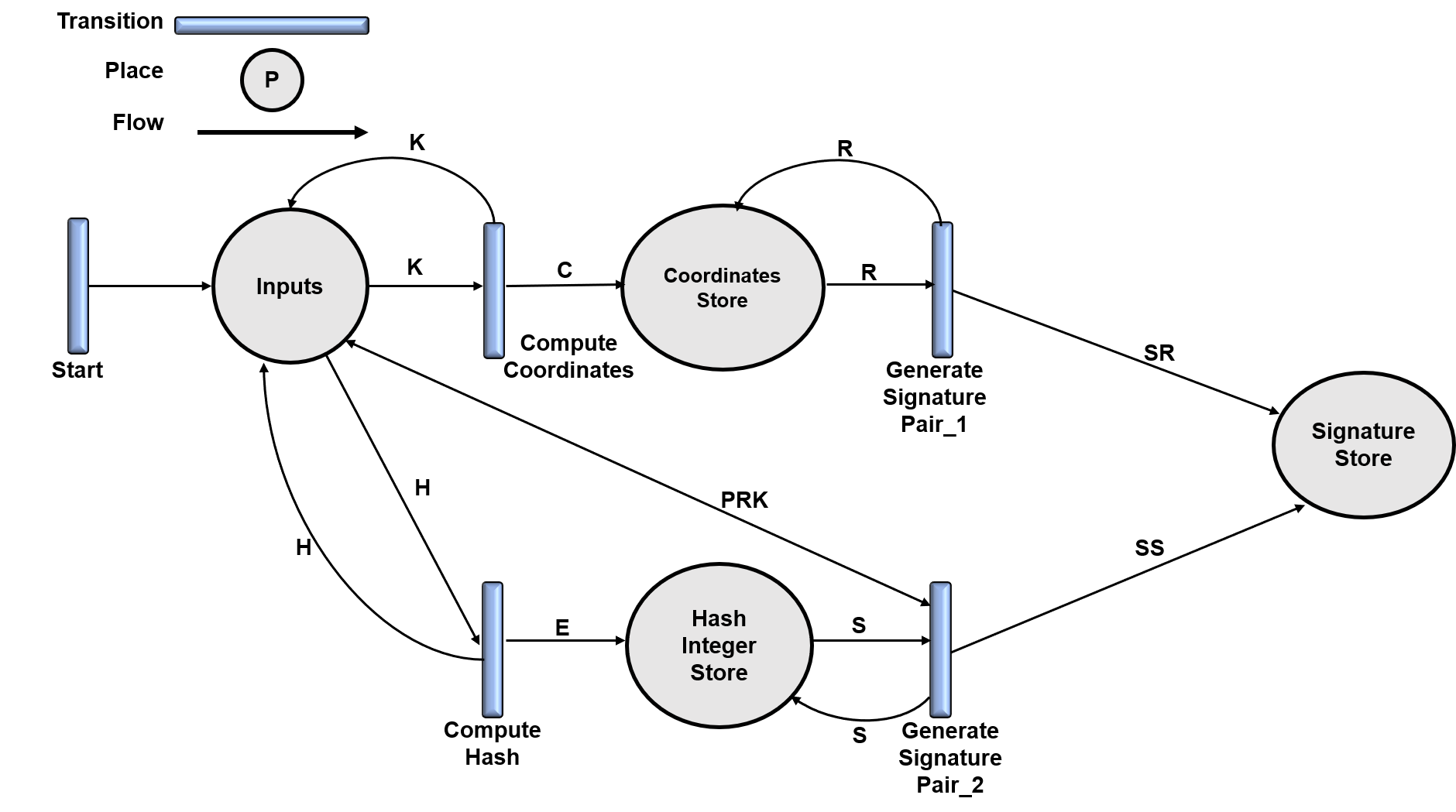}}
  \caption{An ECDSA* Signature Generation - Petri Net}
  \label{petrinet2}
  \vspace{-4mm}%Put here to reduce too much white space after your table 

\end{figure}

The Petri net of the signature generation process in an ECDSA* technique consists of five working transactions and four places, as shown in Fig. \ref{petrinet2}. 

In this Petri net, the transitions are ``Start'', ``Compute Coordinates'', ``Compute Hash'', ``Generate Signature Pair 1'', and ``Generate Signature Pair 2'', while the places are ``Inputs'', ``Coordinates store'', ``Hash Integer Store'', and ``Signature Store''.  The ``Start'' transition begins the ECDSA* signature generation process and stores all initial variables in the ``Inputs'' place. In contrast, the ``Compute Coordinates'' transition accepts the integer value from the ``Inputs'' place. It computes the coordinates used in the signature generation pairs stored in the ``Coordinates Store'' place. The ``Compute Hash'' transition computes the hash value of the inputs specified in the ``Inputs'' place, converts it to an integer value, and stores it in the ``Hash Integer Store'' place. The transitions ``Generate Signature Pair 1'' and ``Generate Signature Pair 2'' take inputs from the ``Coordinates Store'' and ``Hash Integer Store'' places and generate the signature pair, which is then stored in the ``Signature Store'' place.

\subsubsection{An ECDSA* Signature Verification}

\begin{figure}[ht]
  \centerline{\includegraphics[width=9cm, height=6cm]{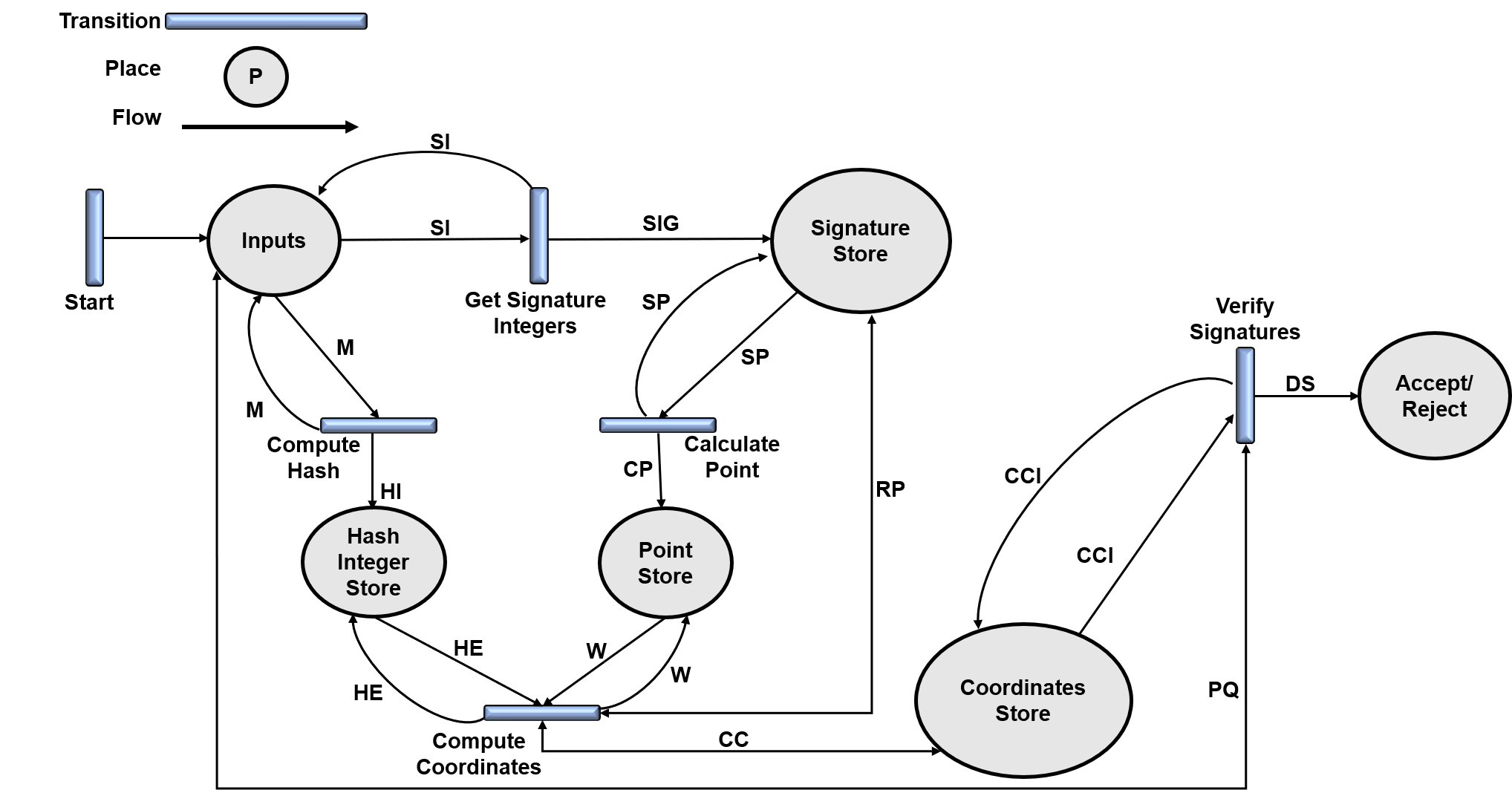}}
  \caption{An ECDSA* Signature Verification - Petri Net}
  \label{petrinet3}
  \vspace{-4mm}%Put here to reduce too much white space after your table 

\end{figure}

The HLPN of the signature verification in an ECDSA* scheme is depicted in Fig. \ref{petrinet3}.

The HLPN contains various transitions, including ``Start'', ``Get Signature Integers'', ``Calculate Hash'', ``Calculate Point'', ``Calculate Coordinates'' and ``Verify Signatures''. On the other hand, the places include the ``Inputs'', ``Signature Store'', ``Hash Integer Store'', ``Point Store'', ``Coordinates Store'', and ``Accept / Reject''. The signature verification procedure begins with the ``Start'' transition, which stores the signature parameters and public key in the ``Inputs'' place. The ``Get Signature Integers'' transition then extracts the signature parameters and stores them in the ``Signature Store'' place for later verification. The ``Compute Hash'' transition accepts the message as an input and converts it to a hash integer, placed in the ``Hash Integer Store'' place.
On the other hand, the ``Calculate Point'' transition accepts the signature parameters from the ``Signature Store'' place and storing them in the ``Point Store'' place. Next, the ``Compute Coordinates'' transition calculated the coordinates by obtaining the inputs from the ``Hash Integer Store'' and ``Point Store'' places and storing them in the ``Coordinates Store'' place. Finally, a ``Verify Signature'' transition accepts the coordinates from the ``Coordinates Store'' place and verifies whether to accept or reject the signature decision.

\begin{figure*}[!hbtp]
\centering
  \includegraphics[width=14cm, height=7cm]{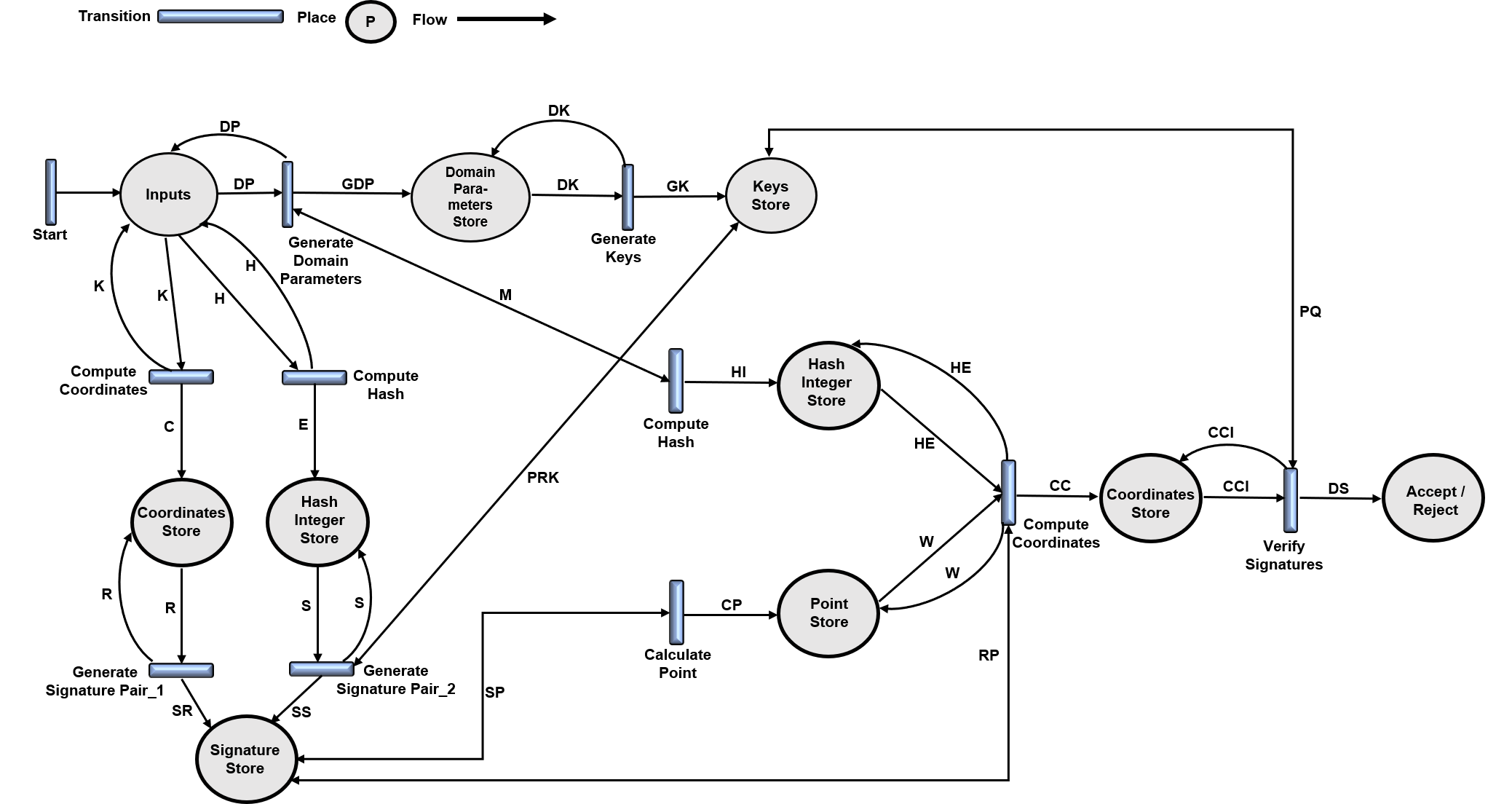}
  \caption{A Complete HLPN of an ECDSA* Technique}
  \label{petrinet4}
    \vspace{-6mm}%Put here to reduce too much white space after your table 

\end{figure*}

The complete HLPN of the ECDSA* technique, including key generation, signature generation, and signature verification, is illustrated in Fig. \ref{petrinet4}, and the working mechanism is then described in detail using a set of Z specification rules from \ref{R1}-\ref{R11}.

\begin{table}[!t]
\tiny
\centering
\caption{Places, Mappings and Description of an ECDSA* Technique}
\label{places}

\begin{adjustbox}{width=0.47\textwidth,center}

%\resizebox{\columnwidth}{!}{
\begin{tabular}{|l|l|l|}
\hline
     \textbf{Places}      &    \textbf{Mappings}  & \textbf{Description}      \\ \hline

\multicolumn{3}{|l|}{\textbf{An ECDSA* Key Generation}} \\ \hline
    
$\varphi$ (Inputs) &   \textit{P}(\textit{p} $\times$ \textit{E} $\times$ \textit{P} $\times$ \textit{n} $\times$ \textit{h}) & \makecell{Holds the standard \\ domain parameters}    \\ \hline

$\varphi$ (Domain Parameters Store)    &   \textit{P}(\textit{p} $\times$ \textit{E} $\times$ \textit{P} $\times$ \textit{n} $\times$ \textit{h})  &  \makecell{Holds the standard \\ domain parameters}   \\ 
    \hline
    
$\varphi$ (Keys Store)    &   \textit{P}(\textit{d} $\times$ \textit{Q})  & \makecell{Store the private \\ and public keys}    \\ \hline

\multicolumn{3}{|l|}{\textbf{An ECDSA* Signature Generation}} \\ \hline

$\varphi$ (Inputs)    & \textit{P}(\textit{m} $\times$ \textit{d} $\times$ \textit{H} $\times$ \textit{P} $\times$ \textit{k})       & \makecell{Holds the input \\ variables used in the \\ signature generation}  \\ \hline

 $\varphi$    (Coordinates Store)  & \textit{P}(\textit{X}) & \makecell{Stores the \textit{x} \\ and \textit{y} coordinates} \\ \hline 

$\varphi$    (Hash Integer Store)  & \textit{P}(\textit{e}) & \makecell{Stores the hash \\ integer value} \\ \hline

$\varphi$    (Signature Store)  & \textit{P}(\textit{r} $\times$ \textit{s}) & \makecell{Stores the signature \\ pair} \\ \hline

\multicolumn{3}{|l|}{\textbf{An ECDSA* Signature Verification}} \\ \hline
     $\varphi$ (Inputs)     &    \textit{P}(\textit{r} $\times$ \textit{s} $\times$ \textit{Q} $\times$ \textit{m})    & \makecell{Holds the signature \\ pair and public key} \\ \hline
     $\varphi$ (Signature Store)     &   \textit{P}(\textit{r} $\times$ \textit{s}) & \makecell{Stores the signature \\ pair } \\ \hline
     $\varphi$  (Hash Integer Store)    &       \textit{P}(\textit{e})   & \makecell{Stores the hash \\ integer value} \\ \hline
     $\varphi$ (Point Store)     &   \textit{P}(\textit{w})  & \makecell{Stores the point \\ value}    \\ \hline
     $\varphi$ (Coordinates Store)     &   \textit{P}(\textit{$u_{1}$} $\times$ \textit{$u_{2}$})  & \makecell{Stores individual \\ coordinate values}    \\ \hline
     $\varphi$ (Accept / Reject)     &   \textit{P}(\textit{X})  & \makecell{Stores \\ Coordinate point} \\ \hline

\end{tabular}
\end{adjustbox}
\vspace{-4mm}%Put here to reduce too much white space after your table 
\end{table}

Table \ref{places} specifies the places, mappings, and descriptions of each place used in an ECDSA* technique. Since an ECDSA* technique includes several steps or actions, from key generation to signature generation and verification, we define algorithms for each step. Each step associated with the algorithms specified for an ECDSA* technique defines the operations of each component, including standard domain parameters, key generation, point coordinate generation, hash functions, signature generation, and signature verification, as well as their associated working flows, such as input and output. For example, three places are utilised in the keys generation Petri net: inputs, domain parameter store, and keys store. The input place is used to store the domain parameters at their initial state. The domain parameters place is used to store the standard domain parameters after being modified, and the keys store place is used to store the private and public keys.

The process of generating an ECDSA* signature comprises four places: inputs, coordinates store, hash integer store, and signature store. The inputs store contains a message that must be hashed and signed, the private key used to sign the signature, the hash function, a random base point, and a random value. The coordinates store place is used to hold the values of the coordinates or points. Further, the hash integer store saves a hash integer value.  Finally, the signature pairs are stored in a signature store place. 

Finally, the ECDSA* signature verification method entails using the following places: inputs, signature store, hash integer store, point store, coordinates store, and accept/reject store. The inputs place contains signature pairs, a public key, and the message to verify. A signature store place is where the signature pairs extracted from the input places are stored. The message's hash integer value is computed and saved in the hash integer store. The point value and coordinates are respectively kept in the places called point store and coordinates store. Finally, the accept/reject store place stores the verification decision.

Table \ref{datatypes} lists the data types used in an ECDSA* scheme.

%For example, \textit{P} is an integer data type that is used to describe the order of prime field. The \textit{E} is a float data type that is used to represent the coordinates of a point on the prime field. \textit{P} is an integer data type that is used in \textit{E} to denote the non-zero random base point. The integer data type \textit{n} is used to represent the ordinal value of \textit{P}. \textit{h} is a float data type that is used to represent the value of the co-factor. The byte data type \textit{d} is used to represent the private key used to sign the signatures. Additionally, \textit{Q} is a byte data type used to represent the public key used to validate signatures. \textit{m} is a byte data type that is used to represent the hashed and signed message. \textit{H} is a byte data type that is used to represent the hash function's value. The integer data type \textit{k} is used to represent the random value. \textit{X} is a float data type that is used to represent the curve's coordinates. The integer data type \textit{e} is used to represent the value translated from the byte hash value \textit{H}. \textit{r} is a float data type that is used to represent the value of the first signature pair. The float data type \textit{s} is used to represent the value of the second signature pair. The \textit{w} data type is a float data type that is used to represent a point value. \textit{$u_{1}$} is a float data type that represents the first coordinate value, and \textit{$u_{2}$} is also a float data type that represents the second coordinate value.

\begin{table}[ht]
\tiny
\centering
\caption{Data Types used in an ECDSA* Scheme}
\label{datatypes}
\begin{adjustbox}{width=0.47\textwidth,center}
\begin{tabular}{|c|c|}
\hline
\textbf{\makecell{Data \\ Types}} & \textbf{Description}  \\ \hline

\textit{p} &  \makecell{An integer type for the representation of order of prime field} \\ \hline
\textit{E} & \makecell{A float type for the representation of coordinates point over the prime field}   \\ \hline
\textit{P} & \makecell{An integer type for the representation of a non-zero random base point in \textit{E}}   \\ \hline
\textit{n} & \makecell{A integer type for the representation of the ordinal value of \\ \textit{P}, which is normally a prime number} \\ \hline
\textit{h} &  \makecell{A float type for the representation of co-factor}  \\ \hline
\textit{d} & \makecell{A byte type for the for the representation of private key}   \\ \hline
\textit{Q} & \makecell{A byte type for the representation of public key}   \\ \hline
\textit{m} & \makecell{A byte type for the representation of messages to be signed} \\ \hline
\textit{H} & \makecell{A byte type for the representation of values of hash function} \\ \hline
\textit{k} & \makecell{An integer type for the representation of the random value} \\ \hline
\textit{X} & \makecell{A float type for the representation of coordinate points} \\ \hline
\textit{e} & \makecell{An integer type for the representation of hash value \\ converted from hash byte value \textit{H}} \\ \hline
\textit{r} & \makecell{A float type for the representation of first pair value of signature} \\ \hline
\textit{s} & \makecell{A float type for the representation of second pair value of signature} \\ \hline
\textit{w} & \makecell{A float type for the representation of point value} \\ \hline
\textit{$u_{1}$} & \makecell{A float type for the representation of first coordinate value} \\ \hline
\textit{$u_{2}$} & \makecell{A float type for the representation of second coordinate value} \\ \hline
\end{tabular}$$$$
\end{adjustbox}
\vspace{-4mm}%Put here to reduce too much white space after your table 
\end{table}

\subsection{Location Proof System} \label{modelling2}
The proposed clone node attack detection scheme based on a LPS is divided into three HLPNs based on the algorithms used, such as calculating location, generating location proof, and verifying location proof.

\subsubsection{Calculate Location}

\begin{figure}[!htp]
  \centerline{\includegraphics[width=8.5cm, height=4cm]{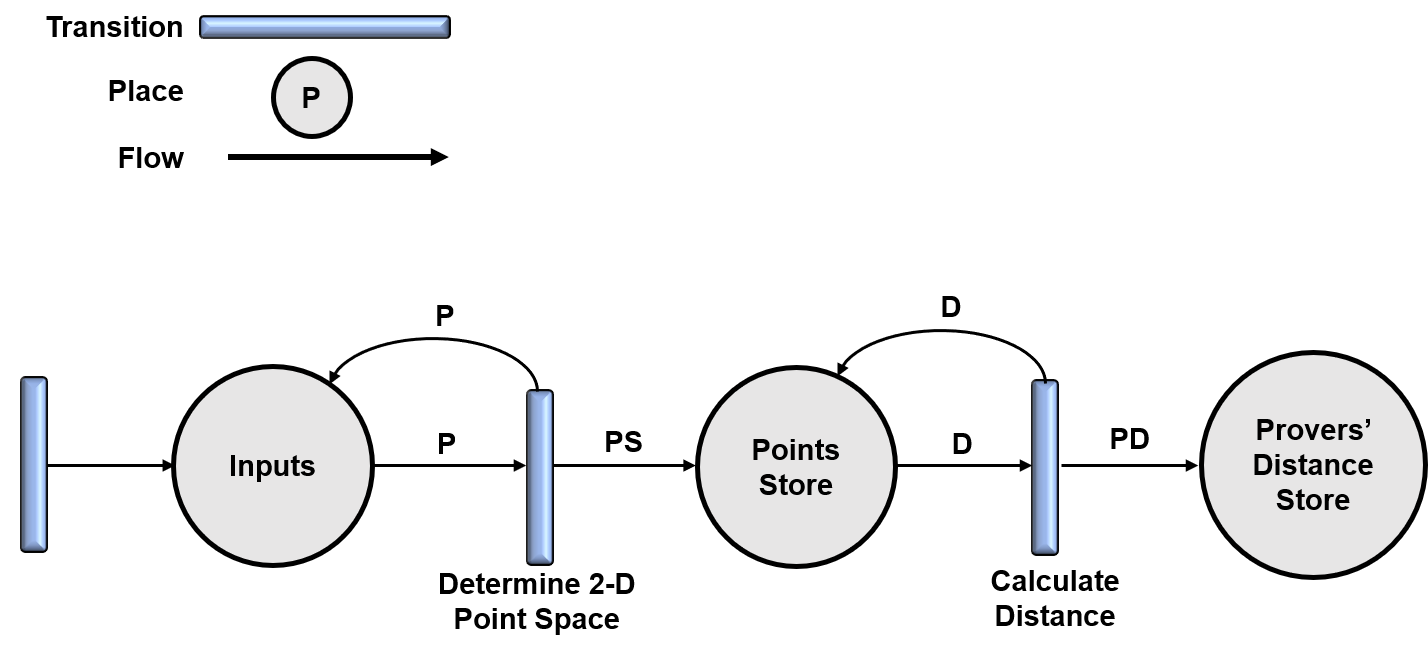}}
  \caption{Calculate Location- Petri Net}
  \label{petrinet5}
  \vspace{-5mm}%Put here to reduce too much white space after your table 
\end{figure}

Fig. \ref{petrinet5} illustrates the HLPN of the location calculation for the proposed location proof scheme in which the location calculation process starts with the ``Start'' transition providing the initial variables to ``Inputs'' place to determine 2-D point space.

The ``Points Store'' place holds the points, also known as coordinates on 2-D space, to determine the position of provers and verifiers. Based on these coordinates, the distance is calculated between provers and verifiers and stored in the ``Proverss' Distance Store''. Transitions are responsible for carrying out the anticipated actions or activities in the algorithms. For instance, as indicated previously, the ``Start' transition initiates the entire process, the ``Determine 2-D Point Space'' transition executed the 2-D space for the provers and verifiers in the proposed scheme, and ``Calculate Distance'' transition calculates the distance as a location between provers and verifiers.

\subsubsection{Generate Location Proof}

\begin{figure}[!h]
  \centerline{\includegraphics[width=8.5cm, height=6cm]{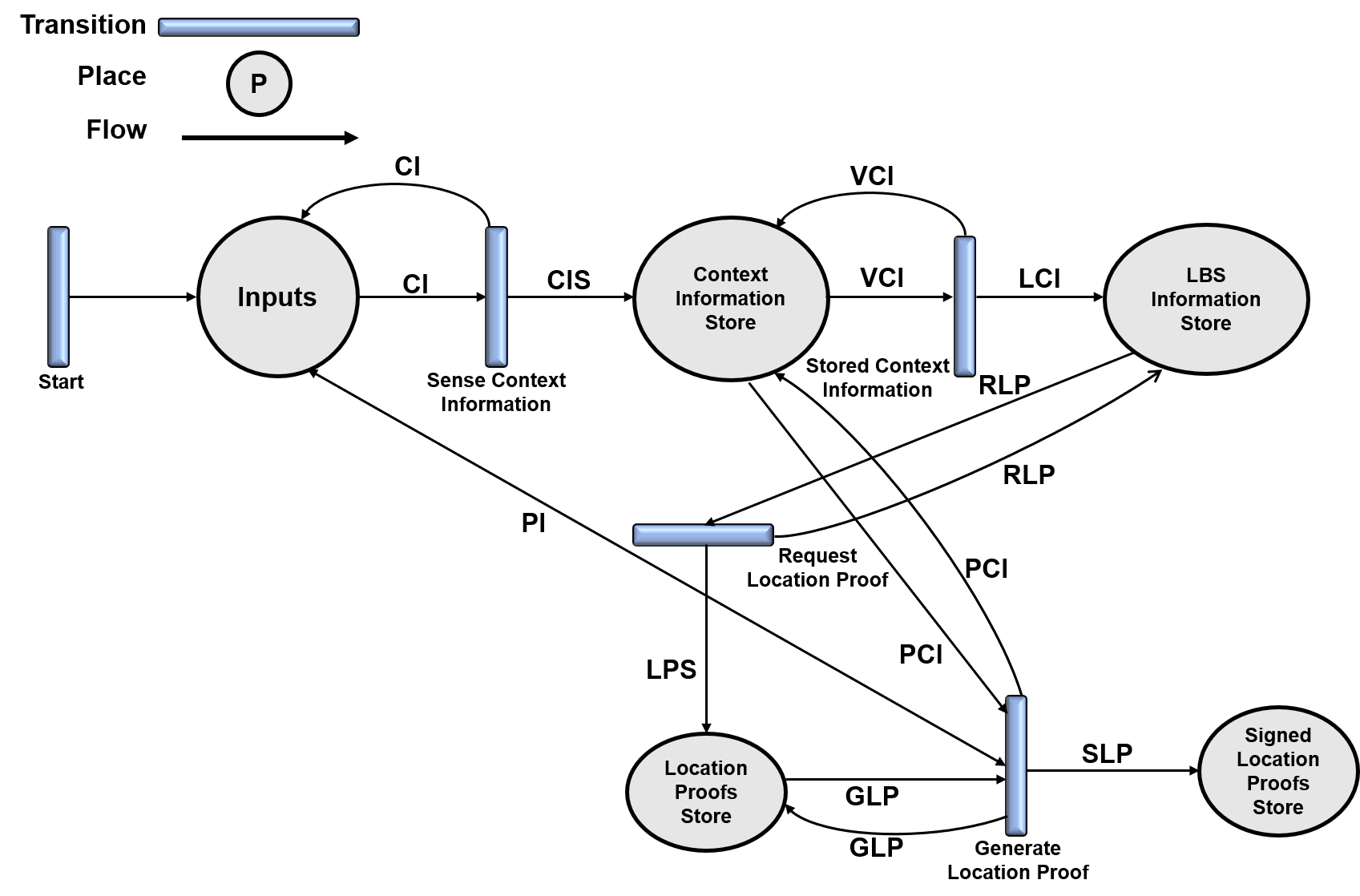}}
  \caption{Generate Location Proof- Petri Net}
  \label{petrinet6}
  \vspace{-4mm}%Put here to reduce too much white space after your table 
\end{figure}

The Petri net of the generate location proof in the proposed scheme is depicted in Fig. \ref{petrinet6}.

The HLPN consists of five places and five transitions. The places includes in the generate location proof HLPN are ``Inputs'', ``Context Information Store'', ``LBS Information Store'', ``Location Proofs Store'' and ``Signed Location Proofs Store'' while transitions are ``Start'', ``Sense Context Information'', ``Stored Context Information'', ``Request Location Proof'' and ``Generate Location Proof''. The location proof generation procedure starts with the ``Start'' transition, which forwards and store the context information, private key and hash function to the ``Inputs'' place. Following that, the ``Sense Context Information'' transition allows both provers and verifiers to sense the context information such as ID, time, location and activity from the deployed environment and store the context information to the ``Context Information Store''. Finally, the verifier performs an additional step in the ``Stored Context Information'' transition, which allows the verifiers to store the sensed context information in the ``LBS Information Store'' place. Besides, the verifiers request the provers with the location proofs performed with the ``Request Location Proof'' transition to send the locations proofs to verify their locations in the 2-D space and stored in the ``Location Proofs Store'' place. Upon receiving the requests from the verifiers at the prover side, the provers generated the location proofs by signing their sensed context information with their private key as generated in the ``Generate Location Proof'' transition and send and store in the ``Signed Location Proofs Store''.

\subsubsection{Verify Location Proof}

\begin{figure}[!h]
  \centerline{\includegraphics[width=8.5cm, height=6cm]{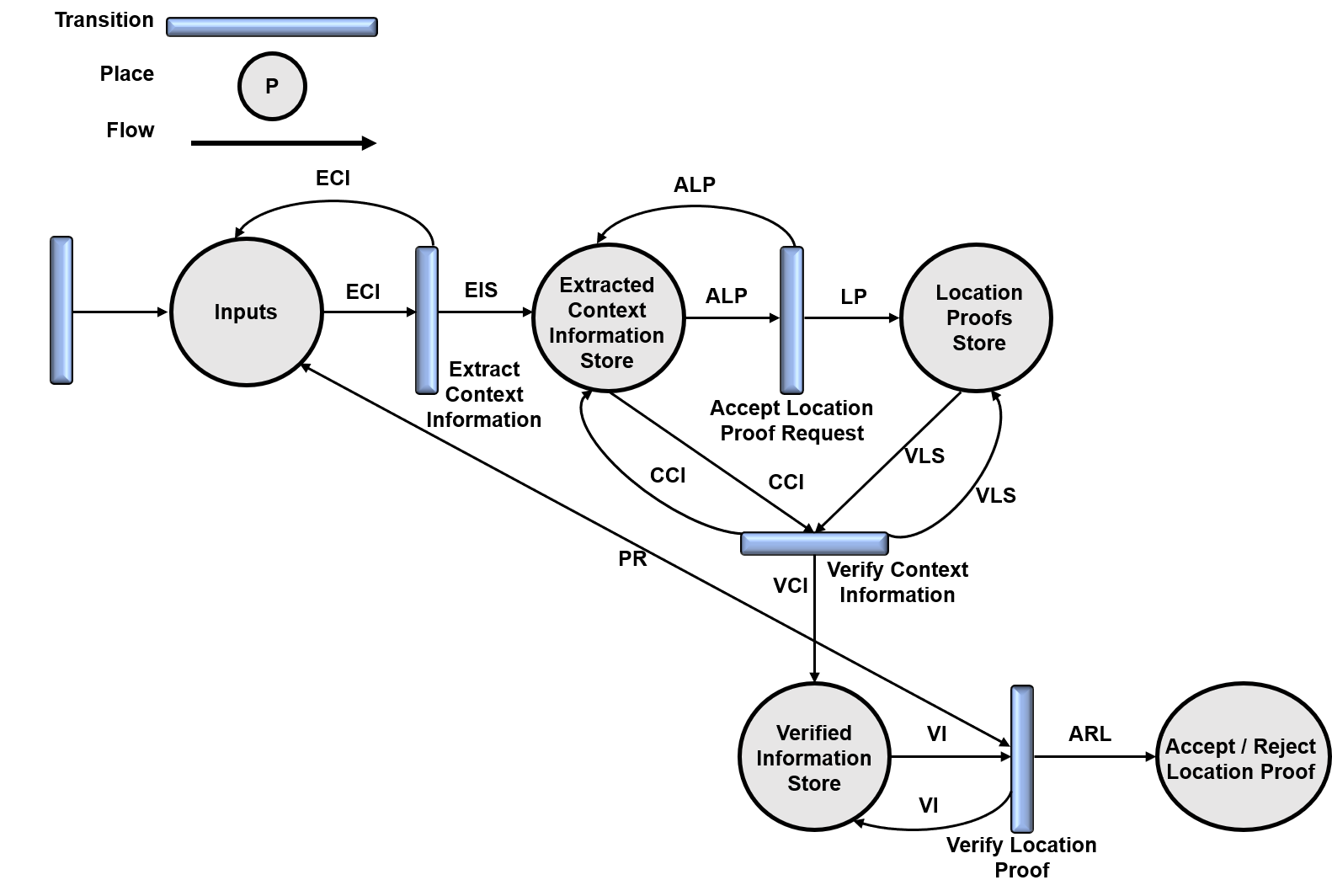}}
  \caption{Verify Location Proof- Petri Net}
  \label{petrinet7}
    \vspace{-4mm}%Put here to reduce too much white space after your table 
\end{figure}

The HLPN of the verifying location proofs in the proposed scheme is depicted in Fig. \ref{petrinet7}.

Similar to HLPN of the generate location proof, the HLPN of verifying location proof is also consists of five places and five transitions. The places includes in the HLPN are ``Inputs'', ``Extracted Context Information Store'', ``Location Proofs Store'', ``Verify Context Information'' and ``Verify Location Proof''. The transitions includes in the HLPN are ``Start'', ``Extract Context Information'', ``Accept Location Proof Request'',``Verify Context Information'' and ``Verify Location Proof''. The process of verifying location proof is initiated with the ``Start'' transition, which created the initial variables such as a set of provers' signatures and their public keys, and a set of verifiers and stored all of them in the ``Inputs'' place. Following the initiation of the verification process, the verifiers extract the context information from the LBS by using the ``Extract Context Information'' transition and stored this information in the ``Extracted Context Information Store'' place. Upon accepting the location proofs from the provers through the ``Accept Location Proof Requests'' transition, the verifiers stored the location proof requests to the ``Location Proof Store''. When the verifiers receive the proofs from the provers, the verifiers verify and match the context information in the ``Verify Context Information'' transition obtained from the provers with the context information from the  LBS and stored in the ``Verified Information Store''. Finally, the ``Verify Location Proof'' transition verify the location proof signatures with the respective public key and determines the acceptance or rejection of location proof in the ``Accept / Reject Location Proof''.

\begin{figure*}[t]
  \centerline{\includegraphics[width=16cm, height=7cm]{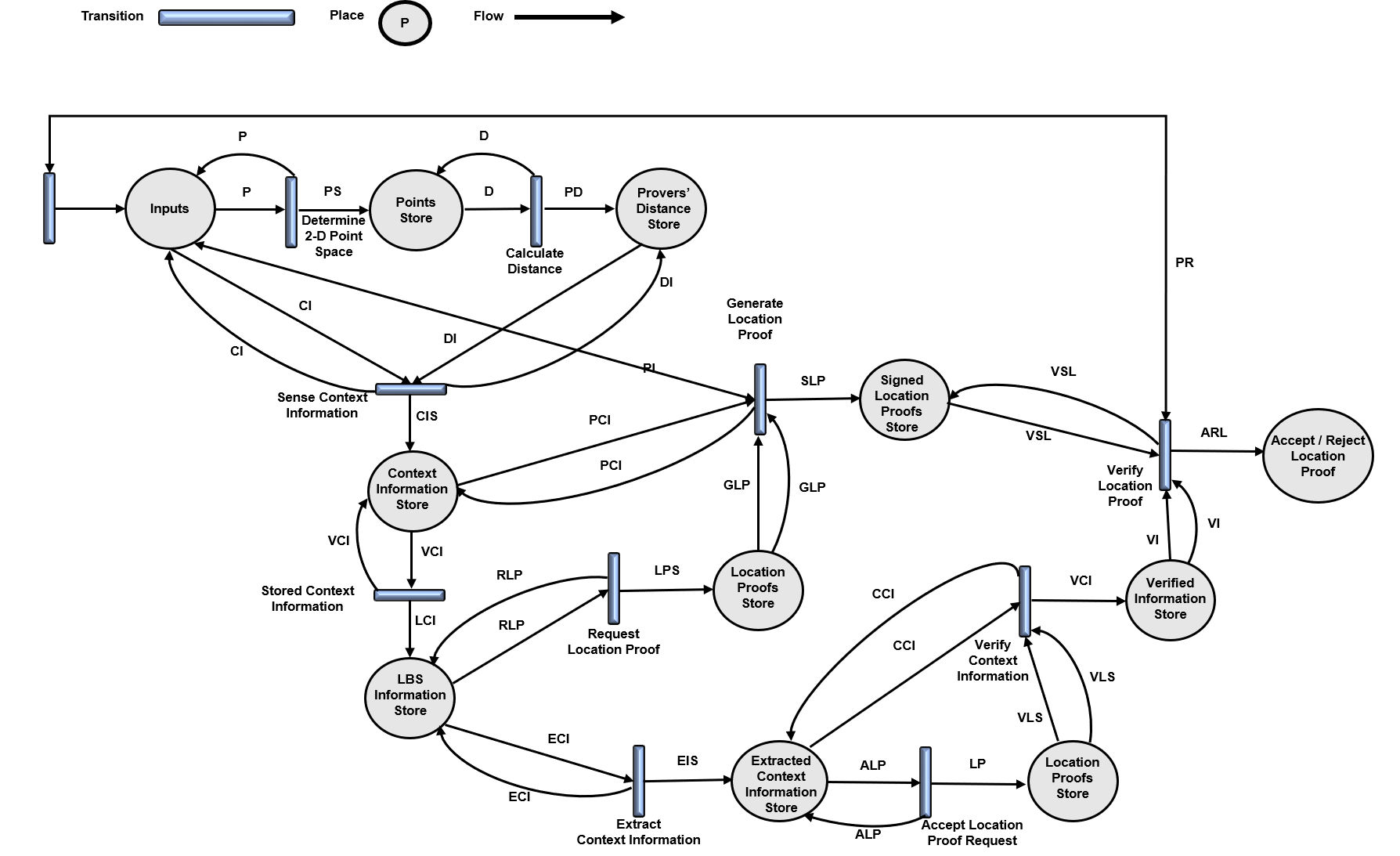}}
  \caption{A Complete HLPN of the LPS}
  \label{petrinet8}
  \vspace{-6mm}%Put here to reduce too much white space after your table 
\end{figure*}

The complete HLPN of the proposed clone node detection scheme on the base of LPS, including location calculation, generate location proof and verify location proof, is illustrated in Fig. \ref{petrinet8}. The working mechanism of each Petri net is described in detail using a set of rules from \ref{R12} - \ref{R21}.

\begin{table}[hbt!]
\tiny
\centering
\caption{Places, Mappings and Description of location proof system}
\label{newplaces}
%\resizebox{\columnwidth}{!}{%
\begin{adjustbox}{width=0.47\textwidth,center}
\begin{tabular}{|l|l|l|}
\hline

 \textbf{Places}      &    \textbf{Mappings}  & \textbf{Description}      \\ \hline
     
\multicolumn{3}{|l|}{\textbf{Calculate Location}} \\ \hline
    $\varphi$ (Inputs)       &  \textit{P}(\textit{$p_{1}$} $\times$ \textit{$p_{2}$} $\times$ \textit{$v_{1}$} $\times$ \textit{$v_{2}$}) &    \makecell{Holds the Coordinates of \\ provers and verifiers}   \\ \hline
    $\varphi$ (Points Store)    &   \textit{P}(\textit{$p_{1}$} $\times$ \textit{$p_{2}$} $\times$ \textit{$v_{1}$} $\times$ \textit{$v_{2}$}) &  \makecell{Holds the Coordinates of \\ provers and verifiers}   \\ \hline
    $\varphi$ (Provers' Distance Store)    &   \textit{P}(\textit{d})  & \makecell{Stores the provers' distances}    \\ \hline

\multicolumn{3}{|l|}{\textbf{Generate Location Proof}} \\ \hline
      $\varphi$ (Inputs)    & \makecell{\textit{P}(\textit{ID} $\times$ \textit{T} $\times$ \textit{Loc} $\times$ \textit{Actv} $\times$ \textit{$K_{Pr}$} \\ $\times$ \textit{H}} )       & \makecell{Holds the context information \\ variables, private keys, \\ and hash function}  \\ \hline
     $\varphi$    (Context Information Store)  & \textit{P}(\textit{CI}) & \makecell{Stores the context information \\ \textit{CI}} \\ \hline 
     
     $\varphi$    (LBS Information Store)  & \textit{P}(\textit{CI}) & \makecell{Stores the context information \\ \textit{CI}} \\ \hline 
     
     $\varphi$    (Location Proofs Store)  & \textit{P}(\textit{Pr}) & \makecell{Stores the location \\ proof requests} \\ \hline
     
     $\varphi$    (Signed Location Proofs Store)  & \textit{P}(\textit{$P_{sign}$} ) & \makecell{Stores the signed \\ location proofs} \\ \hline

\multicolumn{3}{|l|}{\textbf{Verify Location Proof}} \\ \hline
 $\varphi$ (Inputs)     &    \textit{P}($P_{sign}$ $\times$ \textit{$K_{Pb}$} $\times$ \textit{V})    & \makecell{Holds the signatures, \\public keys, and \\ list of verifiers} \\ \hline    

$\varphi$ (Extracted Context Information Store)     &   \textit{P}(\textit{CI}) & \makecell{Stores the extracted context \\ information } \\ \hline

$\varphi$  (Location Proofs Store)    &       \textit{P}(\textit{Pr})   & \makecell{Stores the location proofs} \\ \hline

 $\varphi$ (Verified Information Store)     &   \textit{P}(\textit{CI})  & \makecell{Stores the verified \\ context information}    \\ \hline

$\varphi$ (Accept / Reject Location Proofs Store)     &   \textit{P}(\textit{A} $\times$ \textit{R})  & \makecell{Stores individual \\ coordinate values}    \\ \hline

\end{tabular}
\end{adjustbox}
  \vspace{-4mm}%Put here to reduce too much white space after your table 

\end{table}

Table \ref{newplaces} specifies the places, mappings, and descriptions of each place used in LPS. The proposed scheme consists of a mechanism detecting the clone node attack in IoT networks. Thus, it includes several steps or actions, from location finding, location proof generation and location proof verification. Each action is associated with the algorithms specified for detecting clone nodes attacks, such as deploying nodes in the 2-D space, calculate the location of prover and verifier using the euclidean distance method, sensing context information from the deployed environment, storing context information on LBS, generate location proof and then verify location proof. For each Petri net, we specified different places to store the data or variables before or after performing the operations through transitions.

To calculate the Petri net, the inputs points store and provers' distance store are utilised. In inputs and point store places, the coordinates of provers and verifiers on 2-D space are stored. In the provers' distance store, the location of provers concerning verifiers in the LPS is stored. 

Generating the location proofs comprises four places: inputs, context information store, LBS information store, location proofs store, and signed location proofs store. The inputs place contains the context information such as ID, time, location, activity, private key of prover and hash function. The context information store and LBS information store both saves the context information sensed by the verifiers. All location proofs requests from the provers are stored at the location proofs store place. Finally, the signed location proofs are stored at the signed location proofs store place.

Like the generate location proof process, five places are utilised in the verify location proof process: inputs, extracted context information store, location proofs store, verified information store, and accept or reject location proofs store. The inputs place is utilised to store the signed location proofs, public keys and list of verifiers. The context information extracted from the LBS is stored on the extracted context information store place. All received location proofs received by the verifiers are stored in the location proofs store place. Further, verified context information is stored at the verified information store place. Finally, the decision about acceptance or rejection of location proofs is stored in the accept/reject location proofs store. 

Table \ref{newdatatypes} lists the data types used in the LPS of the proposed clone node attack detection scheme.

 \begin{table}[!h]
\tiny
\centering
\caption{Data Types used in LPS}
\label{newdatatypes}
\begin{adjustbox}{width=0.47\textwidth,center}
\begin{tabular}{|c|c|}
\hline
\textbf{\makecell{Data \\ Types}} & \textbf{Description}  \\ \hline

\textit{$p_{1}$} &  \makecell{A float type for the representation of x-coordinate of a prover} \\ \hline
\textit{$p_{2}$} & \makecell{A float type for the representation of y-coordinate of a prover}   \\ \hline
\textit{$v_{1}$} & \makecell{A float type for the representation of x-coordinate of a verifier}   \\ \hline
\textit{$v_{2}$} & \makecell{A float type for the representation of y-coordinate of a prover}
   \\ \hline
\textit{d} &  \makecell{A float type for the representation of location of a prover}  \\ \hline
\textit{ID} & \makecell{An integer type for the representation of unique identification}   \\ \hline
\textit{T} & \makecell{An integer type for the representation of data sensing time}   \\ \hline
\textit{Loc} & \makecell{A float type for the representation of location in 2-D space} \\ \hline
\textit{Actv} & \makecell{A string type for the representation of provers' activities} \\ \hline
\textit{$K_{Pr}$} & \makecell{A byte type for the representation of private key of prover} \\ \hline
\textit{H} & \makecell{A byte type for the representation of hash value} \\ \hline
\textit{CI} & \makecell{A byte type for the representation of context information} \\ \hline
\textit{Pr} & \makecell{A byte type for the representation of location proof request} \\ \hline
\textit{$P_{sign}$} & \makecell{A byte type for the representation location proof signature} \\ \hline
\textit{$K_{Pb}$} & \makecell{A byte type for the representation of public key of prover} \\ \hline
\textit{V} & \makecell{A byte type for the representation of verifiers list} \\ \hline
\textit{$A$} & \makecell{A string type for the representation of accepted decision} \\ \hline
\textit{$R$} & \makecell{A string type for the representation of rejected decision} \\ \hline

\end{tabular}
\end{adjustbox}
\vspace{-4mm}%Put here to reduce too much white space after your table 

\end{table}

\subsection{Z Specification Rules} \label{rules}

This section presented a set of rules consistent with the syntax and semantics of the Z language for specifying the interaction behaviours of the proposed scheme's components. As previously stated, the proposed scheme is modelled into two parts, each illustrated with its HLPN. The rules \ref{R1} - \ref{R11} specify the modelling part of an ECDSA*, while rules \ref{R12} - \ref{R21} specify the modelling part of a LPS.

An ECDSA* process begins with the generation of cryptographic keys, both public and private. To create the keys, the process accepts a series of standard key parameters referred to as domain parameters, which include \{ \textit{p}, \textit{E}, \textit{P}, \textit{n}, \textit{h}\}. These parameters are defined in the following order: \textit{p} denotes the order in which the prime field exists, \textit{E} denotes an elliptic curve formed over the prime field, \textit{P} denotes a non-zero random base point in \textit{E}, \textit{n} denotes the ordinal value of \textit{P}, which is typically a prime integer, and \textit{h} denotes a co-factor. The entire process of generating domain parameters is demonstrated in Rule \ref{R1}, namely the ``Generate Domain Parameters'' function.

\begin{equation}
\begin{aligned}
\label{R1}
\textbf{R}(\textrm{Generate Domain Parameters}) = \forall dp \in DP, \forall gdp \in  GDP | \\
gdp[1] := \textrm{prime.field.order(dp)} \land \\
gdp[2] := \textrm{elliptic.curve(dp)}\land \\
gdp[3] := \textrm{base.point(dp)}\land \\
gdp[4] := \textrm{ordinal.value(dp)}\land \\
gdp[5] := \textrm{cofactor(dp)}\land \\
gdp[6] := \textrm{Generate Domain Parameters} (gdp[1], gdp[2], gdp[3], \\ gdp[4],gdp[5]) \land \\
GDP'= GDP \cup \left\{gdp[1], gdp[2], gdp[3], gdp[4], gdp[5], gdp[6]\right\}
\end{aligned}
\end{equation}

The procedure for generating keys, such as public and private keys, is described in Rule \ref{R2}. Key generation procedure generates a pair of public and private keys for use in the signing and verification processes. For example, the private key can be generated by selecting a random integer \textit{d} and multiplying it by a non-zero random base point \textit{P} to get the public key \textit{Q}. The ``Generate Keys'' function receives the domain parameters as input and returns the private and public keys to sign and verify signatures.
  
\begin{equation}
\label{R2}
\begin{aligned}
\textbf{R}(\textrm{Generate Keys}) = \forall dk \in DK, \forall gk \in  GK | \\
gk[1] := \textrm{generate.private.key(dk)} \land \\
gk[2] := \textrm{generate.public.key(dk)} \land \\
gk[3] := \textrm{Generate Keys} (gk[1], gk[2]) \land \\
GK'= GK \cup \left\{gk[1], gk[2], gk[3]\right\}
\end{aligned}
\end{equation}

The function ``Compute Coordinates'' in Rule \ref{R3} defines and computes the coordinates that can be used in the signature generation process. This procedure begins by selecting a random integer for the \textit{k} parameter. The coordinates \textit{X} are then calculated by multiplying the random number \textit{k} by the random point \textit{P}.

\begin{equation}
\label{R3}
\begin{aligned}
\textbf{R}(\textrm{Compute Coordinates}) = \forall gdp \in GDP, \forall k \in  K, \forall c \in C | \\
c[1] := \textrm{random.integer(k)} \land \\
c[2] := \textrm{base.point(gdp[3])} \land \\
c[3] := \textrm{Compute Coordinates} (c[1], c[2]) \land \\
C'= C \cup \left\{c[1], c[2], c[3]\right\}
\end{aligned}
\end{equation}

After computing the coordinates, the function ``Compute Hash'' acts as a hash function \textit{H}, taking the message \textit{m} and producing a hash value in the form of a digest string value, which is then transformed into an integer \textit{e}, as shown in Rule \ref{R4}.

\begin{equation}
\label{R4}
\begin{aligned}
\textbf{R}(\textrm{Compute Hash}) = \forall h \in H, \forall e \in  E| \\
e[1] := \textrm{message(h)} \land \\
e[2] := \textrm{compute.hash(e[1])} \land \\
E'= E \cup \left\{e[1], e[2]\right\}
\end{aligned}
\end{equation}

The Rule \ref{R5} outlines the procedure of generating signature pair 1, which is performed by the function ``Generate Signature Pair 1''. This function calculates the signature parameter \textit{r} by modifying the previously calculated \textit{x} in the range of total numbers \textit{n}.

\begin{equation}
\label{R5}
\begin{aligned}
\textbf{R}(\textrm{Generate Signature Pair 1}) = \forall r \in R, \forall sr \in SR | \\
sr[1] := \textrm{mod(r)} \land \\
sr[2] := \textrm{mod(r)} \land \\
sr[3] := \textrm{Generate Signature Pair 1 (sr[1], sr[2])} \land \\
SR'= SR \cup \left\{sr[1], sr[2], sr[3]\right\}
\end{aligned}
\end{equation}

The Rule \ref{R6} computes the function ``Generate Signature Pair 2'', which outlines the process of calculating the second signature pair \textit{s} by taking the inverse of the \textit{k} random integers and multiplying the sum of the integer \textit{e} and the private key \textit{d} by \textit{r}. Finally, the signature is stored with both the signature values computed in Rules \ref{R5} and \ref{R6}. Afterwards, the signature is then sent to the verifier's party for verification.

\begin{equation}
\label{R6}
\begin{aligned}
\textbf{R}(\textrm{Generate Signature Pair 2}) = \forall k \in K, \forall gk \in GK,\forall prk \in PRK, \\ \forall s \in S, \forall SS \in S | \\
ss[1] := \textrm{private.key(gk[1])} \land \\
ss[2] := \textrm{random.integer(k[1])} \land \\
ss[3] := \textrm{hash.integer(s)} \land \\
ss[4] := \textrm{Generate Signature Pair 1(ss[1], ss[2], ss[3])} \land \\
SS'= SS \cup \left\{ss[1], ss[2], ss[3], ss[4]\right\}
\end{aligned}
\end{equation}

The process of signature verification begins with extracting the signatures computed in the function ``Get Signature Integers'', such as \textit{r} and \textit{s}, from the signatures store generated in the Rules \ref{R5} and \ref{R6}. The procedure of determining whether the signature values \textit{r} and \textit{s} are within the interval. The entire procedure for obtaining the signature integers is explained in Rule \ref{R7}.

\begin{equation}
\label{R7}
\begin{aligned}
\textbf{R}(\textrm{Get Signature Integers}) = \forall si \in SI, \forall sig \in SIG | \\
sig[1] := \textrm{signature.integer.1(si)} \land \\
sig[2] := \textrm{signature.integer.2(si)} \land \\
sig[3] := \textrm{Generate Signature Integers(sig[1], sig[2])} \land \\
SIG'= SIG \cup \left\{sig[1], sig[2], sig[3]\right\}
\end{aligned}
\end{equation}

After obtaining the signature integers for verification, the hash function \textit{H} computes the hash value of the message \textit{m} for comparison. Then, as with the hash creation in Rule \ref{R4}, the hash value is transformed to an integer \textit{e} using the function ``Compute Hash'', as computed in Rule \ref{R8}.

\begin{equation}
\label{R8}
\begin{aligned}
\textbf{R}(\textrm{Compute Hash}) = \forall m \in M, \forall hi \in HI | \\
hi[1] := \textrm{message(m)} \land \\
hi[2] := \textrm{Compute Hash(hi[1])} \land \\
HI'= HI \cup \left\{hi[1], hi[2]\right\}
\end{aligned}
\end{equation}

In the function ``Calculate Point'', Rule \ref{R9} provides the procedure for calculating the point integer. By using the modulus of the inverse value of the signature \textit{s}, the algorithm calculated the point \textit{w}.

\begin{equation}
\label{R9}
\begin{aligned}
\textbf{R}(\textrm{Calculate Point}) = \forall sp \in SP, \forall cp \in CP | \\
cp[1] := \textrm{get.integer.point(sp)} \land \\
cp[2] := \textrm{Calculate Point(cp[1])} \land \\
CP'= CP \cup \left\{cp[1], cp[2]\right\}
\end{aligned}
\end{equation}
Rule \ref{R10} outlines the procedure for computing coordinates by calculating the integer value \textit{w}. \textit{$u_{1}$} and \textit{$u_{2}$} are determined by multiplying the numbers \textit{e} and \textit{r} by the value \textit{w}, respectively, using the function ``Compute Coordinates''.

\begin{equation}
\label{R10}
\begin{aligned}
\textbf{R}(\textrm{Compute Coordinates}) = \forall he \in HE, \forall w \in W, rp \in RP, cc \in CC |\\
cc[1] := \textrm{integer.point.1(he)} \land \\
cc[2] := \textrm{integer.point.2(w)} \land \\
cc[3] := \textrm{coordinate.point(rp)} \land \\
cc[4] := \textrm{Compute Coordinates (cc[1], cc[2], cc[3])} \land \\
CC'= CC \cup \left\{cc[1], cc[2], cc[3], cc[4]\right\}
\end{aligned}
\end{equation}

Finally, Rule \ref{R11} calculates the \textit{X} value by multiplying \textit{P} and \textit{Q} by the determined coordinates \textit{$u_{1}$} and \textit{$u_{2}$}, as in the preceding Rule \ref{R10}. For signature verification, the function ``Verify Signatures'' is computed and it is determined whether \textit{X} = $O$ indicates that the signature should be rejected or accepted.

\begin{equation}
\label{R11}
\begin{aligned}
\textbf{R}(\textrm{Verify Signatures}) = \forall cci \in CCI, \forall o \in O, \forall pq \in PQ, ds \in DS | \\
ds[1] := \textrm{point.1(pq)} \land \\
ds[2] := \textrm{point.1(cci)} \land \\
ds[3] := \textrm{calculate.point.1(ds[1], ds[2])} \land \\
ds[4] := \textrm{point2 (pq)} \land \\
ds[5] := \textrm{point2 (cci)} \land \\
ds[6] := \textrm{calculate.point.2(ds[4], ds[5])} \land \\
ds[7] := \textrm{Verify Signatures (ds[3], ds[6])} \land \\
ds[7] = o[1] \rightarrow \textrm{Not Verified} \land  DS'= DS \cup \{ds[1], ds[2], ds[3], \\ ds[4], ds[5], ds[6]\} \lor\\
ds[7] \neq o[1] \rightarrow \textrm{Verified} \land  DS'= DS \cup \{ds[1], ds[2], ds[3], \\ ds[4], ds[5], ds[6]\}\\
\end{aligned}
\end{equation}

The process of detecting clone node attacks on IoT devices in a LPS begins with deploying provers and verifiers in their defined locations inside the mobile network. By measuring the distance between the prover and verifiers in 2-D space, we estimate the location of each device (such as the prover and verifier) as a key element of context information in our proposed location proof framework model. In 2-D space, a prover is represented by \textit{P} = \{$x_{1}$, $y_{1}$\}, whereas a verifier is represented by \textit{V} = \{$x_{1}$, $y_{1}$\}. The process for identifying the coordinates of both the prover and the verifiers in 2-D space is computed using the function ``Determine 2-D Point Space'' as described in Rule \ref{R12}.

\begin{equation}
\label{R12}
\begin{aligned}
\textbf{R}(\textrm{Determine 2-D Point Space}) = \forall p \in P, \forall ps \in PS | \\
ps[1] := \textrm{prover.coordinate.1(p)} \land \\
ps[2] := \textrm{prover.coordinate.2(p)} \land \\
ps[3] := \textrm{verifier.coordinate.1(p)} \land \\
ps[4] := \textrm{verifier.coordinate.2(p)} \land \\
ps[5] := \textrm{Determine 2-D Point Space(ps[1],ps[2],ps[3],ps[4])} \land \\
PS'= PS \cup \left\{ps[1], ps[2], ps[3], ps[4], ps[5]\right\}\\
\end{aligned}
\end{equation}

We used the Euclidean distance algorithm to determine the location of each prover concerning the verifiers based on their estimated distances. The $d_{\left(V, P\right)}$ denotes the distance between verifier and prover. The Euclidean distance calculation procedure begins by taking the prover and verifier's coordinates in 2-D space as input values, such as \textit{P} = \{$x_{1}$, $y_{1}$\} and \textit{V} = \{$x_{1}$, $y_{1}$\}. It determines the distance between each verifier \{$Ver_1$, $Ver_2$, $Ver_3$, $\dots$, $Ver_n$\} by calculating the square root of the difference in the coordinates of each verifier and prover pair. Following the calculation of the distance between their provers, each verifier keeps a list of the euclidean distances between their provers as determining positions at specific points, such as \{$d_{P1}$, $d_{P2}$, $d_{P3}$, $\dots$, $d_{Pn}$\}. The complete process of computing the distance between provers and verifiers is accomplished through the use of the function ``Calculate Distance'', which is stated in Rule \ref{R13}.

\begin{equation}
\label{R13}
\begin{aligned}
\textbf{R}(\textrm{Calculate Distance}) = \forall d \in D, \forall pd \in PD | \\
pd[1] := \textrm{prover.coordinate(d)} \land \\
pd[2] := \textrm{verifier.coordinate(d)} \land \\
pd[3] := \textrm{Calculate Location(pd[1],pd[2])} \land \\
PD'= PD \cup \left\{pd[1], pd[2], pd[3]\right\}\\
\end{aligned}
\end{equation}

After deploying provers and verifiers in 2-D space, both provers and verifiers sense contextual information about the deployed environment. The context information contains the IoT device's identifier \textit{ID}, the data sensing time \textit{T}, the device's accurate position \textit{Loc}, and its activity \textit{Actv}. The activity of an IoT device can take any form, including monitoring, detecting, or transmitting at a certain time \textit{T}. The function ``Sense Context Information'' is responsible for perceiving the context information, as described in Rule \ref{R14}.
\begin{equation}
\label{R14}
\begin{aligned}
\textbf{R}(\textrm{Sense Context Information}) = \forall ci \in CI, \forall pd \in PD, \forall cis \in CIS | \\
cis[1] := \textrm{id(ci)} \land \\
cis[2] := \textrm{time(ci)} \land \\
cis[3] := \textrm{location(pd[3])} \land \\
cis[4] := \textrm{activity(ci)} \land \\
cis[5] := \textrm{Sense Context Information(cis[1], cis[2], cis[3], cis[4])} \land \\
CIS'= CIS \cup \left\{cis[1], cis[2], cis[3], cis[4], cis[5]\right\}\\
\end{aligned}
\end{equation}

Rule \ref{R15} computes the function ``Stored Context Information'', which denotes the process of storing context information at location-based services (LBS). The combination of these pieces of information is referred to as \textit{CI}, and it is maintained and stored at LBS in the form of a list as \{$CI_{1}$, $CI_{2}$, $CI_{3}$, \dots, $CI_{n}$\}.

\begin{equation}
\label{R15}
\begin{aligned}
\textbf{R}(\textrm{Stored Context Information}) = \forall vci \in VCI, \forall cis \in CIS, \\ \forall lci \in LCI | \\
lci[1] := \textrm{extract.context.information(cis[5])} \land \\
lci[2] := \textrm{store.context.information(vci)} \land \\
lci[3] := \textrm{Stored Context Information(lci[1], lci[2])} \land \\
LCI'= LCI \cup \left\{lci[1], lci[2], lci[3]\right\}\\
\end{aligned}
\end{equation}

After sensing and storing context information on the LBS, the verifier generates and sends a location proof to the prover in order to authenticate its location and determine whether or not the prover has been compromised. To build a location proof, the verifier first requests that the prover generate a proof using sensed context information such as \textit{CI}. The procedure for requesting location proofs from verifiers is defined in the Rule a verifier first requests that the prover generate a proof using sensed context information such as \textit{CI} and is determined using the function ``Request Location Proof''.

\begin{equation}
\label{R16}
\begin{aligned}
\textbf{R}(\textrm{Request Location Proof}) = \forall rlp \in RLP, \forall lps \in LPS | \\
lps[1] := \textrm{extract.context.information(rlp)} \land \\
lps[2] := \textrm{Request Location Proof (lps[1])} \land \\
LPS'= LPS \cup \left\{lps[1], lps[2]\right\}\\
\end{aligned}
\end{equation}

The Rule \ref{R17} details the procedure for generating location proofs, which is accomplished through the use of the function ``Generate Location Proof''. First, the proof is generated by hashing and signing the \textit{CI} context information with the prover's private key $K_{Pr}$. Next, the signature $P_{sign}$ is generated using the ECDSA* technique for signature generation.

\begin{equation}
\label{R17}
\begin{aligned}
\textbf{R}(\textrm{Generate Location Proof}) = \forall pi \in PI, \forall pci \in PCI, \forall glp \in GLP, \\ \forall slp \in SLP, | \\
slp[1] := \textrm{location.proof(glp)} \land \\
slp[2] := \textrm{extract.context.information(pci)} \land \\
slp[3] := \textrm{private.key(pi)} \land \\
slp[4] := \textrm{hash(pi)} \land \\
slp[5] := \textrm{Generate Location Proof (slp[1], slp[2], slp[3], slp[4])} \land \\
SLP'= SLP \cup \left\{slp[1], slp[2], slp[3], slp[4], slp[5]\right\}\\
\end{aligned}
\end{equation}

After the provers in the LPS generate the location proofs, the next task is to verify them at the verifier's side to assess the clone node attack. This document outlines the process of evaluating location proofs for IoT devices that claim to be at a specific place using context information \textit{CI}. In addition, the Rule \ref{R18} outlines the process for extracting context information from the LBS using the function ``Extract Context Information''.

\begin{equation}
\label{R18}
\begin{aligned}
\textbf{R}(\textrm{Extract Context Information}) = \forall eci \in ECI, \forall rlp \in RLP,  \\ \forall eis \in EIS | \\
eis[1] := \textrm{extract.context.information(rlp)} \land \\
eis[2] := \textrm{extract.context.information(eci)} \land \\
eis[3] := \textrm{Extract Context Information (eis[1], eis[2])} \land \\
EIS'= EIS \cup \left\{eis[1], eis[2], eis[3]\right\}\\
\end{aligned}
\end{equation}

After gathering information from the LBS, the verifier takes location proofs from provers and verifies them for clone node detection. Accepting location proofs from provers is stated in the Rule \ref{R19} via the function ``Accept Location Proof Request''.

\begin{equation}
\label{R19}
\begin{aligned}
\textbf{R}(\textrm{Accept Location Proof Request}) = \forall alp \in ALP, \forall lp \in LP | \\
lp[1] := \textrm{location.proof.request(alp)} \land \\
lp[2] := \textrm{Accept Location Proof Request (lp[1], lp[2])} \land \\
LP'= LP \cup \left\{lp[1], lp[2]\right\}\\
\end{aligned}
\end{equation}

The procedure of verifying the context information obtained in the accepting location proof (Rule \ref{R19} and Rule \ref{R18}) with the extracted information from LBS is described in Rule \ref{R20}. If the information is successfully matched using the ``Verify Context Information'' function, the verifier proceeds with the accepted location proof and performs the actual location proof-verification task.

\begin{equation}
\label{R20}
\begin{aligned}
\textbf{R}(\textrm{Verify Context Information}) = \forall cci \in CCI, \forall vls \in VLS, \\ \forall vci \in VCI | \\
vci[1] := \textrm{location.proof.request(vls)} \land \\
vci[2] := \textrm{extracted.context.information(cci)} \land \\
vci[3] := \textrm{Verify Context Information (vci[1], vci[2])} \land \\
VCI'= VCI \cup \left\{vci[1], vci[2], vci[3]\right\}\\
\end{aligned}
\end{equation}

The verification process begins by obtaining inputs such as the provers' signatures in the structure \{$P_{1_sign}$, $P_{2_sign}$, $P_{3_sign}$, \dots, $P_{n_sign}$\}, as well as their respective public keys in the structure \{$K_{1_Pb}$, $K_{2_Pb}$, $K_{3_Pb}$, \dots, $K_{n_Pb}$\}. To validate the prover's position proof, the verifiers \{$Ver_1$, $Ver_2$, $Ver_3$, \dots, $Ver_n$\} evaluate the contextual information \textit{CI} from the LBS and execute ECDSA* batch verification on the signatures after confirming the availability of stored information on the LBS. Since the batch verification process is carried out by several verifiers selected according to the trust model, the LBS manages and controls the list of verifiers. The verifiers validated the signature $P_{sign}$ using the prover's public key $K_{Pb}$. After successfully confirming the signatures collected from each designated verifier, the verifier will confirm the IoT device's validity in the network and accept the proof with location confirmation and other credentials. If the signature cannot be verified correctly, the verifier notifies the LBS of the prover's compromise in the IoT network. The function "Verify Location Proof" is responsible for the entire verification process of location proofs, as stated in the Rule \ref{R21}.

\begin{equation}
\label{R21}
\begin{aligned}
\textbf{R}(\textrm{Verify Location Proof}) = \forall pr \in PR, \forall vi \in VI, \\ \forall arl \in ARL | \\
arl[1] := \textrm{verified.information(vi)} \land \\
arl[2] := \textrm{public.key(pr)} \land \\
arl[3] := \textrm{Verify Location Proof (arl[1], arl[2])} \land \\
arl[3] \neq arl[1] \rightarrow \textrm{Not Verified } \land  \\ ARL'= ARL \cup \left\{arl[1], arl[2], arl[3]\right\} \lor \\
arl[3] = arl[1] \rightarrow \textrm{Verified } \land \\ ARL'= ARL \cup \left\{arl[1], arl[2], arl[3]\right\}\\
\end{aligned}
\end{equation}

\subsection{Analysis} \label{analysis}

We modelled the ECDSA* and LPS techniques in the proposed clone node attack detection scheme using the PIPE+ tool \cite{dingle2009pipe2, bonet2007pipe}. PIPE+ is a commonly used modelling tool for developing and analysing Petri nets built using the bounded model checking (BMC) technique. A model is a network of vertices and edges containing two distinct nodes (or components): places and transitions. \textit{P} places are denoted by circles that contain tokens representing the state of the modelled system, whilst \textit{T} transitions are denoted by bars that reflect the events to determine how tokens ``flow'' inside the Petri Net.

Transitions are further subdivided into two categories: timed and immediate. When events are connected to delay times, a timed transition \textit{$T_{t}$} is created, which is represented as a white transition. When no delay period exists, an immediate transition \textit{$T_{i}$} occurs, as indicated by a black transition.

To facilitate understanding and application of our proposed scheme from a modelling standpoint, we model each component (or algorithms) of techniques (e.g., an ECDSA* and LPS) separately. 

Please refer to \ref{HLPNsModels} for Pipe+ illustrations of each HLPN model designed for each component. The illustrations of HLPN models for both techniques demonstrated the complete execution of the specified rules and parameters due to the HLPNs' formal modelling.
 
\begin{figure*}[!t]
  \centerline{\includegraphics[width=14cm, height=6cm]{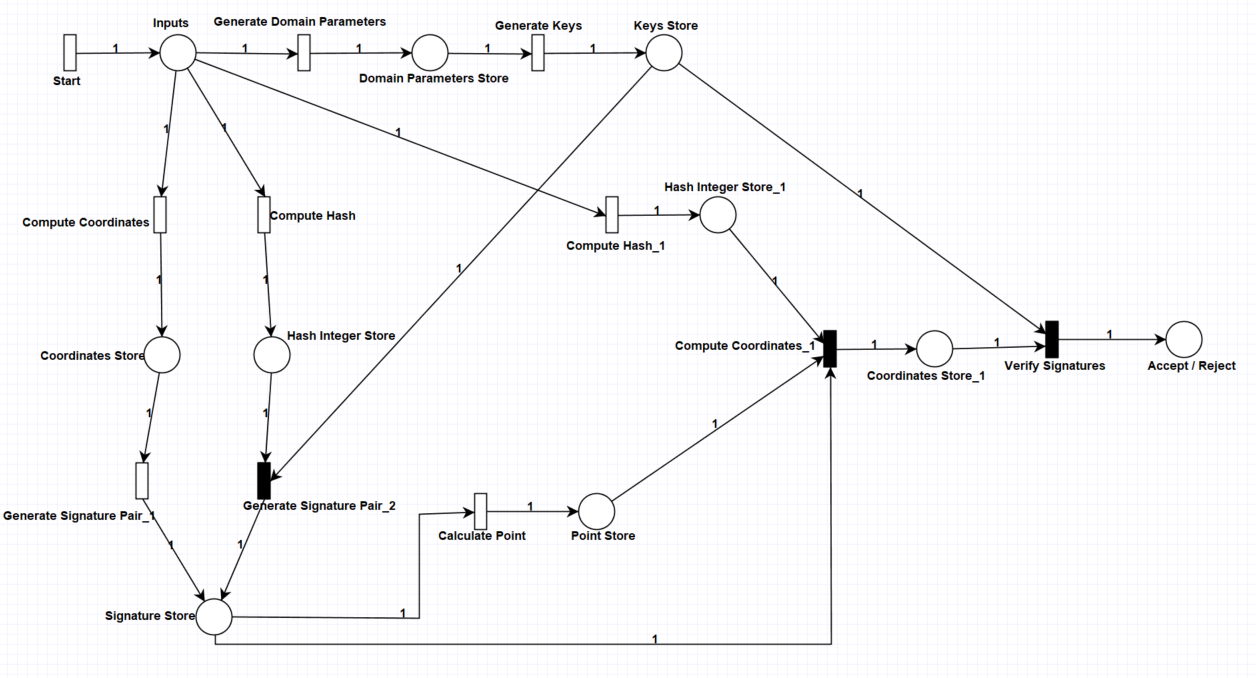}}
  \caption{A Pipe+ View of Complete HLPN of an ECDSA* Scheme}
  \label{pipe4}
    \vspace{-4mm}%Put here to reduce too much white space after your table 

\end{figure*}

\begin{figure*}[!hbt]
  \centerline{\includegraphics[width=18cm, height=6cm]{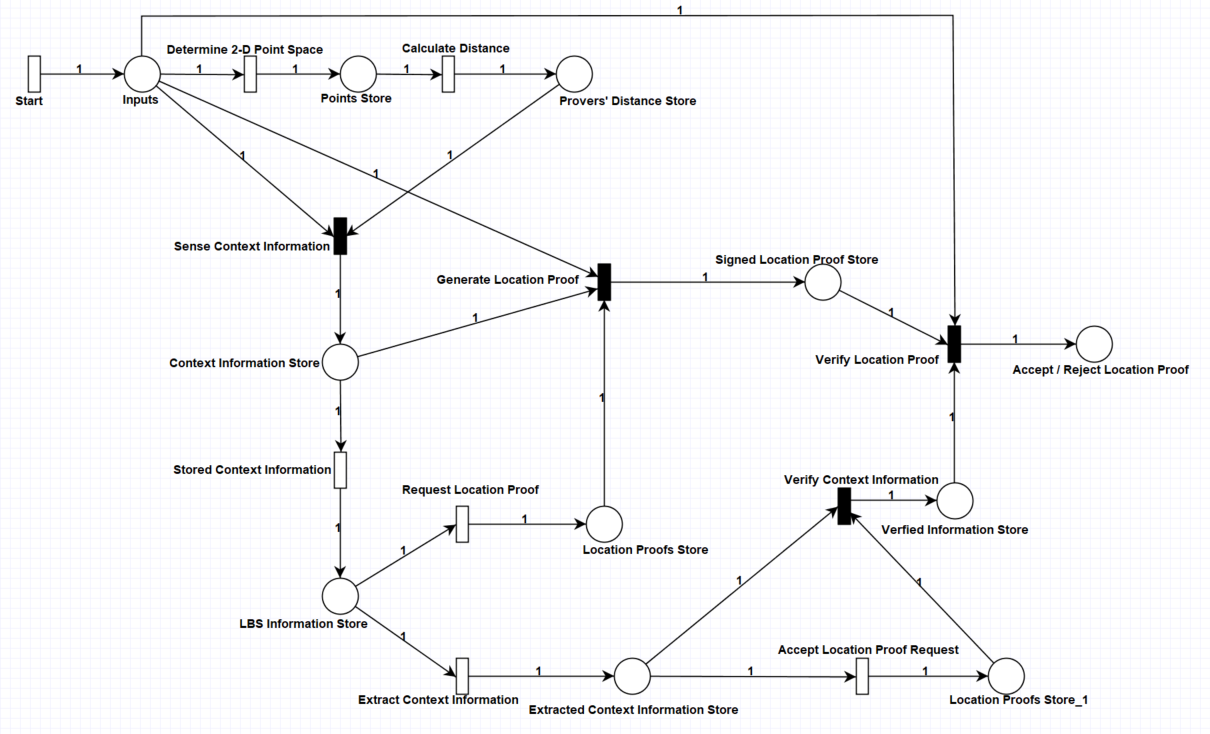}}
  \caption{A Pipe+ View of Complete HLPN of LPS}
  \label{pipe8}
    \vspace{-4mm}%Put here to reduce too much white space after your table 

\end{figure*}

However, to provide users with an idea of the HLPN models, we have only included the complete Pipe+ illustrations of these HLPN models. Figures \ref{pipe4} and Fig. \ref{pipe8} illustrate the entire Pipe+ view of HLPN models for ECDSA* and LPS methods, respectively.

%Fig. \ref{pipe8} depicts the whole HLPN model of the LPS in the proposed clone node attack detection scheme, which comprises 11 places and 11 transitions. 

%The places are as follows: inputs, points store, provers' distance store, context information store, LBS information store, extracted context information store, location proofs store, signed location proof store, location proof store 1, verified information store, and accept / reject location proof. Places are represented as \textit{P} = \{ ``Inputs'', ``Points Store'', ``Provers' Distance Store'', ``Context Information Store'', ``LBS Information Store'', ``Extracted Context Information Store'', ``Location Proofs Store'',``Signed Location Proofs Store'',``Location Proofs Store 1'',``Accept / Reject Location Proof''\}. Transitions such as start, determine 2-D point space, calculate distance, store context information, extract context information, request location proof, and accept location proof request are used as timed transitions and are denoted by \textit{$T_{t}$} = \{ ``Start'', ``Determine 2-D Point Store'', ``Calculate Distance'', ``Store Context Information'', ``Extract Context Information'',``Request Location Proof'', ``Accept Location Proof Request''\} whereas the transitions referred to as sense context information, generate location proof, verify context information, and verify location proof are used as immediate transitions and are denoted by \textit{$T_{i}$} = \{ ``Sense Context Information'', ``Generate Location Proof'',``Verify Context Information'',``Verify Location Proof''\}.

We employed two evaluation factors, incidence marking and confidence intervals (minimum and maximum threshold values), to obtain simulation results for these HLPN models, as discussed in the following sections \ref{inci} and \ref{confi}, respectively.

\subsubsection{Incidence Marking}\label{inci}

In incidence marking, we compute the various types of matrices, including forwards, backwards, combined, and inhibition, to define the link between places and transitions. To specify the relationship between places and transitions in HLPNs, forwards and backwards incidence matrices are employed. The arc between points \textit{P} and \textit{T} leads to value (\textit{P}, \textit{T}) $>$ 0 only if the desired input is transmitted from place to transition. The matrix of forwards incidences represents the total number of tokens generated from places and fires in the transition direction. In comparison, the backwards incidence matrix represents the total number of tokens available at the place \textit{P} that used to enable the transaction \textit{T}. The 1’s indicate interactions between places and transitions in forwards and backwards incidence matrices, whereas the 0’s indicate no interaction between places and transitions.

The combined incidence matrix  \textit{I} integrates the incidence results, such as 0’s, 1’s and -1’s. For example, both 0’s and 1’s represent the forwards and backwards incidence matrices in terms of summarising the token created at numerous places and determining their difference. However, -1 indicates the direction in which the relation between the transition and the place is backwards. Finally, we calculated the inhibition matrix \textit{H} of the Petri Nets presented, which characterises any weak connections between the points in the given graph. In our example, the Os in the inhibition matrix denotes all places’ interconnectivity, thereby validating the correctness and reachability of the techniques in our proposed scheme.

The Petri nets, in combination with the analysis of the results of the preceding matrices, are used to simplify the interpretation of the models and to aid in further improving the verifiability and correctness of any simulation. Additionally, these matrices can achieve the most accurate and trustworthy results of any algorithm, technique, or methodology. We used the incidence marking feature in the PIPE+ tool to obtain and analyse the results of ECDSA* and LPS techniques under these matrices.

The results analysis of the ECDSA* key generation process using various incidence markings (forwards, backwards, combined, and inhibition matrices) are provided in Table \ref{table:petri1}.

Please see the \ref{IncidenceMarkingResults} for an analysis of incidence markings of other methods in the ECDSA* and LPS techniques.

%Table \ref{table:petri1} presents the results of forwards, backwards, combined, and inhibition matrices for ECDSA* Key Generation Petri net. 

\begin{table}[!h]
\tiny
\centering
\caption{High Level Petri Net Incidences - ECDSA* Key Generation}
%{\tiny\renewcommand{\arraystretch}{0.2}
%\resizebox{!}{.07\paperheight}{%
\begin{adjustbox}{width=0.47\textwidth,center}
\begin{tabular}{|c|c|c|c|c|c|c|c|}
\hline

\multicolumn{4}{|l|}{\textbf{Forwards Incidence Matrix \textit{$I^{+}$}}} & \multicolumn{4}{l|}{\textbf{Backwards Incidence Matrix \textit{$I^{-}$}}} \\ \hline
&    Start   &  \makecell{Generate \\ Domain  \\ Parameters}     & \makecell{Generate \\ Keys}    &  & Start &   \makecell{Generate \\ Domain  \\ Parameters} & \makecell{Generate \\ Keys}  \\ \hline
       
   $\varphi$ $(Inputs)$   &     1  & 0               & 0 &  $\varphi$ $(Inputs)$  & 0     &1  &0 \\ \hline
   
      $\varphi$ $\makecell{(Domain \\ Parameters \\Store)}$ &    0  & 1   & 0   &   $\varphi$ $\makecell{(Domain \\ Parameters \\Store)}$    &    0   &  0 &    1\\ \hline

      $\varphi$ $\makecell{(Keys \\ Store)}$ &   0   &    0& 1   &   $\varphi$ $\makecell{(Keys \\ Store)}$    &    0   &  0 &   0 \\ \hline

\multicolumn{4}{|l|}{\textbf{Combined Incidence Matrix \textit{I}}} & \multicolumn{4}{l|}{\textbf{Inhibition Matrix \textit{H}}} \\ \hline

&    Start   &  \makecell{Generate \\ Domain  \\ Parameters}     & \makecell{Generate \\ Keys}    &  & Start &   \makecell{Generate \\ Domain  \\ Parameters} & \makecell{Generate \\ Keys}  \\ \hline
       
   $\varphi$ $(Inputs)$   &    1   & -1               & 0 &  $\varphi$ $(Inputs)$  &0      & 0 & 0\\ \hline
   
      $\varphi$ $\makecell{(Domain \\ Parameters \\Store)}$ &    0  & 1   &-1    &   $\varphi$ $\makecell{(Domain \\ Parameters \\Store)}$    &    0   &  0 & 0   \\ \hline

      $\varphi$ $\makecell{(Keys \\ Store)}$ &    0  &    0&  1  &   $\varphi$ $\makecell{(Keys \\ Store)}$    &    0   &  0 & 0   \\ \hline

\end{tabular}
\end{adjustbox}

\label{table:petri1}
\vspace{-4mm}%Put here to reduce too much white space after your table 

\end{table}

\subsubsection{Confidence Interval}\label{confi}

Confidence interval is another critical factor in analysing HLPNs as it determines the necessary confidence intervals for the observed transitions’ firing rates and specifies the difference between them using the minimum and maximum threshold values. In our simulation setup, the minimum and maximum threshold values are first calculated to determine the confidence interval for each HLPN model using the ECDSA* and LPS techniques. The minimum threshold values are determined by passing 100 firings and five replications through the Petri nets. In comparison, the maximum threshold values are determined by passing 1000 firings and 50 replications through the Petri nets. The simulation results indicate the minimum and maximum threshold values derived for the 95\% confidence interval regarding the average number of tokens produced at each place and the acceptable threshold of error for each processing state.

We calculate the confidence interval using the probability value for each processing state and corresponds to the confidence interval between 1 and $\alpha$. The confidence interval, also a percentage, tells the reader about the probability of accurate findings within the specified range. We used the parameter $\alpha=0.05$ in our simulation results to establish the lowest and maximum thresholds, using (1-$\alpha$), which provides a 95\% confidence value in the chance of accurate outcomes.

We estimated the average number of token throws at places specified in the HLPNs and determined the minimum and maximum threshold values for the confidence interval for each Petri net. For example, Table \ref{Token1} presented the average number of tokens at each place in the ECDSA* key generation HLPN and the confidence interval’s derived minimum and maximum threshold values. The simulation results for the maximum and minimum HLPN threshold values demonstrate the proposed scheme’s precision regarding the accessibility of numerous places and states once the stated rules are used. Moreover, the confidence interval values in our HLPNs also indicate the maximum level of confirmation of the proposed clone node attack detection scheme’s correctness.

\begin{table}[!h]
\centering
\tiny
\caption{Average Number of Token with Minimum and Maximum Threshold Values - ECDSA* Key Generation}
\begin{adjustbox}{width=0.47\textwidth,center}
\begin{tabular}{|c|c|c|c|c|}
\hline
\textbf{\makecell{Places}} &\textbf{ \makecell{Average Number\\ of Tokens}} & \textbf{\makecell{Minimum \\ Threshold \\ Values}} & \textbf{\makecell{Average Number \\ of Tokens}} & \textbf{\makecell{Maximum \\ Threshold\\  Values}}  \\ \hline
Inputs &      3.86     &   4.5        &   11.35065        &     15.17      \\ \hline
\makecell{Domain \\ Parameters \\ Store} &     1.9802      &    2.228       &  18.33566         &    9.32       \\ \hline
 Key Store&       14.11881    &     1.85      &      150.55633     &    7.897       \\ \hline
\end{tabular}
\end{adjustbox}
\label{Token1}
\vspace{-4mm}%Put here to reduce too much white space after your table 
\end{table}

Please refer to \ref{ConfidenceInterval} for the average number of tokens with minimum and maximum threshold values for other methods in the ECDSA* and LPS techniques.

\section{Verification of Proposed Models Using SMT-Lib and Z3 Solver} \label{verification}

This section details our proposed HLPNs model verification process. First, we utilise the Z specification language to illustrate the behaviour of our system. After that, the HLPN models and properties are transformed to SMT-Lib. A Z3 solver is then utilised to validate or invalidate the correctness claims.

Using SMT-Lib and Z3 solver, we validated both ECDSA* and the proposed clone node attack detection scheme. The Z3-SMT Solver is a powerful tool for evaluating and analysing algorithms, applications, and systems based on one or more theories. Using the theories mentioned above, we utilised array theory \cite{de2008z3} to validate and satisfy the logic rules (or formulae) in our proposed scheme.

\subsection{Properties}

To validate an ECDSA* and our proposed clone node detection approaches, we defined the following properties of the proposed algorithms.

\subsubsection{ECDSA* Key Generation}

An ECDSA* Key Generation method is detailed in Rule \ref{formalrule1}, which creates cryptographic keys, including public and private keys, using standard elliptic curve cryptography domain parameters. Following is the order in which the system selects and asserts the domain parameters: prime field order, elliptic curve, random base point, prime number and co-factor. The key generation method produced the public and private keys that are used in signing and verifying processes. Our conclusion in response to this argument is unsat.

\begin{equation} \label{formalrule1}
\begin{split}
(assert (not(or (=(select prime.field.order 1) \\ (select
elliptic.curve 2) \\ (select base.point 3)(select ordinal.value
4)\\(select cofactor 5)) \\ (= (select generate.private.key 6) \\ (select 
generate.public.key 7))))) (check-sat)
\end{split}
\end{equation}

\subsubsection{ECDSA* Signature Generation}

 An ECDSA* Signature Generation procedure selects and asserts properties such as a random integer and a base point in the system used to compute the elliptic curve’s coordinates. Following this, a hash function is chosen and claimed to compute the message hash and generate the asserted integer value. Finally, using the hash integer value and private key, the signature attributes are asserted and computed. The property for generating ECDSA* signatures is shown in Rule \ref{formalrule2}. The outcome we reached in response to this assertion is unsat.

\begin{equation} \label{formalrule2}
\begin{split}
(assert(not(or (=(select random.integer 1) \\ (select base.point 2)) \\  (=(select message 3) \\ (select compute.hash 4)) \\ (=(select mod 5)) \\ (=(select private.key 6) \\ (select hash.integer 7))))) (check-sat)
\end{split}
\end{equation}

\subsubsection{ECDSA* Signature Verification}
In Rule \ref{formalrule3}, the property for ECDSA* Signature Verification is described. This property verifies and approves a signature established in Rule \ref{formalrule2}. This property selects and asserts the required signature attributes for the message. A hash function is applied to the message to transform it into an integer value known as the hash value. When verifiers accept signature attributes, a point value and corresponding coordinates are asserted and computed to indicate whether the signature is accepted or rejected. The outcome we reached in response to this assertion is unsat.

\begin{equation} \label{formalrule3}
\begin{split}
(assert(not(or 
 (= (select signature.integer 1) \\ (select  signature.integer2)) \\ (=(select message 3)) \\ (=(select get.integer.point 4)) (= \\ (select integer.point.1 5)(select integer.point2 6) \\(select coordinate.point 7)) (= \\(select calculate point.1 verified )\\(select calculate point.2 verified))))) (check-sat)
\end{split}
\end{equation}

\subsubsection{Calculate Location}

The property for determining the location of provers in the proposed clone node attack detection mechanism is shown in Rule \ref{formalrule4}. The location calculation process starts with selecting and asserting the coordinates of the provers and verifiers on 2-D space. By following this, the distance or location of each prover is calculated by using the Euclidean Distance algorithm concerning the verifiers that generate the location request in the system. The outcome we reached in response to this assertion is unsat.

\begin{equation} \label{formalrule4}
\begin{split}
(assert(not(or (=(select prover.coordinate 1) \\(select prover.coordinate 2)) \\(=(select verifier.coordinate 3) \\(select  verifier.coordinate 4)) \\ (=(select prover.coordinate 5)\\ (select  verifier.coordinate 6))))) (check-sat)
\end{split}
\end{equation}

\subsubsection{Generate Location Proof}

The Rule \ref{formalrule5} describes the process of generating location proofs for IoT devices demonstrating their presence at a specified location in terms of LBS. The proof generation procedure begins by selecting and asserting the prover’s and verifier’s contextual information about their deployed environment. The context information contains the IoT device’s unique identifier, the date and time of data sensing, as well as the device’s exact location and activity. Next, the verifier selects the location request and sends it to the prover to validate. After receiving the location proof request, the prover uses her private key to generate the signature and sends it to the verifier. The outcome we reached in response to this assertion is unsat.

\begin{equation} \label{formalrule5}
\begin{split}
(assert(not(or (=(select id 1) \\ (select time 2) \\(=(select location 3)\\(select activity 4)))\\ (=(select extract.context.information 5)\\ (select store.context.information 6)) \\ (=(select \\ location.proof 7)\\(select private.key 8) (select hash 9))))) (check-sat)
\end{split}
\end{equation}

\subsubsection{Verify Location Proof}

In Rule \ref{formalrule6}, the property for verifying location proofs for IoT devices claiming to be at a given location with context information is demonstrated. The verification process begins by obtaining the signatures of provers and their associated public keys as inputs. After confirming the availability of stored information on the LBS, the verifiers examine the contextual information CI from the prover and perform ECDSA* batch verification on the signatures. The outcome we reached in response to this assertion is unsat.
    
\begin{equation} \label{formalrule6}
\begin{split}
(assert(not(or \\ (=(select extract.context.information 1) \\(select  extract.context.information 2)) \\ (=(select location.proof.request 3)) \\ (=(select extracted.context.information 4)) \\ (=(select verified.information 6) \\ (select public.key 7)) \\ (=(select verified.location.proof verified))))) (check-sat)
\end{split}
\end{equation}

\subsubsection{Results}

For verification, the proposed model and transition rules are defined and converted to SMT. We used the``QF AUFLIA” logic of SMT-Lib to transform the rules. This logic includes quantified free logic equations and free sort functions that can be easily integrated into many theories, including integer arrays. Next, we used the Z3 solver to determine whether or not the proposed model meets the specified properties. The system modelling results and defined properties imply that the proposed scheme is correct in terms of reliability and validity in the underlying scenario. The Z3 solver takes the properties as input parameters, computes them, and delivers the results as Boolean satisfiers (sat, unsat). Unsatisfactory results (sat) result in imprecision or lack of specified attributes. However, satisfactory findings (unsat) indicate that the proposed scheme is correct in system modelling and properties.

The verification results of the ECDSA* technique and LPS technique in the proposed clone node attack detection scheme are expressed in terms of the time required by the solver to verify the given properties of the proposed algorithms. Table \ref{tabexecution} displays the execution time (in seconds) for the specified properties. The verification results for the stated properties are generated in a finite amount of time. Hence, confirm and satisfy the properties in terms of correctness.

\begin{table}[!h]
\centering
\caption{Execution Time of the Specified Properties}

\begin{tabular}{|l|l|}
\hline
\textbf{Specified Properties} & \textbf{Execution Time } \\ \hline
    ECDSA* Key Generation &  0.0357 sec         \\ \hline
       ECDSA* Signature Generation   &  0.0598 sec         \\ \hline
      ECDSA* Signature Verification    &      0.0620 sec     \\ \hline
        Calculate Location  &  0.0293 sec         \\ \hline
                Generation Location Proof  &  0.0575 sec         \\ \hline
        Verify Location Proof  &  0.0762 sec         \\ \hline

\end{tabular}
\label{tabexecution}
\vspace{-4mm}%Put here to reduce too much white space after your table 

\end{table}

\section{Modelling and Analysis of Proposed Scheme Using Coloured Petri Nets} \label{coloured}

This section describes the modelling and analysis procedure for the proposed scheme using CPNs. As with section \ref{modelling}, we divided the proposed scheme into two parts: an enhanced ECDSA* and a LPS for modelling and analysis using CPN. The sections \ref{CPN1} and \ref{CPN2} describe the process of developing a CPN model for an enhanced ECDSA* and a LPS, respectively. We then used timed and untimed methods to analyse and demonstrate the CPN models.

\subsection{An Enhanced ECDSA* Scheme}\label{CPN1}

This section demonstrates how an ECDSA* approach can be modelled using CPNs. A CPN model is characterised as follows: The places include colour sets and are described as data types. Arc inscriptions are specified expressions evaluating the current placements of input places during the transition. A transition is acceptable in CPNs if the variables used in the transition arcs have consistent bindings from all inputs. The resulting transition arc is a subset of a multiset. When a transition is fired, it removes the related multiset of tokens from each input point where the transition was applied. Also, it uses the multiset of tokens to generate each outcome for which the expression evaluates. In arc expressions, variables must be declared using \texttt{var}.

The figures \ref{PetriNet1_WOT}-\ref{PetriNet3_WT} illustrates the CPN models used in the proposed scheme for the timed and untimed ECDSA* key generation, ECDSA* signature generation, and ECDSA* signature verification processes, respectively.

\subsubsection{ECDSA* Key Generation}

\begin{figure}[!h]
  \centerline{\includegraphics[width=8.8cm, height=4cm]{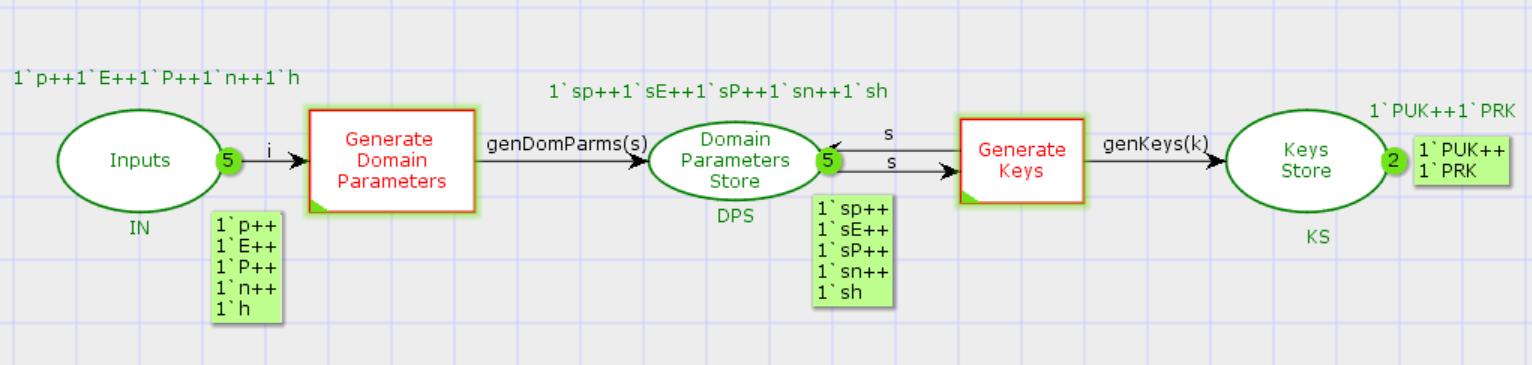}}
  \caption{A CPN Model of an ECDSA* Key Generation Process}
  \label{PetriNet1_WOT}
\end{figure}

Fig. \ref{PetriNet1_WOT} illustrates a CPN model of an ECDSA* Key Generation process. We used the comprehensive types declaration feature in CPN to declare Inputs as a token with all desirable characteristics. The following are the colour set declarations for the CPN model of an ECDSA* Key Generation: \\
\texttt{colset IN = with p | E | P | n | h;} \\
\texttt{colset DPS = with sp | sE | sP | sn | sh;} \\
\texttt{colset KS = with PUK | PRK;} \\

The initial marking for the ``Inputs'' place is specified by a multiset inscription such as \texttt{1'p++1'E++1'P++1'n++1'h}. The variables in the green box and circle show the total number of current tokens and their colour set. For the ``Inputs'' place, the inscription \texttt{IN} identifies the input colour as an option. The domain parameters \{\textit{p}, \textit{E}, \textit{P}, \textit{n}, \textit{h}\}, are required to generate the public and private keys. The multiset inscription such as \texttt{1'sp++1'sE++1'sP++1'sn++1'sh} specifies the initial marking for the ``Domain Parameters Store'' place and shows the details and number of current tokens in the place. The inscription \texttt{DPS} proclaims the colour set for the ``Domain Parameters Store'' place, which contains domain parameters like \{\textit{sp}, \textit{sE}, \textit{sP}, \textit{sn}, \textit{sh}\}. Following that, the multiset inscription \texttt{1'PUK++1'PRK} specified the initial marking for the ``Keys Store'' place, and the associated green box and circle represent the details and number of currently residing tokens in the place during simulation. The inscription \texttt{KS} specifies the colour scheme for the ``Keys Store'' place, which contains the created keys such as \{\textit{PUK}, \textit{PRK}\}.

A CPN model of an ECDSA* Key Generation has three variables, each with its own declaration: \\
\texttt{var i: IN;} \\
\texttt{var s: DPS;} \\
\texttt{var k: KS;}

The variables \textit{i}, \textit{s} and \textit{k} are defined for the colour set declarations \texttt{IN}, \texttt{DPS} and \texttt{KS}, respectively.

The following user-defined functions leverage a built-in CPN functionality to use inputs to initiate transitions and store the result in many places ahead. The functions for the ECDSA* Key Generation CPN model are as follows.\\
\texttt{fun genDomParms (i) = (s, IN)}; \\
\texttt{fun genKeys (s) = (k, DPS)};

During CPN transitions, the guard function can be used to control the execution of associated transitions. In the ECDSA* Key Generation CPN model, one transition is labelled ``Generate Domain Parameters'', and the other ``Generate Keys''. An arc inscription over the ``Inputs'' place states that the value of \textit{i}  can be set to any of the domain parameters. The ``Generate Domain Parameters'' transition executes and stores the result of the function call \texttt{genDomParms(i)} in the ``Domain Parameters Store'' place. The arc inscription for the ``Domain Parameters Store'' place is a variable \textit{s}, which can be tied to any of the stored domain parameters. The transition ``Generate Keys'' executes and stores the result of the function call \texttt{genKeys(s)} in the ``Keys Store'' place. The arc inscription for the ``Keys Store'' place is a variable \textit{k}, which can be tied to any of the following transitions.

\begin{figure}[!h]
  \centerline{\includegraphics[width=8.8cm, height=4cm]{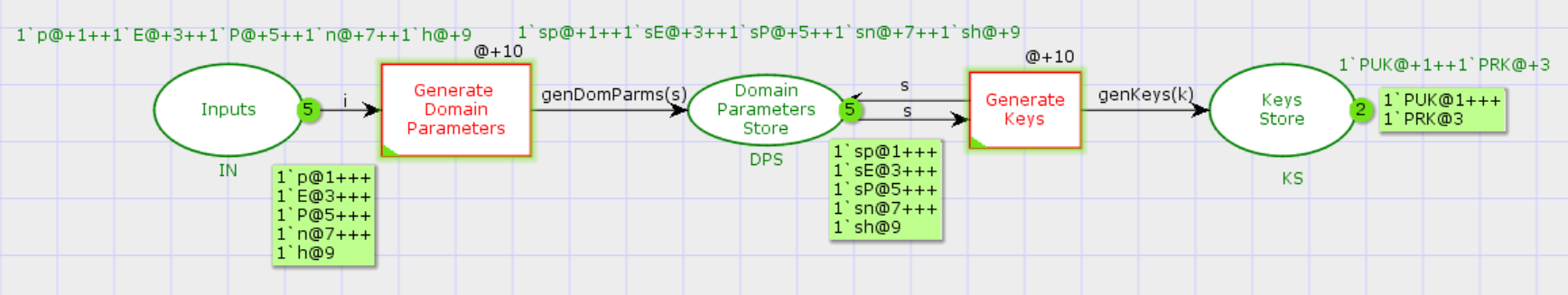}}
  \caption{A timed CPN Model of an ECDSA* Key Generation Process}
  \label{PetriNet1_WT}
\end{figure}

Fig. \ref{PetriNet1_WT} shows how to include time into a CPN model for ECDSA* Key Generation. This information allows us to evaluate the system's efficiency and processes. Timed models are beneficial for validating real-time systems when accurate timing of events and outputs is required rather than just output completion. Time is introduced with CPNs by a colour set token and a second value containing the time stamp. This process is done by setting a timer for the desired colour set. Depending on the context, this timestamp may be interpreted as the processing or execution time of the impacted transitions. CPNs allow us to utilise both timed and untimed tokens, so we do not have to declare each token type individually. Thus, if a network supports both timed and untimed activities, the untimed transitions must fire before the timed transitions.

For the ECDSA* Key Generation CPN model, the following timed colour set declarations are provided: \\
\texttt{colset IN = with p | E | P | n | h timed;} \\
\texttt{colset DPS = with sp | sE | sP | sn | sh timed;} \\
\texttt{colset KS = with PUK | PRK timed;} \\

With these enhancements, the domain parameters in the ``Inputs'' place now include a timestamp showing the initial arrival time. Each domain parameter is allocated a unique timestamp, increment by \texttt{1'p@+1++1'E@+3++1'P@+5++1'n@+7\\++1'h@+9} respectively. Therefore, until the global clock has read a time that is higher than or equal to the time stamp on a domain parameter, transition \textit{i} will not be fired. A similar convention is employed for all stored parameters, including incremental time stamps, such \texttt{1'sp@+1++1'sE@+3++1'sP@+5++1'sn\\@+7++1'sh@+9}.

\subsubsection{ECDSA* Signature Generation}

\begin{figure*}[!t]
  \centerline{\includegraphics[scale=0.5]{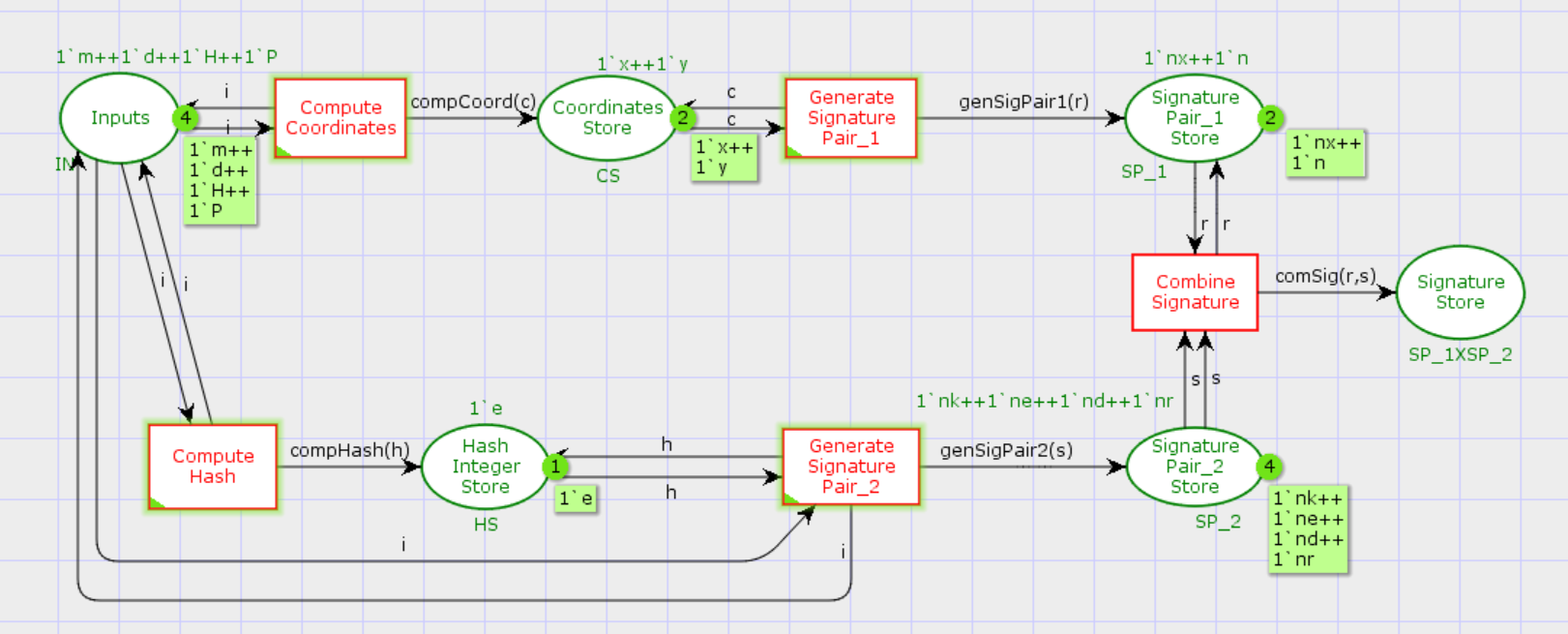}}
  \caption{A CPN Model of an ECDSA* Signature Generation Process}
  \label{PetriNet2_WOT}
\end{figure*}

A CPN model of an ECDSA* Signature Generation process in the proposed scheme is illustrated in Fig. \ref{PetriNet2_WOT}. The following are the extensive type colour set declarations for the CPN model of an ECDSA* Signature Generation: \\
\texttt{colset IN: = with m | d | H | P;} \\
\texttt{colset CS = with x | y;} \\
\texttt{colset $SP_{1}$ = with nx | n;} \\
\texttt{colset $SP_{2}$ = with nk | ne | nd | nr;} \\
\texttt{colset $SP_{1}$X$SP_{2}$ = product $SP_{1}$ * $SP_{2}$;} \\

%\begin{sidewaysfigure}
%\includegraphics[width=18cm, height=11cm]{PetriNet2_WOT.PNG}
 % \caption{A CPN Model of an ECDSA* Signature Generation Process}
 % \label{PetriNet2_WOT}
%\end{sidewaysfigure}

The multiset inscription \texttt{1'm++1'd++1'H++1'P} specifies the initial marking for the ``Inputs'' place. In green box and circle, these variables show how many total current tokens are present and their colour set declaration, respectively. The inscription \texttt{IN} declares the colour set for the ``Inputs'' place, which contains the prescribed input parameters such as \{\textit{m}, \textit{d}, \textit{H}, \textit{P}\} required to commence the signature generation procedure in an ECDSA*. The multiset inscription \texttt{1'x++1'y} specifies the initial marking for the ``Coordinates Store'' place, while the accompanying green box and circle represent the features and currently placed tokens in the place during simulation process, respectively. The inscription \texttt{CS}  indicates the colour set for the ``Coordinates Store'', which is responsible for storing the ECC coordinates points such as \textit{x} and \textit{y}. The multiset inscription \texttt{1'e}, in conjunction with the inscription  \texttt{HS}, defines the initial marking for the '`Hash Integer Store'' place, which stores the hash integer value \textit{e}. In the green box and circle, these variables show how many total current tokens are present and their colour set declaration. The following multiset inscriptions, including \texttt{1'nx++1'n} and \texttt{1'nk++1'ne++1'nd++1'nr}, as well as the inscriptions \texttt{$SP_{1}$} and \texttt{$SP_{2}$}, specify the initial marking for the ``Signature Pair 1 Store'' and ``Signature Pair 2 Store'' places, respectively. These places are used to hold signature values such as \textit{r} and \textit{s}, and the matching green box and circle indicate the features and count of located tokens in those locations during simulation. Finally, an inscription \texttt{$SP_{1}$X$SP_{2}$} declares the colour set for the ``Signature Store'' place, which is responsible for merging and storing the signature data.

The following six variables are declared in a CPN model of an ECDSA* Signature Generation. \\
\texttt{var i: IN;} \\
\texttt{var c: CS;} \\
\texttt{var h: HS;} \\
\texttt{var r: $SP_{1}$;} \\
\texttt{var s: $SP_{2}$;} \\
\texttt{var ss: $SP_{1}$ X $SP_{2}$;}

A variable \textit{i} is declared for the colour set \texttt{IN}, a variable \textit{c} is declared for the colour set \texttt{CS}, a variable \textit{h} is declared for the colour set \texttt{HS}, a variable \textit{r} is declared for the colour set \texttt{$SP_{1}$}, a variable \textit{s} is declared for the colour set \texttt{$SP_{2}$} and a variable \textit{ss} is declared for the colour set \texttt{$SP_{1}$ X $SP_{2}$}.

The following are user-defined functions for the ECDSA* Signature Generation CPN model.\\
\texttt{fun compCoord(i) = (c, IN)}; \\
\texttt{fun compHash (i) = (h, IN)}; \\
\texttt{fun genSigPair1 (c) = (r, CS)}; \\
\texttt{fun genSigPair2 (h) = (s, HS)}; \\
\texttt{fun comSign (r,s) = (s, $SP_{1}$, r, $SP_{2}$}); \\

%\begin{sidewaysfigure}
%\includegraphics[width=18cm, height=11cm]{PetriNet2_WT.PNG}
 % \caption{A timed CPN Model of an ECDSA* Signature Generation Process}
  %\label{PetriNet2_WT}
%\end{sidewaysfigure}

Five transitions are used in the ECDSA* Signature Generation CPN model: ``Compute Coordinates'', ``Compute Hash'', ``Generate Signature Pair 1'', ``Generate Signature Pair 2'' and ``Combine Signature''. An arc inscription at the "Inputs" place is a variable \textit{i}, which can be tied to any of the input parameters. The transition ``Compute Coordinates'' executes and stores the result of the function call \texttt{comCoord(i)} in the ``Compute Coordinates'' place. The variable \textit{c} contains an arc inscription of the Coordinates Store" place and can be tied to any of the coordinates in the coordinate store.  The transition ``Compute Hash'' executes and stores the result of the function call \texttt{compHash(i)} in the ``Hash Integer Store'' place. The arc inscription of the ``Hash Integer Store'' location is a variable \textit{h}, which can be associated with an integer value in the hash integer store. 
The transition ``Generate Signature Pair 1'' executes and stores the result of the function call \texttt{genSigPair1(c)} in the ``Signature Pair 1 Store'' place. Similarly, the transition ``Generate Signature Pair 2'' executes and stores the result of the function call \texttt{genSigPair2(h)} in the ``Signature Pair 2 Store'' place. The arc inscriptions for the ``Signature Pair 1 Store'' and ``Signature Pair 2 Store'' places are the variables \textit{r} and \textit{s}, which can be tied to the signature value stored in them. Finally, the transition ``Combined Signature'' executes and stores the result of the function call \texttt{comSig(r,s)} in the ``Signature Store'' place. The arc inscription for the ``Signature Store'' place are the variables \textit{s} and \textit{r}, which can be tied to any of the final signature values.

\begin{figure*}[ht]
  \centerline{\includegraphics[scale=0.5]{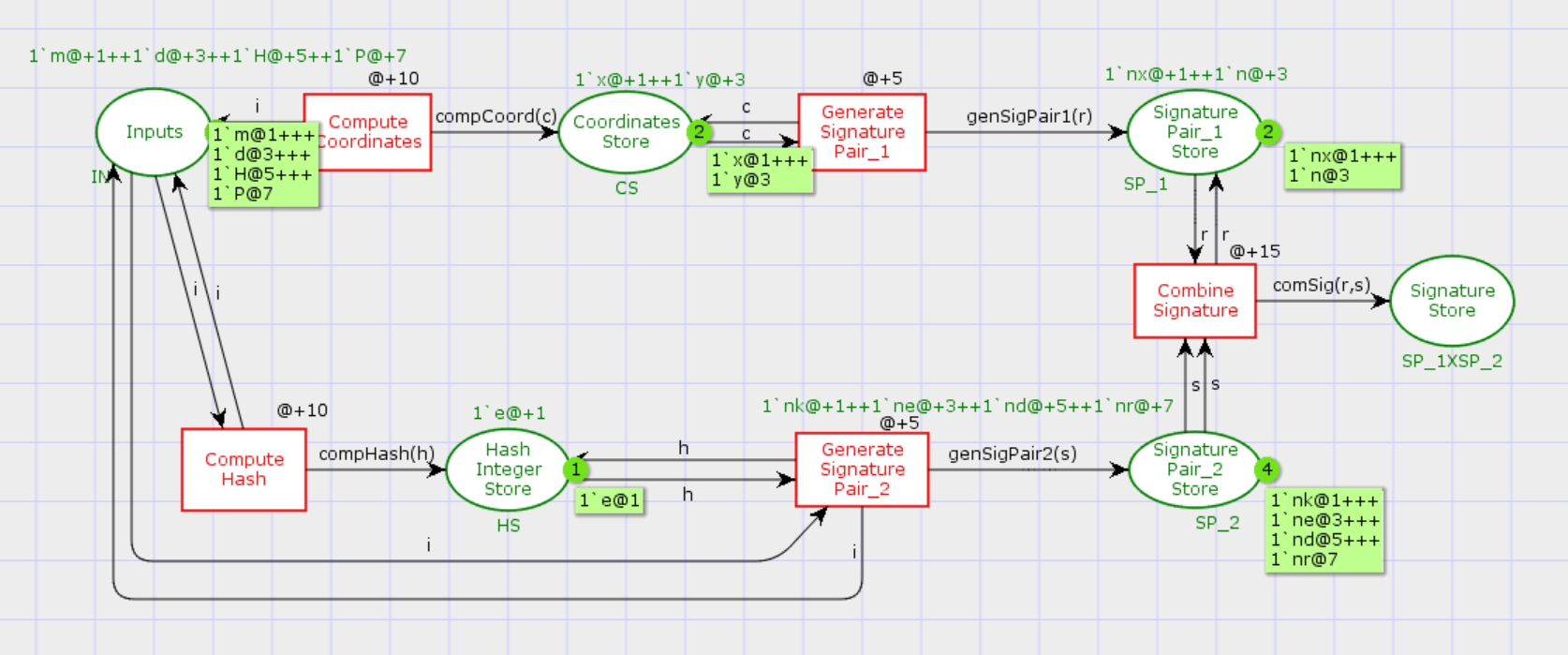}}
  \caption{A timed CPN Model of an ECDSA* Signature Generation Process}
  \label{PetriNet2_WT}
\end{figure*}

A timed CPN model of an ECDSA* Signature Generation process in the proposed scheme is depicted in Fig. \ref{PetriNet2_WT}. For the ECDSA* Signature Generation CPN model, the timed colour set declarations are as follows:\\
\texttt{colset IN = with m | d | H | P timed;} \\
\texttt{colset CS = with x | y  timed;} \\
\texttt{colset $SP_{1}$ = with nx | n timed;} \\
\texttt{colset $SP_{2}$ = with nk | ne | nd | nr timed;} \\
\texttt{colset $SP_{1}$X$SP_{2}$ = product $SP_{1}$ * $SP_{2}$ timed;} \\

With these enhancements, the input parameters in the ``Inputs'' section include a time stamp showing their arrival time. Each input parameter contains an incremental time stamp, such as\texttt{1'm@+1++1'd@+3++1'H@+5++1'P@+7}. Consequently, the transition  \textit{i} will not be activated until the global clock reads a time more than or equivalent to the time stamp on the set parameters. Each coordinate in the ``Coordinates Store'' has a unique time stamp assigned incrementally, such as \texttt{1'x@+1++1'y@+3} using a variable \textit{c} on the arc that fire to the ``Generate Signature Pair 1'' transition. A hash value in the ``Hash Integer Store'' is assigned a unique time stamp incrementally, for example, \texttt{1'e@+1}, using the variables \textit{h} on arc that fire to the ``Generate Signature Pair 2'' transition. Both signature pairs generated in the ``Generate Signature Pair 1'' and ``Generate Signature Pair 2'' processes have a unique time stamp assigned incrementally, as \texttt{1'nx@+1++1'n@+3} and \texttt{1'nk@+1++1'ne@+3++1'nd@+5++1'nr@+7}, respectively, using the variables \textit{r} and \textit{s} on their arcs.

\subsubsection{ECDSA* Signature Verification}

\begin{figure*}[!t]
  \centerline{\includegraphics[scale=0.5]{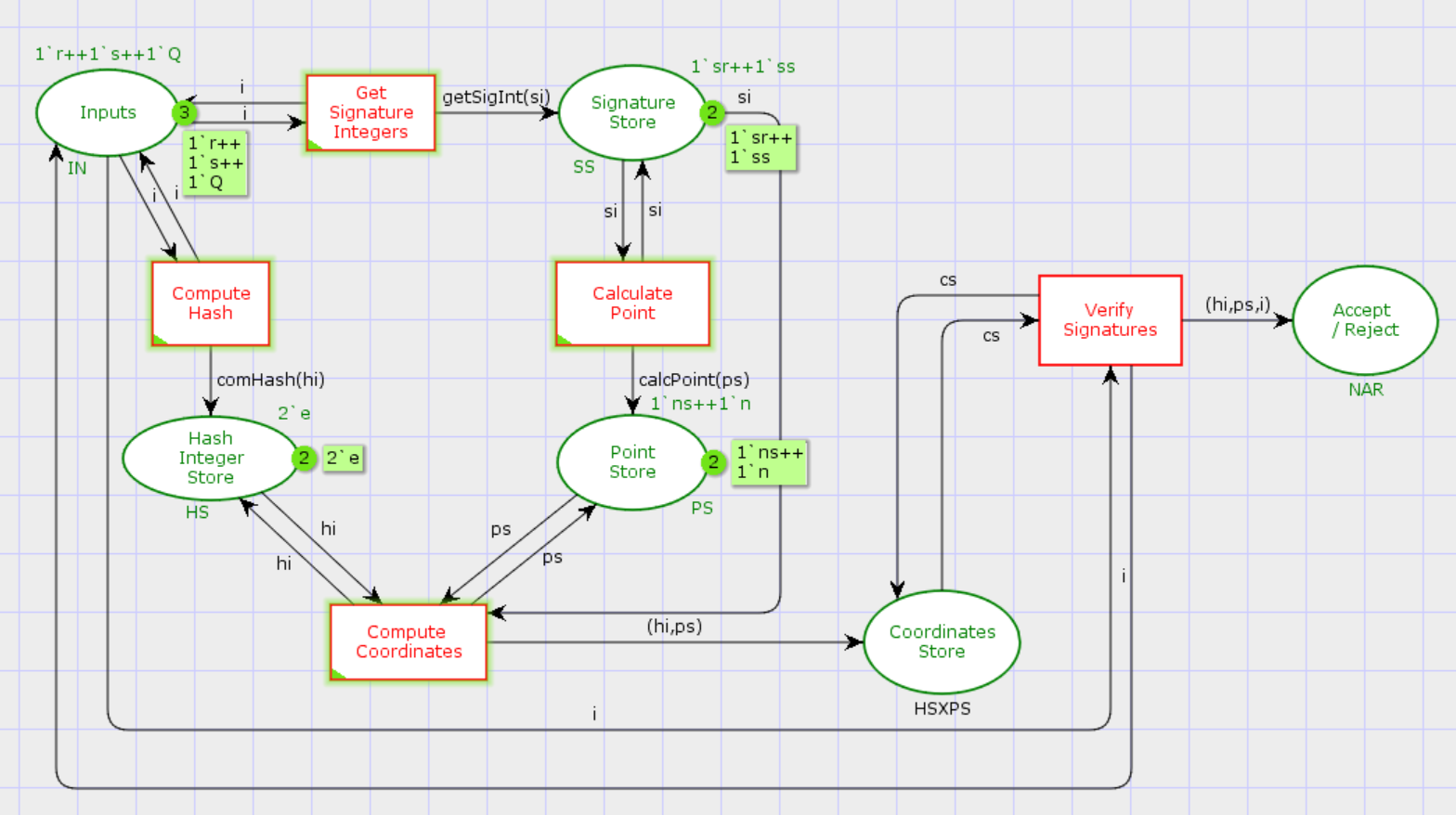}}
  \caption{A CPN Model of an ECDSA* Signature Verification Process}
  \label{PetriNet3_WOT}
\end{figure*}

A CPN model of an ECDSA* Signature Verification process in the proposed scheme is shown in Fig. \ref{PetriNet3_WOT}. A CPN model of an ECDSA* Signature Verification has the following user-defined colour set declarations: \\
\texttt{colset IN = with r | s | Q;} \\
\texttt{colset SS = with sr | ss;} \\
\texttt{colset HS = with e;} \\
\texttt{colset PS = with ns | n;} \\
\texttt{colset HS X PS = product HS * PS;} \\
\texttt{colset NAR = with product HS * PS * IN;} \\

The multiset inscription  \texttt{1'r++1's++1'Q} establishes the initial marking for the ``Inputs'' place.  In the green box and circle, these variables show how many total current tokens are present and their colour set declaration. An arc inscription \texttt{IN} specifies the colour scheme for the ``Input'' place that contains the input parameters \{\textit{r}, \textit{s}, \textit{Q}\} required to commence the signature verification procedure in an ECDSA* technique. The multiset inscription, for example, \texttt{1'sr++1'ss}, establishes the initial marking for the ``Signature Store'' place. The inscription \texttt{SS} indicates the colour set for the ``Signature Stor'' place, which is responsible for storing signature values such as \textit{r} and \textit{s}. The multiset inscription \texttt{2'e} with an inscription \texttt{HS} specifies the initial marking for the '`Hash Integer Store'' place by storing the hash integer value \textit{e}, while the matching green box and circle represent the details and count of current tokens located in the location during simulation. Additionally, the multiset inscription \texttt{1'ns++1'n} with an inscription \texttt{PS} defines the initial marking for the `'Point Store'' place by storing the point value \textit{w}. Additionally, an inscription \texttt{HS X PS } establishes the initial marking for the ``Coordinates Store'' location by storing the product of the \texttt{HS} and \texttt{PS} values. Finally, an inscription \texttt{NAR} establishes the initial marking for the ``Coordinates Store'' location by storing the product of the \texttt{HS}, \texttt{PS} and \texttt{IN} values depicting the decision about signature acceptance or rejection.

A CPN model of an ECDSA* Signature Verification contains six variables, the declarations for which are listed below: \\
\texttt{var i: IN;} \\
\texttt{var si: SS;} \\
\texttt{var hi: HS;} \\
\texttt{var ps: PS;} \\
\texttt{var cs: HS X PS ;} \\

A variable \textit{i} is declared for the colour set \texttt{IN}, a variable \textit{si} is declared for the colour set \texttt{SS}, a variable \textit{hi} is declared for the colour set \texttt{HS}, a variable \textit{ps} is declared for the colour set \texttt{PS}, a variable \textit{cs} is declared for the colour set \texttt{HS X PS} and a variable \textit{ar} is declared for the colour set \texttt{AR}.

The following are user-defined functions for the ECDSA* Signature Verification CPN model.\\
\texttt{fun getSigInt(i) = (si, IN)}; \\
\texttt{fun compHash (i) = (hi, IN)}; \\
\texttt{fun calcPoint (si) = (ps, SS)}; \\

Five transitions are used in the ECDSA* Signature Verification CPN model: ``Get Signature Integers'', ``Compute Hash'', ``Calculate Point'', ``Compute Coordinates'' and ``Verify Signatures''. An arc inscription at the ``Inputs'' place is the variable \textit{i}, which can be tied to any of the input parameters. The transition ``Get Signature Integers'' executes and stores the result of the function call \texttt{getSigInt(i)} in the ``Signature Store'' place. The variable \textit{si} contains the arc inscription of the ``Signature Store'' place and can be tied to any of the signature values in the signature store. The transition ``Compute Hash'' executes and stores the result of the function call \texttt{compHash(i)} in the ``Signature Store'' place. The arc inscription of ``Hash Integer Store'' place is simply the variable \textit{hi}, which can be bound to an integer value in the hash integer store.  The transition ``Calculate Point'' executes and stores the result of the function call \texttt{calcPoint(si)} in the ``Point Store'' place. The arc inscription for the ``Point Store'' place is the variable \textit{ps}, which can be tied to any of the point values in the point store. The transition ``Compute Coordinates'' executes and stores the result of the \texttt{hi, ps} in the ``Coordinates Store'' place. The arc inscription of ``Coordinates Store'' place is the variable \textit{cs}, which can be tied to either the hash integer store value or the point store inscriptions in the coordinates store. The transition ``Verify Signatures'' executes and stores the result of the \texttt{hi, ps, i} in the ``Accept / Reject'' place, which can be tied to either the accept or reject final signature choice.

\begin{figure*}[!t]
  \centerline{\includegraphics[scale=0.5]{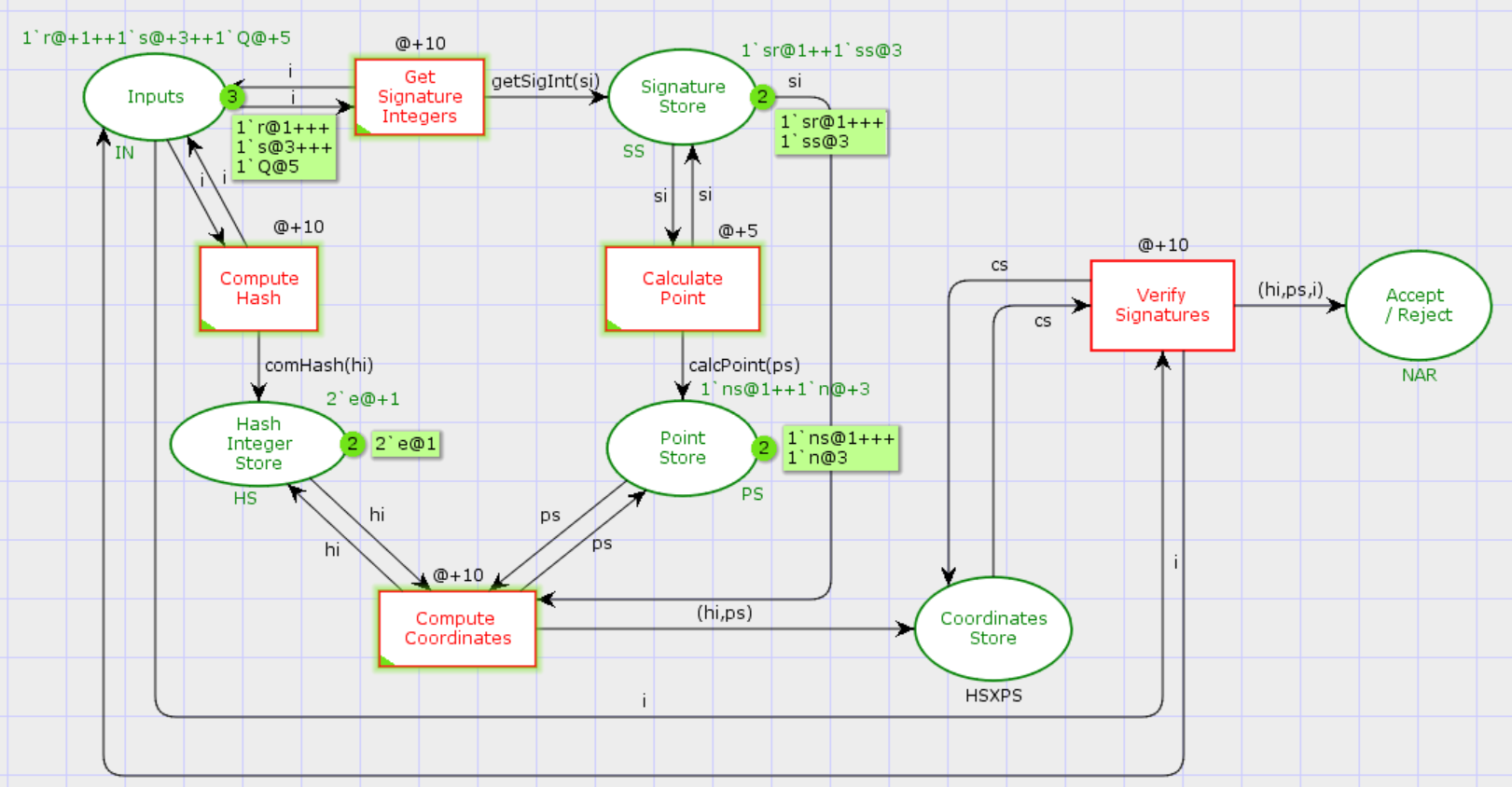}}
  \caption{A timed CPN Model of an ECDSA* Signature Verification Process}
  \label{PetriNet3_WT}
\end{figure*}

Fig. \ref{PetriNet3_WT} illustrates the timed CPN model of an ECDSA* Signature Verification process. The following are the timed colour set declarations for the CPN model of an ECDSA* Signature Verification:\\
\texttt{colset IN = with r | s | Q timed;} \\
\texttt{colset SS = with sr | ss timed;} \\
\texttt{colset HS = with e timed;} \\
\texttt{colset PS = with ns | n timed;} \\
\texttt{colset HSXPS = product HS * PS timed;} \\
\texttt{colset NAR = with product HS * PS * IN timed;}

With these modifications, the input parameters in the ``Inputs'' place now include a timestamp showing their arrival time. Each input parameter has an incremental time stamp, such as \texttt{1'r@+1++1's@+3++1'Q@+5}. Therefore, the transition \textit{i} will remain inactive until the world clock reads a time more than or equal to the time stamp on the set parameters. Each signature value in the ``Signature Store'' place is allocated a unique time stamp incrementally, as follows: \texttt{1'sr@+1++1'ss@+3} using the variables \textit{si} on arc that fire to the ``Calculate Point'' transition. A hash value in the ``Hash Integer Store'' is assigned a unique time stamp incrementally, as in \texttt{2'e@+1}, using the variables \textit{hi,ps} on arc that fire to the ``Compute Coordinates'' transition. A signature decision in the ``Accept / Reject'' place has a unique timestamp assigned incrementally with a variable \textit{hi, ps, i} on an arc that fires to the next transitions if any.

\subsection{Location Proof System}\label{CPN2}

This section outlines the process of modelling a LPS utilising CPNs in a proposed scheme.

\subsubsection{Calculate Location}

\begin{figure}[ht]
  \centerline{\includegraphics[width=8.8cm, height=3cm]{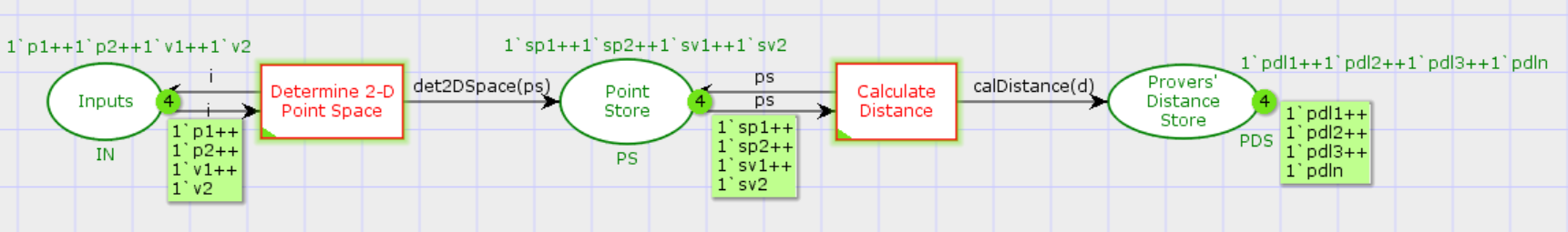}}
  \caption{A CPN Model of a Calculate Location Process in LPS}
  \label{PetriNet4_WOT}
\end{figure}

Fig. \ref{PetriNet4_WOT} depicts a CPN model to calculate the location process of the LPS in the proposed scheme. We refer to this as utilising the rich type declaration facility of CPNs by declaring Inputs as a token with all desired characteristics. The following are the declarations for the colour sets used in the CPN model of calculating location:\\
\texttt{colset IN = with p1 | p2 | v1 | v2;} \\
\texttt{colset PS = with sp1 | sp2 | sv1 | sv2;} \\
\texttt{colset PDS = with pdl1 | pdl2 | pdl3 | pdln;} \\

On the top of the ``Inputs'' place, a multiset inscription \texttt{1'p1++1'p2++1'v1\\++1'v2} defines the initial marking for it.  In the green box and circle, these variables show how many total current tokens are present and their colour set declaration. An inscription \texttt{IN} specifies the colour scheme for the ``Inputs'' place. It contains the coordinates in 2-D space for provers and verifiers such as \{\textit{$p_{1}$}, \textit{$p_{2}$}, \textit{$v_{1}$}, \textit{$v_{2}$}\} that are required to determine the prover's location relative to the verifier. The multiset inscription, for example, \texttt{1'sp1++1'sp2++1'sv1++1'sv2}, establishes the initial marking for the ``Point Store'' place, while the accompanying green box and circle display the details and count of current tokens existing in the location during simulation. An inscription \texttt{PS} specifies the colour scheme for the ``Point Store'' place. It contains the prover and verifier's stored coordinates, such as \{\textit{$sp_{1}$}, \textit{$sp_{2}$}, \textit{$sv_{1}$}, \textit{$sv_{2}$}\}. The multiset inscriptions such as \texttt{1'pdl++1'pdl2++1'pdl3++1'pdln} specified the initial marking for the ``Provers' Distance Store'' place, while the accompanying green box and circle represent the details and count of currently residing tokens in the location during simulation. An inscription \texttt{PDS} specifies the colour scheme for the ``Provers' Distance Store''. It contains the provers distance list, such as \{\textit{$pdl_{1}$}, \textit{$pdl_{2}$}, \textit{$pdl_{3}$} $\dots$ \textit{$pdl_{n}$}\}.

A CPN model of a Calculate Location in the LPS contains three variables, the declarations for which are listed below: \\
\texttt{var i: IN;} \\
\texttt{var ps: PS;} \\
\texttt{var d: PDS;} 

A variable \textit{i} is declared for the colour set \texttt{IN}, a variable \textit{ps} is declared for the colour set \texttt{PS} and a variable \textit{d} is declared for the colour set \texttt{PDS}. 

The following are user-defined functions for the Calculate Location CPN model.\\
\texttt{fun det2DSpace(i) = (ps, IN)}; \\
\texttt{fun calDistance(ps) = (d, PS)}; \\

Two transitions are used in the Calculate Location CPN model: ``Determine 2-D Point Space'' and ``Calculate Distance''. The arc inscription for the ``Inputs'' place is simply the \textit{i} variable. The transition ``Determine 2-D Point Space'' executes and stores the result of the function call \texttt{det2DSpace(i)} in the ``Point Store'' place. Therefore, for the provers and verifiers, \textit{i} can be tied to any of the coordinates in the 2-D space. The arc inscription for the ``Point Store'' place is just the variable \textit{ps}, which can be tied to any of the stored coordinates. The transition ``Calculate Distance'' executes and stores the result of the function call \texttt{calDistance(ps)} in the ``Provers's Distance Store'' place. The arc inscription for the ``Provers' Distance Store'' place is the variable \textit{d}, which can be tied to any of the following places.

\begin{figure}[!h]
  \centerline{\includegraphics[width=8.8cm, height=3cm]{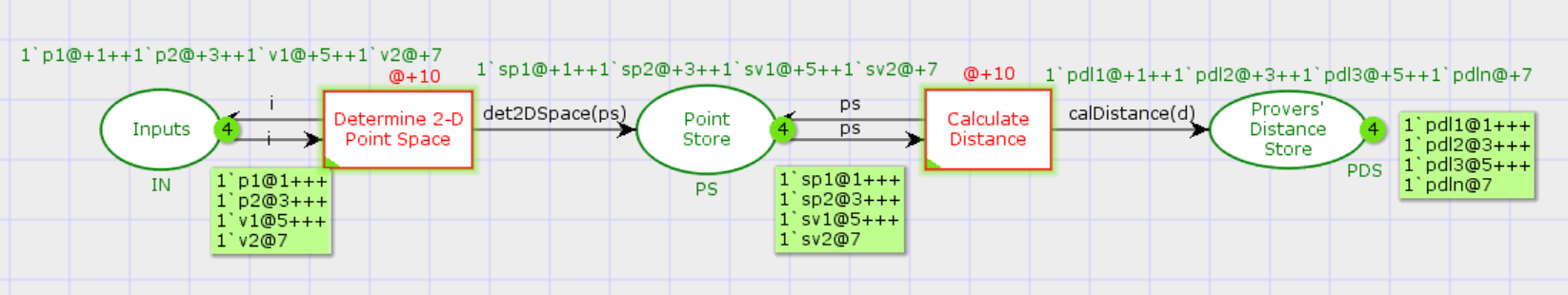}}
  \caption{A timed CPN Model of a Calculate Location Process in LPS}
  \label{PetriNet4_WT}
\end{figure}

The timed CPN model of the Calculate Location process for the LPS in the proposed scheme is depicted in Fig. \ref{PetriNet4_WT}. The following are the timed colour set declarations for the Calculate Location CPN model: \\
\texttt{colset IN = with p1 | p2 | v1 | v2 timed;} \\
\texttt{colset PS = with sp1 | sp2 | sv1 | sv2 timed;} \\
\texttt{colset PDS = with colset PDS = with pdl1 | pdl2 | pdl3 | pdln timed;} \\

\begin{figure*}[!t]
  \centerline{\includegraphics[scale=0.5]{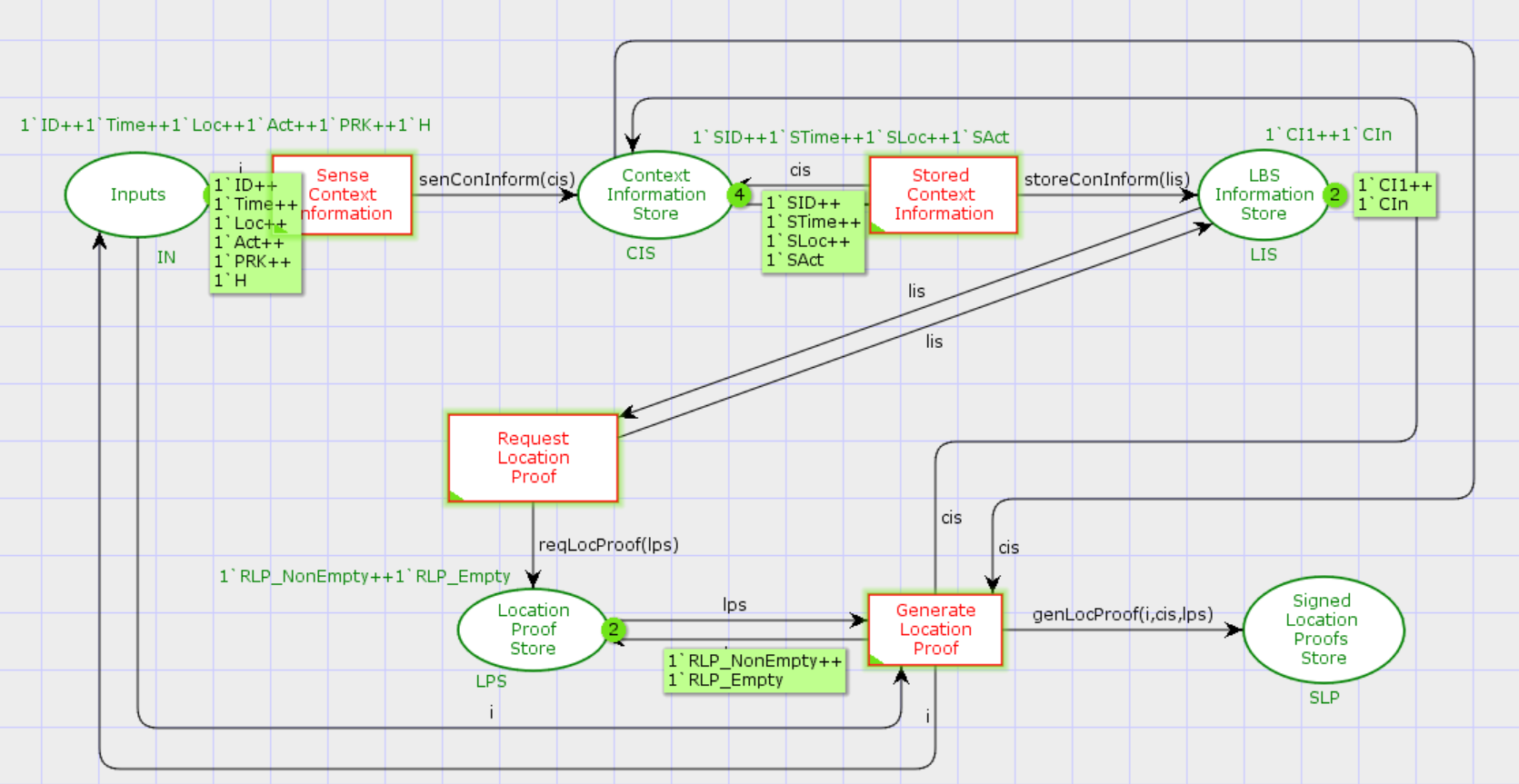}}
  \caption{A CPN Model of a Generate Location Proof Process in LPS}
  \label{PetriNet5_WOT}
\end{figure*}

With these updates, the input parameters in the ``Inputs'' place now include a timestamp showing their arrival time. Each input parameter contains an incremental time stamp, such as \texttt{1'p1@+1++1'p2@+3++1'v1@+5++1'v2@+7}. Therefore, the transition \textit{i} will remain inactive until the world clock reads a time more than or equal to the time stamp on the set parameters. Each coordinate in the ``Point Store'' for the prover and verifier has a unique timestamp assigned incrementally, as follows: \texttt{1'sp1@+1++1'sp2@+3++1'sv1@+5++1'sv2@+7} with the variables \textit{ps} on arc that fire to the ``Calculate Distance'' transition. In the ``Provers' Distance Store'' place, each prover distance has a unique timestamp assigned incrementally, such as \texttt{1'pdl@+1} with the variables \textit{d} on an arc that fire to the next forthcoming transition.

\subsubsection{Generate Location Proof}

Fig. \ref{PetriNet5_WOT} depicts a CPN model of a Generate Location Proof process in the proposed scheme. We refer to this as utilising the rich type declaration facility of CPNs by declaring Inputs as a token with all desired characteristics. The following are the declarations for the colour sets used in the CPN model of produce location proof: \\
\texttt{colset IN = with ID | Time | Loc | Act | PRK | H;} \\
\texttt{colset CIS = with SID | STime | SLoc | SAct;} \\
\texttt{colset LIS = with CI1 | CIn;} \\
\texttt{colset LPS = with $RLP_{NonEmpty}$ | $RLP_{Empty}$;} \\
\texttt{colset SLP = product IN * CIS * LPS;} \\

On the top of the ``Inputs'' place, the multiset inscription such as \texttt{1'ID++1'Time++1'Loc++1'Act++1'PRK++1'H} defines the initial marking for it.  In the green box and circle, these variables show how many total current tokens are present and their colour set declaration. The inscription \texttt{IN} specifies the colour scheme for the ``Inputs'' place. It comprises context information sensed by both provers and verifiers, a private key used to sign location proofs, and a hash function, and is thus represented as\{\textit{ID}, \textit{Time}, \textit{Loc}, \textit{Act}, \textit{PRK}, \textit{H}\}. The multiset inscription, for example, \texttt{1'SID++1'STime++1'SLoc++1'SAct}, defines the initial marking for the ``Context Information Store'' place, while the associated green box and circle indicate the details and count of current tokens existing in the location during simulation. The inscription \texttt{CIS} specifies the colour scheme for the ``Context Information Store'' place. It contains context information that the prover and verifier have stored, such as \{\textit{SID}, \textit{STime}, \textit{SLoc}, \textit{SAct}\}. The multiset inscriptions such as \texttt{1'CI1++1`CIn} specified the initial marking for the ``LBS Information Store'' place, and the accompanying green box and circle indicate the details and count of currently holding tokens in the location during simulation. The inscription \texttt{LIS} declares the colour set and provides a list of context information to be utilised in verifying location proofs such as \{\textit{$CI_{1}$}, \textit{$CI_{2}$}, \textit{$CI_{3}$} $\dots$ \textit{$CI_{n}$}\}. Additionally, the multiset inscriptions such as \texttt{1'$RLP_{NonEmpty}$++1`$RLP_{Empty}$} specified the initial marking for the ``Location Proof Store'' place, while the matching green box and circle display the details and count of current tokens existing in the location during simulation. The inscription \texttt{LPS} specifies the colour set and includes a list of request proofs collected from provers for verification, such as \{\textit{$LP_{1}$}, \textit{$LP_{2}$}, \textit{$LP_{3}$} $\dots$ \textit{$LP_{n}$}\}. Finally, the multiset inscriptions such as \texttt{1'CI++1`$PRK_{N}$++1`$Hash_{N}$++1`$P_{Sig}$} defined the initial marking for the '`Signed Location Proofs Store'' place, while the matching green box and circle display the details and count of current tokens residing in the location during simulation. The inscription \texttt{SLP} of this place declares the colour set and includes information necessary to verify the location proofs, such as context information, the provers' private keys, the hash function, and the generated location proof, which are represented as the product of the inscriptions \texttt{IN}, \texttt{CIS} and \texttt{LPS}.

\begin{figure*}[!t]
  \centerline{\includegraphics[scale=0.5]{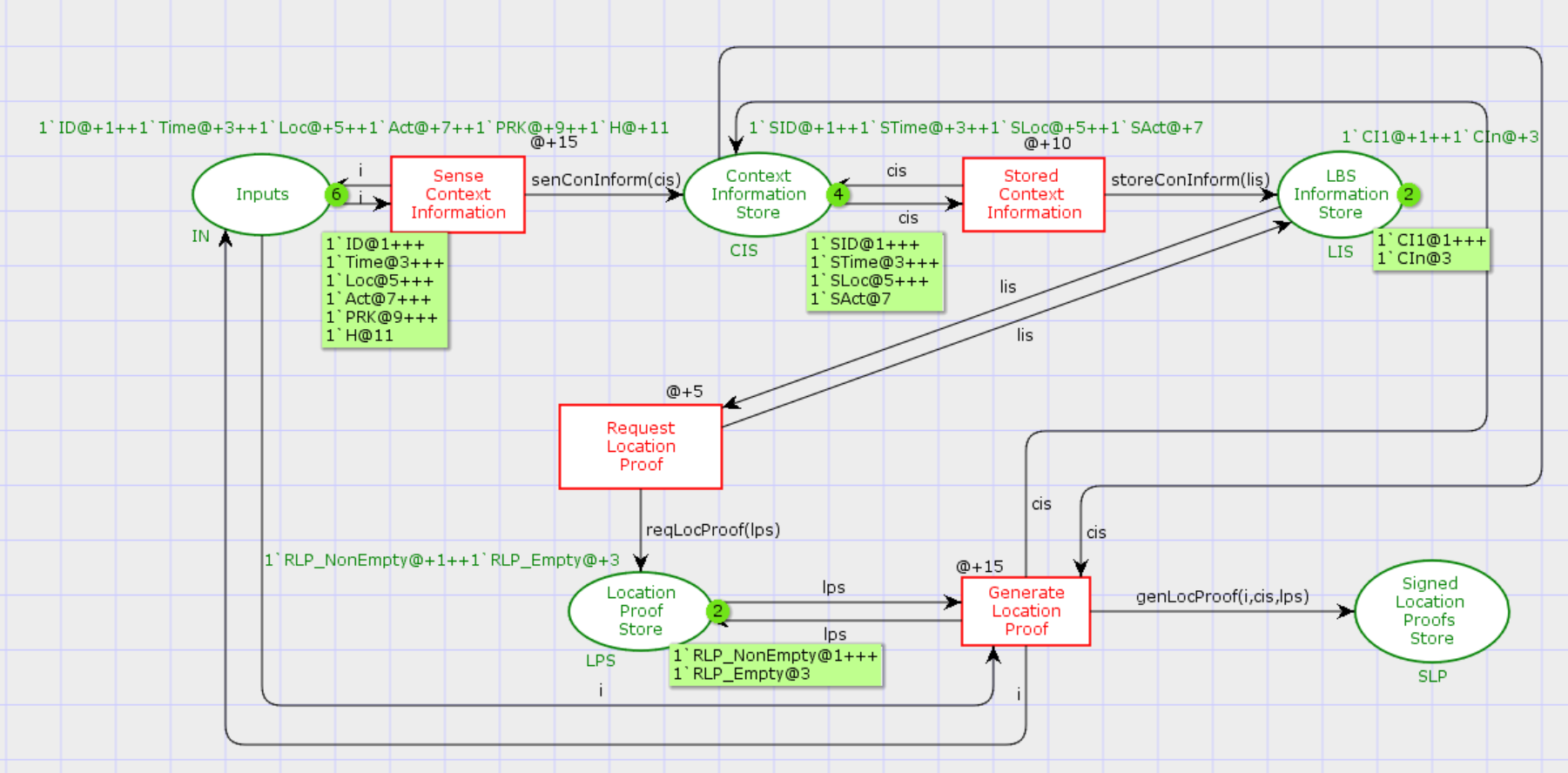}}
  \caption{A timed CPN Model of a Generate Location Proof Process in LPS}
  \label{PetriNet5_WT}
\end{figure*}

A CPN model of Generate Location Proof in the LPS contains five variables, each of which has its own declaration: \\
\texttt{var i: IN;} \\
\texttt{var cis: CIS;} \\
\texttt{var lis: LIS;} \\ 
\texttt{var lps: LPS;} \\
\texttt{var slp: SLP;}

A variable \textit{i} is declared for the colour set \texttt{IN}, a variable \textit{cis} is declared for the colour set  \texttt{CIS}, and a variable \textit{lis} is declared for the colour set \texttt{LIS}, a variable \textit{lps} is declared for the colour set \texttt{LPS} and a variable \textit{slp} is declared for the colour set \texttt{SLP}.

The following are user-defined functions for the Generate Location Proof CPN model.\\
\texttt{fun senConInformation(i) = (cis, IN)}; \\
\texttt{fun storeConInformation(cis) = (lis, CIS)}; \\
\texttt{fun reqLocProof(i, cis, lis) = (i, IN, cis, CIS, lps, LPS)}; \\

Four transitions are used in the Generate Location Proof CPN model: ``Sense Context Information'' and ``Stored Context Information'', ``Request Location Proof'' and ``Generate Location Proof''. The arc inscription of the ``Inputs'' place is merely the variable \textit{i}, which can be tied to any of the context information perceived by the provers and verifiers to reach the ``Sense Context Information'' transition. The transition ``Sense Context Information'' executes and stores the result of the function call \texttt{senConInform(i)} in the ``Context Information Store'' place. The arc inscription for the ``Context Information Store'' place is simply the variable \textit{cis} and can be associated with any of the context information saved. The arc inscription for the ``LBS Information Store'' place is simply the variable \textit{lis}, which can be connected to any of the subsequent transitions; for example, in this case, the subsequent transition is ``Request Location Proof''.
The transition ``Stored Context Information'' executes and stores the result of the function call \texttt{StoreConInform(cis)} in the ``LBS Information Store'' place. The transition ``Request Location Proof'' executes and stores the result of the function call \texttt{reqLocation Proof(lis)} in the ``Location Proof Store'' place. Finally, the transition ``Generate Location Proof'' executes and stores the result of the function call \texttt{genLocProof(i,cis,lps)} in the `` Signed Location Proof Store'' place.

The timed CPN model of the Generate Location Proof process for the LPS in the proposed scheme is depicted in Fig. \ref{PetriNet5_WT}. The following are the timed colour set declarations for the CPN model of Generate Location Proof: \\
\texttt{colset IN = with ID | Time | Loc | Act | PRK | H timed;} \\
\texttt{colset CIS = with SID | STime | SLoc | SAct timed;} \\
\texttt{colset LIS = with CI1 | CIn timed;} \\
\texttt{colset LPS = with $RLP_{NonEmpty}$ | $RLP_{Empty}$ timed;} \\
\texttt{colset SLP = product IN * CIS * LPS timed;} \\

\begin{figure*}[!t]
  \centerline{\includegraphics[scale=0.5]{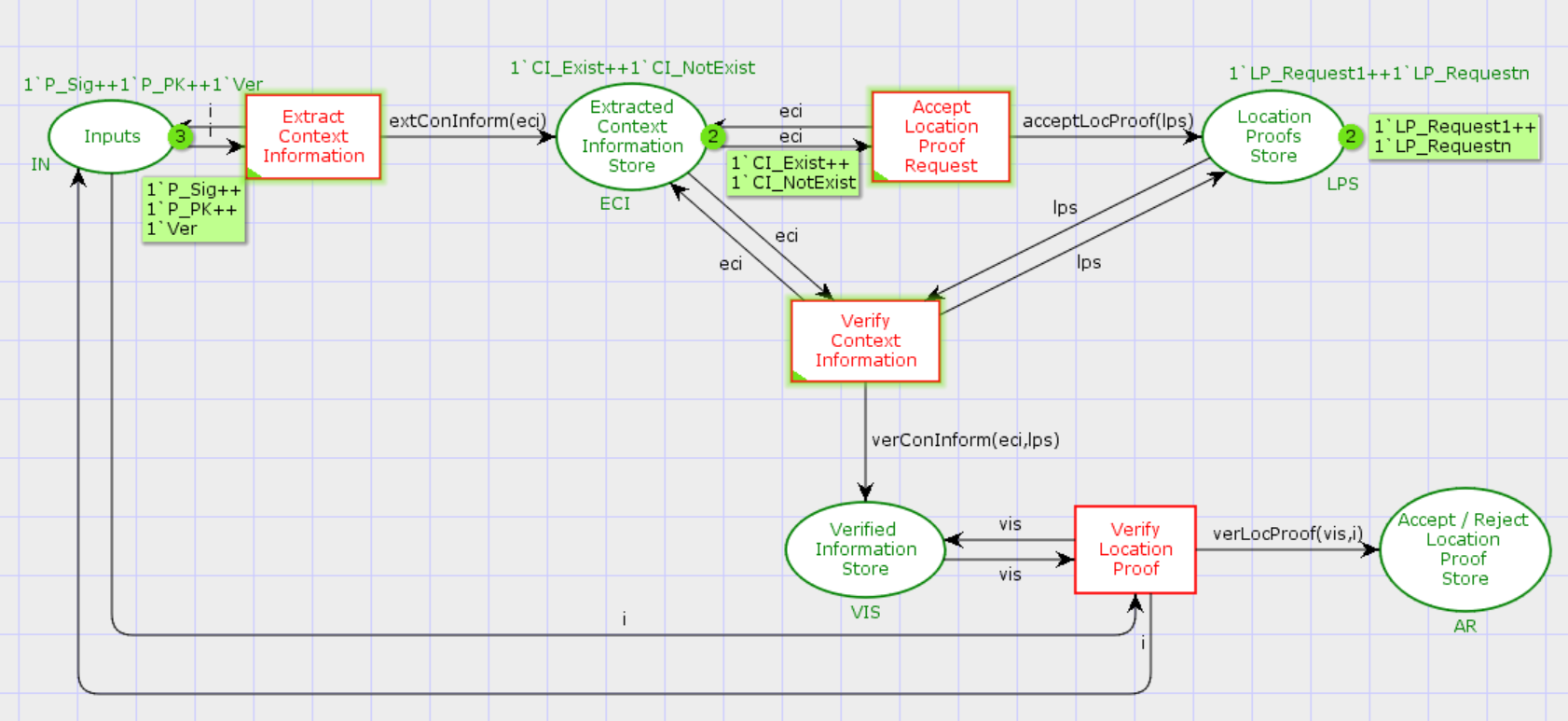}}
  \caption{A CPN Model of a Verify Location Proof Process in LPS}
  \label{PetriNet6_WOT}
\end{figure*}

With these updates, the input parameters in the ``Inputs'' place now include a timestamp showing their arrival time. Each input parameter has an incremental time stamp, such as \texttt{1'ID@+1++1'Time@+3++1'Loc@+5++1'Act@+7++1`PRK\\@+9++1`H@+11}. The transition \textit{i} will remain inactive until the world clock reads a time more than or equal to the time stamp on the set parameters. Following that, each piece of context information stored in the ``Context Information Store'' place by the verifiers is incrementally assigned a unique timestamp, as follows: \texttt{1'ID@+1++1'Time@+3++1'Loc@+5++1'Act@+7} with the variables \textit{cis} on arc that fire to the ``Stored Context Information'' transition. The context information recorded at the LBS and in the "LBS Information Store" has an incremental time stamp, such as  \texttt{1'CI1@+1++1`CIn@+3} with a variable \textit{lis} on an arc that fires to the next transition ``Request Location Proof''. Additionally, the location proof requests carry an incremental time stamp, such as \texttt{1'$RLP_{NonEmpty}$@+1++1`$RLP_{Empty}$@+3}, using a variable \textit{lps} on arc fire to the next transition. Finally, all information associated with signed proofs, such as context information, private key, hash function, and signed proof, and assigned to arc variables \textit{i, cis, slp} via an arc that fires to the next transition, if any.

\subsubsection{Verify Location Proof}

Fig.  \ref{PetriNet6_WOT} illustrates a CPN model of Verify Location Proof process in the proposed scheme. By defining Inputs as a token with all of the desired characteristics, we are utilising CPN powerful type declaration facility. The following are the colour set declarations for the CPN model of verifying location proof: \\
\texttt{colset IN = with $P_{Sig}$ | $P_{PK}$ | Ver;} \\
\texttt{colset ECI = with $CI_{Exist}$ | $CI_{NotExist}$;} \\
\texttt{colset LPS = with $LP_{Request1}$ | $LP_{Requestn}$;} \\
\texttt{colset VIS = product ECI * LPS;} \\
\texttt{colset AR = with VIS * IN;} \\

On the top of the ``Inputs'' place, a multiset inscription such as \texttt{1'$P_{Sig}$++1'\\$P_{PK}$++1'Ver} defines the initial marking for it.  In the green box and circle, these variables show how many total current tokens are present and their colour set declaration. The inscription \texttt{IN} specifies the colour scheme for the ``Inputs'' place. This place contains the signature lists. \{\textit{$P_{Sig_1}$}, \textit{$P_{Sig_2}$}, \textit{$P_{Sig_3}$} $\dots$ \textit{$P_{Sig_n}$}\}, public keys lists \{\textit{$P_{PK_1}$}, \textit{$P_{PK_2}$}, \textit{$P_{PK_3}$} $\dots$ \textit{$P_{PK_n}$}\} and verifiers list \{\textit{$Ver_{1}$}, \textit{$Ver_{2}$}, \textit{$Ver_{3}$} $\dots$ \textit{$Ver_{n}$}\} selected through the use of a trusted model. The multiset inscription \texttt{1'$CI_{Exist}$++1'$CI_{NotExist}$} specifies the initial marking for the ``Extracted Context Information Store'' place, while the associated green box and circle display the details and count of current tokens living in the location during simulation. The inscription \texttt{ECI} specifies the colour scheme for the ``Extracted Context Information Store'' place. It contains context information that the prover and verifier have stored, such as  \{\textit{SID}, \textit{STime}, \textit{SLoc}, \textit{SAct}\}, and has established the presence of such information. The multiset inscriptions such as \texttt{1'$LP_{Request1}$++1`$LP_{Requestn}$} specified the initial marking for the ``Location Proofs Store'' place, and the accompanying green box and circle display the details and count of currently residing tokens in the location during simulation. This place's inscription \texttt{LPS} states the colour set and includes a list of location proof requests that must be validated, such as  \{\textit{$LP_{Request_1}$}, \textit{$LP_{Request_2}$}, \textit{$LP_{Request_3}$} $\dots$ \textit{$LP_{Request_n}$}\}. Additionally, the multiset inscriptions  \texttt{1'VCI1++1`VCIn} specified the initial marking for the ``Verified Information Store'' place, while the accompanying green box and circle indicate the details and count of current tokens residing in the location throughout simulation. The inscription \texttt{VIS} states the colour set and provides a list of validated context information collected from the LBS for verification and represented as the product of the inscriptions \texttt{ECI} and \texttt{LPS}.  Finally, the inscription \texttt{AR} of this location declares the colour set and includes information necessary to verify the location proofs, such as the prover's signature, the prover's public key, and the decision result regarding the location proof's acceptance or rejection, and is thus represented as the product of the inscriptions \texttt{VIS} and \texttt{IN}.

A CPN model of Verify Location Proof in the LPS contains five variables, each of which has its own declaration: \\
\texttt{var i: IN;} \\
\texttt{var eci: ECI;} \\
\texttt{var lps: LPS;} \\ 
\texttt{var vis: VIS;} \\
\texttt{var ar: AR;}

A variable \textit{i} is declared for the colour set \texttt{IN}, a variable \textit{eci} is declared for the colour set \texttt{ECI}, and a variable \textit{lps} is declared for the colour set \texttt{LPS}, a variable \textit{vis} is declared for the colour set \texttt{VIS} and a variable \textit{ar} is declared for the colour set \texttt{AR}.

The following are user-defined functions for the Generate Location Proof CPN model.\\
\texttt{fun extConInform(i) = (eci, IN)}; \\
\texttt{fun accetLocProof(eci) = (lps, ECI)}; \\
\texttt{fun verConInform(eci,lps) = (eci, ECI, lps, LPS)}; \\
\texttt{fun verLocProof(vsi,i) = (vis, VIS, i, IN)}; \\

\begin{figure*}[!t]
  \centerline{\includegraphics[scale=0.5]{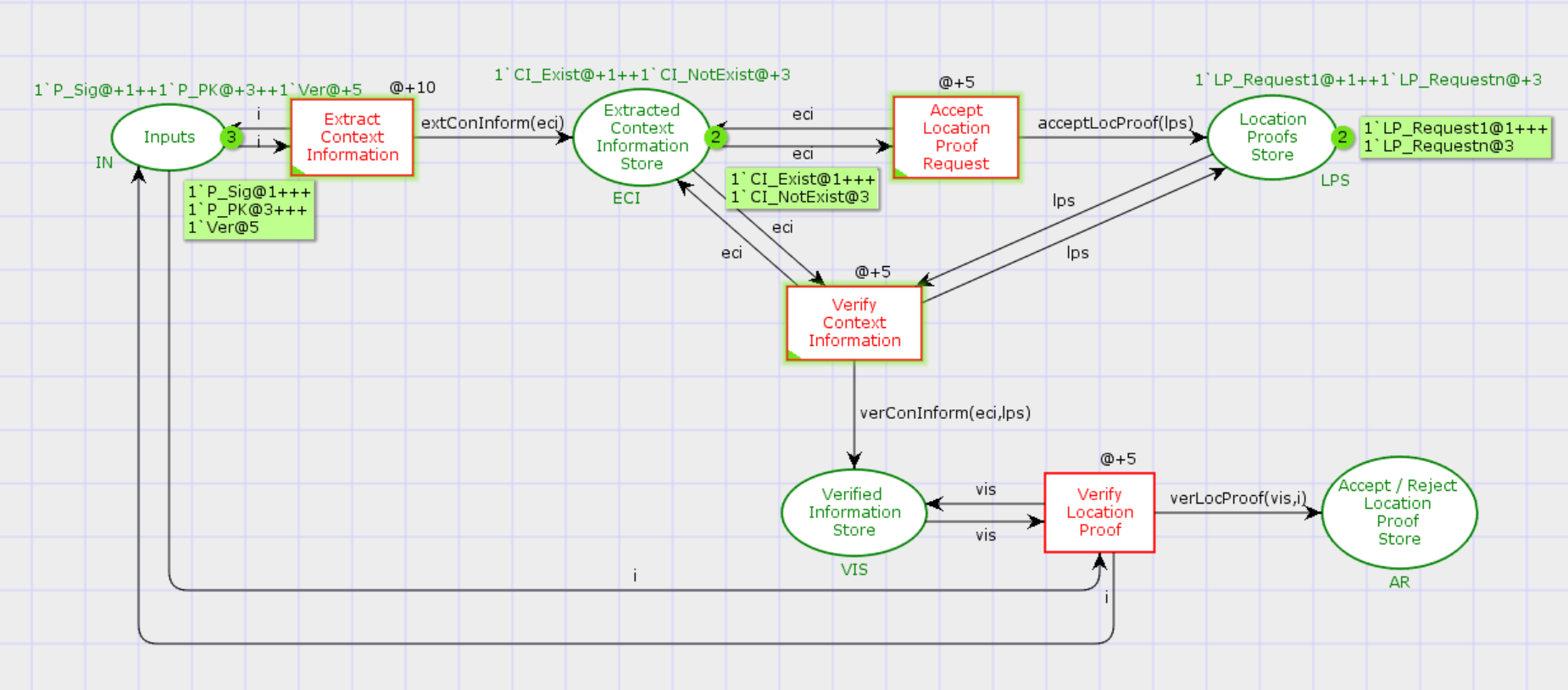}}
  \caption{A timed CPN Model of a Verify Location Proof Process in LPS}
  \label{PetriNet6_WT}
\end{figure*}

Four transitions are used in the Verify Location Proof CPN model: ``Extract Context Information'', ``Accept Location Proof Request'', ``Verify Context Information'' and ``Verify Location Proof''. The arc inscription for the ``Inputs'' place is simply the variable \textit{i}, which can be bound to any of the accessible information, such as the prover's signature, public key, or list of verifiers, and all of which reach the transition ``Extract Context Information''. The transition ``Extract Context Information'' executes and stores the result of the function call \texttt{extConInform(i)} in the ``Extracted Context Information Store'' place. The arc inscription for the ``Extracted Context Information Store'' place is just the variable \textit{eci}, which can be tied to any of the LBS-acquired context information. The arc inscription for the ``Location Proofs Store'' place is simply the variable \textit{lps}, which can be connected to any of the subsequent transitions; for example, in this case, the subsequent transition is `Verify Context Information''. The transition ``Accept Location Proof Request'' executes and stores the result of the function call \texttt{acceptLocProof(eci)} in the ``Location Proofs Store'' place. The transition ``Verify Context Information'' executes and stores the result of the function call \texttt{verConInform(eci,lps)} in the ``Verified Information Store'' place.
Finally, the transition ``Verify Location Proof'' executes and stores the result of the function call \texttt{verLocProof(vis,i)} in the ``Accept / Reject Location Proof Store'' place. The arc inscription for the ``Verified Information Store'' place is a variable \textit{vis}, and the arc inscription for the ``Accept / Reject Location Proof Store'' place are the variables \textit{vis, i}, both of which can be bound to their respective following transitions.

The timed CPN model of the Verify Location Proof procedure for the LPS in the proposed scheme is depicted in Fig. \ref{PetriNet6_WT}. The following are the timed colour set declarations for the CPN model of Verify Location Proof: \\
\texttt{colset IN = with $P_{Sig}$ | $P_{PK}$ | Ver timed;} \\
\texttt{colset ECI = with $CI_{Exist}$ | $CI_{NotExist}$ timed;} \\
\texttt{colset LPS = with $LP_{Request1}$ | $LP_{Requestn}$ timed;} \\
\texttt{colset VIS = product ECI * LPS timed;} \\
\texttt{colset AR = with VIS * IN timed;} \\

With these enhancements, the input parameters in the ``Inputs'' place now include a timestamp showing their arrival time. Each input parameter has an incremental time stamp, such as \texttt{1'$P_{Sig}$@+1++1'$P_{PK}$@+3++1'Ver@+5}. Thus, the transition \textit{i} will remain inactive until the world clock reads a time more than or equal to the time stamp on the set parameters. Following that, each piece of extracted context information from the LBS that is stored in the "Extracted Context Information Store" is assigned an incremental time stamp, such as \texttt{1'$CI_{Exist}$@+1++1'$CI_{NotExist}$@+3}, using the variables \textit{eci} on an arc that fire to the ``Accept Location Proof Request'' transition. The location proof requests received from the provers and kept in the location proofs store include a unique timestamp assigned incrementally, such as \texttt{1'$LP_{Request1}$@+1++1`$LP_{Requestn}$\\@+3} with a variable \textit{lps} on an arc that fires to the following transition ``Verify Context Information''. Additionally, the validated data with a variable \textit{vis} on arc firing to the next transition. Finally, all information about the verification of location proofs, such as signed location proofs, public keys, and decisions regarding location proofs, with the variables \textit{vis, i} via an arc that fires to the next transition, if any.

\subsection{Analysis}

We utilised CPN Tools 4.0.1 \cite{ratzer2003cpn}, the most recent version available, to model and simulate both the ECDSA* technique and LPS technique in the proposed clone node attack detection scheme. CPN Tools is a free, open-source software tool kit that provides comprehensive support for editing, creation, and simulation of CPN models, as well as numerous features for interactive and automatic simulation, monitoring, and process communication. We used the monitoring and data logging capabilities of the CPN Tools to determine the size of the states' markings to keep track of the number of tokens in various places employed in our models. 

The data collector monitors measure two types of statistics: (i) calculate an average (ii) calculate a time average. A monitor tool that computes the mean (of either constant or discrete-parameter variables) is a discrete-parameter statistics calculator. On the other hand, a continuous-time statistic monitor is known as a time-average monitor. Depending on the processes in our proposed scheme, we utilise both types of monitors to perform various calculations, including count (number of observations), minimum, maximum, sum, and average.

Statistics obtained from the data logging facility, such as the total, average, and maximum, are calculated using the values that are returned by the observation function. The sum of the data values recorded by this monitor represented the number of duplicate data packets received throughout a simulation. The percentage of duplicated data packets to the overall number of data packets received is calculated as the average of the data values. Moreover, In simulation-based performance analysis, several scenarios or configurations of a system are frequently compared. As a result, system performance is often determined by multiple variables. In particular, the number of packets that are successfully transmitted, the minimum and maximum periods between the arrival of data packets, and even the diversity of inter-arrival times all impact the timed CPN models.

Tables \ref{CPNResult1_WOT} and \ref{CPNResult1_WT}  summarised the state marking size for the untimed CPN and the timed CPN for an ECDSA* Key Generation process, respectively, utilising various calculated statistics such as token counts, sum, average, minimum, and maximum values.

\begin{table}[!h]
\centering
\tiny
\caption{Marking Size of the States - An Untimed CPN Model of ECDSA* Key Generation}
\begin{adjustbox}{width=0.47\textwidth,center}
\begin{tabular}{|c|c|c|c|c|c|}
\hline
\multicolumn{6}{|c|}{\textbf{Untimed CPN Model Statistics}} \\ \hline
 \textbf{Places}  & \textbf{Count}  &  \textbf{Sum} & \textbf{Average}  & \textbf{Min}  & \textbf{Max}  \\ \hline
  Inputs  &  51 &  56 & 1.098039  & 0  & 5  \\ \hline
  Domain Parameters Store & 51  &  454 & 8.901961  & 5  & 10  \\ \hline
  Keys Store & 51  &  1178 & 23.098039  & 2  & 47  \\ \hline
\end{tabular}
\end{adjustbox}
\label{CPNResult1_WOT}
\vspace{-4mm}%Put here to reduce too much white space after your table 
\end{table}

\begin{table}[!h]
\centering
\tiny
\caption{Marking Size of the States - A Timed CPN Model of ECDSA* Key Generation}
\begin{adjustbox}{width=0.47\textwidth,center}
\begin{tabular}{|c|c|c|c|c|}
\hline
\multicolumn{5}{|c|}{\textbf{Timed CPN Model Statistics}} \\ \hline
 \textbf{Places}  & \textbf{Count}  &  \textbf{Average}  & \textbf{Min}  & \textbf{Max}  \\ \hline
 Inputs  & 7  &  0.510204 & 0  & 5   \\ \hline
  Domain Parameters Store &  52  &  9.489796 & 5  & 10   \\ \hline
  Keys Store & 47 & 22.000000  & 2  &    47  \\ \hline
\end{tabular}
\end{adjustbox}
\label{CPNResult1_WT}
\vspace{-4mm}%Put here to reduce too much white space after your table 
\end{table}

Tables \ref{CPNResult2_WOT} and \ref{CPNResult2_WT}  summarised the state marking size for the untimed CPN and the timed CPN for the ECDSA* Signature Generation process, respectively, utilising various calculated statistics such as token counts, sum, average, minimum, and maximum values.

\begin{table}[!h]
\centering
\tiny
\caption{Marking Size of the States - An Untimed CPN Model of ECDSA* Signature Generation}
\begin{adjustbox}{width=0.47\textwidth,center}
\begin{tabular}{|c|c|c|c|c|c|}
\hline
\multicolumn{6}{|c|}{\textbf{Untimed CPN Model Statistics}} \\ \hline
 \textbf{Places}  & \textbf{Count}  &  \textbf{Sum} & \textbf{Average}  & \textbf{Min}  & \textbf{Max}  \\ \hline
  Inputs  &  51 &  204 & 4.000000  & 4  & 4  \\ \hline
  Coordinates Store & 51  &  377 & 7.392157  & 2  & 13  \\ \hline
  Hash Integer Store & 51  &  280 & 5.490196  & 1  & 12  \\ \hline
  Signature Pair 1 Store & 51  &  330 & 6.470588  & 2  & 10  \\ \hline
Signature Pair 2 Store & 51  &  393 & 7.705882  & 4  & 10  \\ \hline
Signature Store & 51 &  118 &2.313725  & 0  & 5  \\ \hline
  
\end{tabular}
\end{adjustbox}
\label{CPNResult2_WOT}
\vspace{-4mm}%Put here to reduce too much white space after your table 
\end{table}

\begin{table}[!h]
\centering
\tiny
\caption{Marking Size of the States - A Timed CPN Model of ECDSA* Signature Generation}
\begin{adjustbox}{width=0.47\textwidth,center}
\begin{tabular}{|c|c|c|c|c|}
\hline
\multicolumn{5}{|c|}{\textbf{Timed CPN Model Statistics}} \\ \hline
 \textbf{Places}  & \textbf{Count}  &  \textbf{Average}  & \textbf{Min}  & \textbf{Max}  \\ \hline
 Inputs  & 18  &  4.000000 & 4  & 4   \\ \hline
 Coordinates Store &  43 & 6.242424  & 2  & 11    \\ \hline
Hash Integer Store & 9 & 2.818182  &  1 &    4  \\ \hline
Signature Pair 1 Store & 36 & 13.575758  & 2  &    34  \\ \hline
Signature Pair 2 Store & 8 & 3.969697  & 3  &    7  \\ \hline
Signature Store & 4 & 0.787879  & 0  &    2  \\ \hline

\end{tabular}
\end{adjustbox}
\label{CPNResult2_WT}
\vspace{-4mm}%Put here to reduce too much white space after your table 
\end{table}

Tables \ref{CPNResult3_WOT} and \ref{CPNResult3_WT} summarised the state marking size for the untimed CPN and the timed CPN for the ECDSA* Signature Verification process, respectively, utilising various calculated statistics such as token counts, sum, average, minimum, and maximum values.

\begin{table}[!h]
\centering
\tiny
\caption{Marking Size of the States - An Untimed CPN Model of ECDSA* Signature Verification}
\begin{adjustbox}{width=0.47\textwidth,center}
\begin{tabular}{|c|c|c|c|c|c|}
\hline
\multicolumn{6}{|c|}{\textbf{Untimed CPN Model Statistics}} \\ \hline
 \textbf{Places}  & \textbf{Count}  &  \textbf{Sum} & \textbf{Average}  & \textbf{Min}  & \textbf{Max}  \\ \hline
  Inputs  &  51 &  153 & 3.000000  & 3  & 3  \\ \hline
  Signature Store & 51  &  123 & 2.411765 & 0  & 5  \\ \hline
  Hash Integer Store & 51  &  325 & 6.372549  & 2  & 11  \\ \hline
  Point Store & 51  &  429 & 8.411765  & 2  & 15  \\ \hline
Coordinates Store & 51  &  175 & 3.431373  & 0  & 8  \\ \hline
Accept / Reject & 51  &  354 & 6.941176  & 0  & 11  \\ \hline

\end{tabular}
\end{adjustbox}
\label{CPNResult3_WOT}
\vspace{-4mm}%Put here to reduce too much white space after your table 

\end{table}

\begin{table}[!h]
\centering
\tiny
\caption{Marking Size of the States - A Timed CPN Model of ECDSA* Signature Verification}
\begin{adjustbox}{width=0.47\textwidth,center}
\begin{tabular}{|c|c|c|c|c|}
\hline
\multicolumn{5}{|c|}{\textbf{Timed CPN Model Statistics}} \\ \hline
 \textbf{Places}  & \textbf{Count}  &  \textbf{Average}  & \textbf{Min}  & \textbf{Max}  \\ \hline
 Inputs  & 32  &  3.000000 & 3  & 3   \\ \hline
 
Signature Store &  34& 1.589474  & 0  & 3    \\ \hline

Hash Integer Store & 25 & 7.621053  & 2  &    13  \\ \hline

Point Store &22 & 5.294737  & 2  &    10  \\ \hline

Coordinates Store & 21 & 6.452632  & 0  &    12  \\ \hline

Accept / Reject & 9 & 3.178947  & 0  &    7  \\ \hline
\end{tabular}
\end{adjustbox}
\label{CPNResult3_WT}
\vspace{-4mm}%Put here to reduce too much white space after your table 
\end{table}

\begin{table}[!h]
\centering
\tiny
\caption{Marking Size of the States - An Untimed CPN Model of Calculate Location for LPS}
\begin{adjustbox}{width=0.47\textwidth,center}
\begin{tabular}{|c|c|c|c|c|c|}
\hline
\multicolumn{6}{|c|}{\textbf{Untimed CPN Model Statistics}} \\ \hline
 \textbf{Places}  & \textbf{Count}  &  \textbf{Sum} & \textbf{Average}  & \textbf{Min}  & \textbf{Max}  \\ \hline
  Inputs  &  51 &  204 & 4.000000  & 4  & 4  \\ \hline
  Point Store & 51  &  834 & 16.352941 & 4  & 28  \\ \hline
  Provers' Distance Store & 51  &  696 & 13.647059  & 1  & 27  \\ \hline
\end{tabular}
\end{adjustbox}
\label{CPNResult4_WOT}
\vspace{-4mm}%Put here to reduce too much white space after your table 
\end{table}

\begin{table}[!h]
\centering
\tiny
\caption{Marking Size of the States - A Timed CPN Model of Calculate Location for LPS}
\begin{adjustbox}{width=0.47\textwidth,center}
\begin{tabular}{|c|c|c|c|c|}
\hline
\multicolumn{5}{|c|}{\textbf{Timed CPN Model Statistics}} \\ \hline
 \textbf{Places}  & \textbf{Count}  &  \textbf{Average}  & \textbf{Min}  & \textbf{Max}  \\ \hline
 Inputs  & 17  &  4.000000 & 4  & 4   \\ \hline
Point Store &  52& 11.371429  & 4  & 19    \\ \hline
Provers' Distance Store  & 37 & 16.800000  & 4  &    39  \\ \hline
\end{tabular}
\end{adjustbox}
\label{CPNResult4_WT}
\vspace{-4mm}%Put here to reduce too much white space after your table 
\end{table}

For the clone node attack detection scheme, Tables \ref{CPNResult4_WOT} and \ref{CPNResult4_WT} summarised the state marking size for the untimed CPN and the timed CPN for the Calculate Location process, respectively, utilising various calculated statistics such as token counts, sum, average, minimum, and maximum values.

\begin{table}[!h]
\centering
\tiny
\caption{Marking Size of the States - An Untimed CPN Model of Generate Location Proof for LPS}
\begin{adjustbox}{width=0.47\textwidth,center}
\begin{tabular}{|c|c|c|c|c|c|}
\hline
\multicolumn{6}{|c|}{\textbf{Untimed CPN Model Statistics}} \\ \hline
 \textbf{Places}  & \textbf{Count}  &  \textbf{Sum} & \textbf{Average}  & \textbf{Min}  & \textbf{Max}  \\ \hline
  Inputs  &  51 &  306 & 6.000000  & 6  & 6  \\ \hline
  
  Context Information Store & 51  &  628 & 12.313725& 4  & 19  \\ \hline
  
  LBS Information Store & 51  &  408 & 8.000000  & 2  & 16  \\ \hline
  
   Location Proof Store & 51  &  511 & 10.019608  & 2  & 18  \\ \hline
  
    Signed Location Proofs Store& 51  &  340 & 6.666667  & 4  & 9  \\ \hline
\end{tabular}
\end{adjustbox}
\label{CPNResult5_WOT}
\vspace{-4mm}%Put here to reduce too much white space after your table 
\end{table}

\begin{table}[!h]
\centering
\tiny
\caption{Marking Size of the States - A Timed CPN Model of Generate Location Proof for LPS}
\begin{adjustbox}{width=0.47\textwidth,center}
\begin{tabular}{|c|c|c|c|c|}
\hline
\multicolumn{5}{|c|}{\textbf{Timed CPN Model Statistics}} \\ \hline
 \textbf{Places}  & \textbf{Count}  &  \textbf{Average}  & \textbf{Min}  & \textbf{Max}  \\ \hline
 Inputs  & 14  &  6.000000 & 6  & 6   \\ \hline
 
Context Information Store &  26& 8.266667  & 4  & 12    \\ \hline

LBS Information Store  & 40 & 6.733333  & 2  &    14  \\ \hline

 Location Proof Store  & 32 & 11.8000000  & 2  &    28  \\ \hline
 
 Signed Location Proofs Store  & 6 & 6.333333  & 4  &    8  \\ \hline
\end{tabular}
\end{adjustbox}
\label{CPNResult5_WT}
\vspace{-4mm}%Put here to reduce too much white space after your table 
\end{table}

Tables \ref{CPNResult5_WOT} and \ref{CPNResult5_WT} summarised the state marking size for the untimed CPN and the timed CPN for the Generate Location Proof, respectively, utilising various calculated statistics such as token counts, sum, average, minimum, and maximum values.

Finally, Tables \ref{CPNResult6_WOT} and \ref{CPNResult6_WT} summarised the state marking size for the untimed CPN and the timed CPN for the Verify Location Proof, respectively, utilising various calculated statistics such as token counts, sum, average, minimum, and maximum values.\\

\begin{table}[!h]
\centering
\tiny
\caption{Marking Size of the States - An Untimed CPN Model of Verify Location Proof for LPS}
\begin{adjustbox}{width=0.47\textwidth,center}
\begin{tabular}{|c|c|c|c|c|c|}
\hline
\multicolumn{6}{|c|}{\textbf{Untimed CPN Model Statistics}} \\ \hline
 \textbf{Places}  & \textbf{Count}  &  \textbf{Sum} & \textbf{Average}  & \textbf{Min}  & \textbf{Max}  \\ \hline
  Inputs  &  51 &  153 & 3.000000  & 3  & 3  \\ \hline
  
  Extracted Context Information Store & 51  &  484 & 9.490196& 2  & 14  \\ \hline
  
 Location Proofs Store & 51  &  417 & 8.176471  & 2  & 14  \\ \hline
 
   Verified Information Store & 51  &  286 & 5.607843  & 0  & 11  \\ \hline
   
    Accept / Reject Location Proof Store& 51  &  292 & 5.725490  & 0  & 15  \\ \hline
\end{tabular}
\end{adjustbox}
\label{CPNResult6_WOT}
\vspace{-4mm}%Put here to reduce too much white space after your table 
\end{table}

\begin{table}[!h]
\centering
\tiny
\caption{Marking Size of the States - A Timed CPN Model of Verify Location Proof for LPS}
\begin{adjustbox}{width=0.47\textwidth,center}
\begin{tabular}{|c|c|c|c|c|}
\hline
\multicolumn{5}{|c|}{\textbf{Timed CPN Model Statistics}} \\ \hline
 \textbf{Places}  & \textbf{Count}  &  \textbf{Average}  & \textbf{Min}  & \textbf{Max}  \\ \hline
 
 Inputs  & 16  &  3.000000 & 3  & 3   \\ \hline
 
 Extracted Context Information Store &  47& 6.941176  & 2  & 11    \\ \hline

Location Proofs Store  & 38 & 9.441176  & 2  &    20  \\ \hline

 Verified Information Store  & 25 & 6.764706  & 0  &    18  \\ \hline

Accept / Reject Location Proof Store & 7 & 1.705882  & 0  &    5  \\ \hline

\end{tabular}
\end{adjustbox}
\label{CPNResult6_WT}
\vspace{-4mm}%Put here to reduce too much white space after your table 
\end{table}

The CPN model describes the complex network in great detail, explains the way in which it works, and details the various ways information flows through it. Moreover, the CPN models are designed to ensure the integrity of and to demonstrate the presence of deadlocks in a system. From the above CPNs models and the result analysis of the marking size of places in these models, it is clear that all of the model's states are accessible. Each model transition is enabled/fired during the simulation. Additionally, these models have no bottlenecks, indicating that they are entirely free of deadlocks. There may be a range of tokens for each state, as the execution pattern or the number of times a transition is fired will influence this number.

\section{Conclusion}\label{conclusion}

This paper has studied, examined, and analysed our existing proposed clone node attack detection scheme for IoT networks. We modelled the proposed scheme using HLPNs, which have the advantage of giving a stable mathematical description and analysing its functional and structural characteristics. We specified several properties defining the proposed scheme's behaviour using the Z specification language, and if the models satisfy particular properties, they are considered correct. Further, we validated our models using a model checking technique, utilising the SMT-Lib and Z3 solver. Verification results demonstrated in this work indicate that the models used in the proposed framework are adequate for behavioural aspects. We extended our work by modelling the proposed scheme using CPNs, which extend Petri Nets with high-level programming language-like properties, making them more effective and ideal for modelling complex systems. Finally, we verified the logical validity and performance of the proposed approach using both untimed and timed models of CPNs.

\section*{Declaration of Competing Interest}

The authors of this paper declare that they have no known competing financial interests or personal relationships that could have influenced the work reported in this paper.

\section*{Funding}

This research did not receive any specific grant from funding agencies in the public, commercial, or
not-for-profit sectors

 \bibliographystyle{elsarticle-num} 
\bibliography{PaperFile.bib}





% \end{thebibliography}

\appendix

\section{Pipe+ Illustrations of HLPN Models} \label{HLPNsModels}

This appendix includes Pipe+ illustrations for the following HLPN models: ECDSA* key generation, ECDSA* signature generation, ECDSA* signature verification, calculate location, generate location proof, and verify location proof.

\begin{figure}[!h]
  \centerline{\includegraphics[width=9cm,height=2cm]{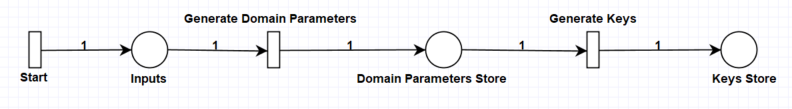}}
  \caption{A Pipe+ View of HLPN Model of an ECDSA* Key Generation}
  \label{pipe1}
\end{figure}

Fig. \ref{pipe1} depicts the HLPN model of an ECDSA* Key Generation, which consists of three transitions and three places. All transitions in this HLPN are timed.

%The include places are inputs, domain parameters store, and keys store, denoted by \textit{P} = \{ ``Inputs'', ``Domain Parameters Store'', ``Keys Store''\}, respectively. The transitions used here, such as start, generate domain parameters, and generate keys, are timed transitions and denoted by \textit{$T_{t}$} = \{ ``Start'', ``Generate Domain Parameters'', ``Generate Keys''\}.

\begin{figure}[!h]
  \centerline{\includegraphics[width=8.5cm, height=4cm]{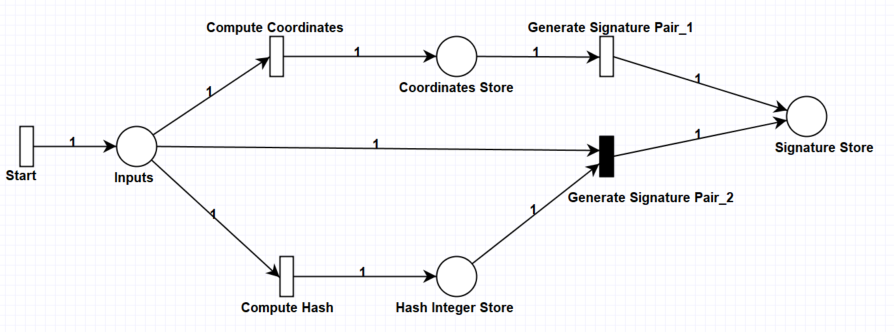}}
  \caption{A Pipe+ View of HLPN Model of an ECDSA* Signature Generation}
  \label{pipe2}
\end{figure}

Fig. \ref{pipe2} shows the HLPN model for ECDSA* Signature Generation. This HLPN has four locations and five transitions. Only "Generate Signature Pair 2" is utilised as an immediate transition.

%The places are inputs, coordinates store, hash integer store, and signature store, denoted by \textit{P} = \{ ``Inputs'', ``Coordinates Store'', ``Hash Integer Store'', ``Signature Store''\}. The transitions start, compute coordinates, compute hash, and generate signature pair 1 are all timed transitions and are denoted by \textit{$T_{t}$} = \{ ``Start'', ``Compute Coordinates'', ``Compute Hash'', ``Generate Signature Pair 1''\}, respectively, whereas the transition generate signature pair 2 is an immediate transition denoted by \textit{$T_{i}$} = \{ ``Generate Signature Pair 2''\}.

\begin{figure}[!h]
  \centerline{\includegraphics[width=9cm, height=5cm]{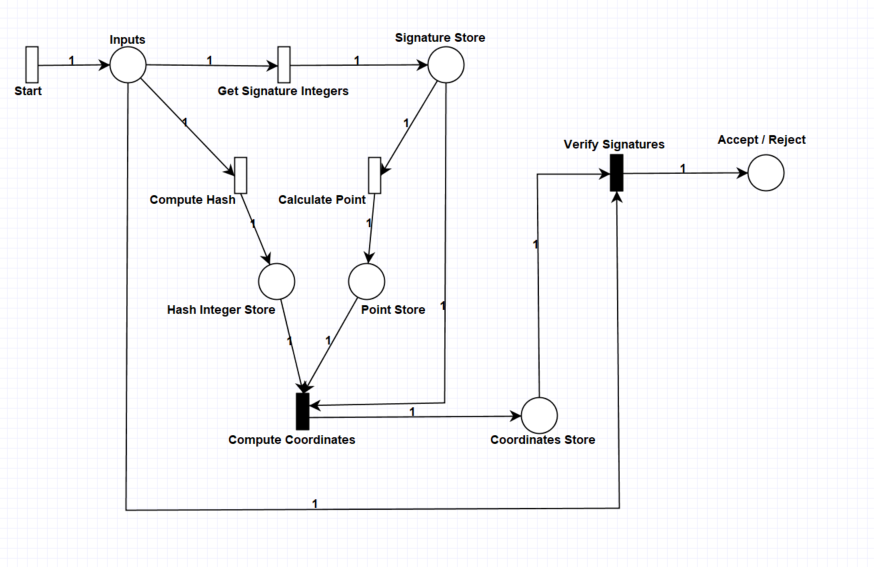}}
  \caption{A Pipe+ View of HLPN Model of an ECDSA* Signature Verification}
  \label{pipe3}
\end{figure}

Fig. \ref{pipe3} shows the HLPN model of an ECDSA* Signature Verification. This HLPN has six places and six transitions. Two of the six transitions are immediate: "Compute Coordinates" and "Verify Signatures".

\begin{figure}[!h]
  \centerline{\includegraphics[width=9cm, height=2cm]{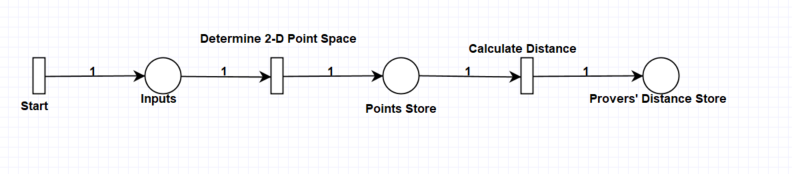}}
  \caption{A Pipe+ View of HLPN Model of the Calculate Location in LPS}
  \label{pipe5}
  \vspace{-4mm}%Put here to reduce too much white space after your table 

\end{figure}

Fig. \ref{pipe5} depicts the HLPN of Calculate Location in LPS, which has three places and three transitions.  All transitions in this HLPN are timed.

%The places are inputs, points stores, and provers' distance stores, and are denoted by \textit{P} = \{ ``Inputs'', ``Points Store'', ``Provers' Distance Store''\}, respectively. The transitions used in this scenario are timed transitions denoted by \textit{$T_{t}$} = \{ ``Start'', ``Determine 2-D Point Space'', ``Calculate Distance''\}.

\begin{figure}[!h]
  \centerline{\includegraphics[width=9cm,height=5cm]{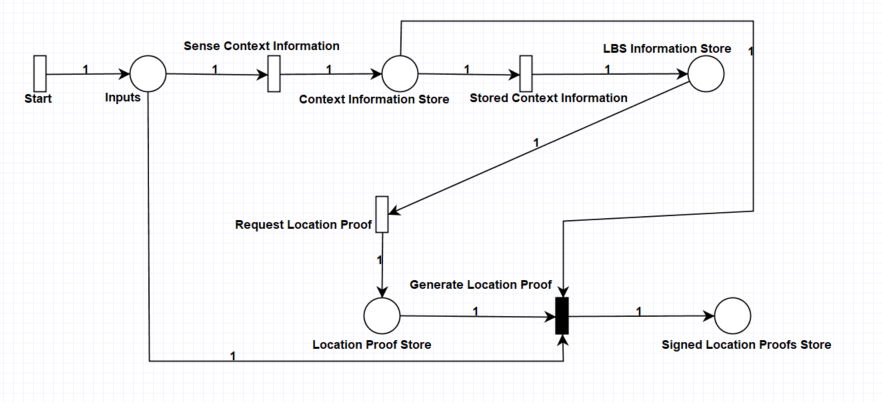}}
  \caption{A Pipe+ View of HLPN Model of Generate Location Proof in LPS}
  \label{pipe6}
  \vspace{-4mm}%Put here to reduce too much white space after your table 

\end{figure}

Fig. \ref{pipe6} illustrates the HLPN model for Generate Location Proof in LPS, which has five places and five transitions. A "Generate Location Proof" is the only immediate transition.

%Inputs, context information store, LBS information store, location proof store, and signed location proof store are all examples of places. They are represented as \textit{P} = \{ ``Inputs'', ``Context Information Store'', ``LBS Information Store'',``Location Proof Store'',``Signed Location Proof Store''\}, respectively. The transitions used, such as start, sense context information, stored context information, and request location proof, are timed transitions and denoted by \textit{$T_{t}$} = \{ ``Start'', ``Sense Context Information'', ``Stored Context Information'',``Request Location Proof''\} whereas the generate location proof transition is used as an immediate transition and is denoted by \textit{$T_{i}$} = \{ ``Generate Location Proof''\}.

\begin{figure}[!h]
  \centerline{\includegraphics[width=9cm,height=6cm]{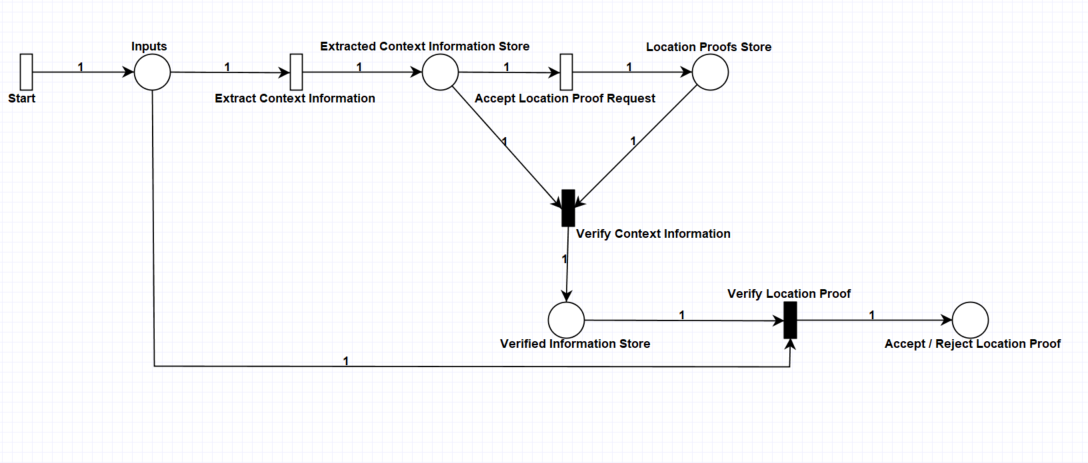}}
  \caption{A Pipe+ View of HLPN Model of the Verify Location Proof in LPS}
  \label{pipe7}
\end{figure}

Fig. \ref{pipe7} illustrates the HLPN model of the verify location proof in LPS, which comprises five places and five transitions. From the five transitions, two are used as immediate transitions: "Verify Context Information", "Verify Location Proof".

%Inputs, extracted context information store, location proof store, verified information store, and accept/reject location proof are all places that are represented as \textit{P} = \{ ``Inputs'', ``Extracted Context Information Store'', ``Location Proof Store'',``Verified Information Store'',``Accept/ Reject Location Proof''\}, respectively. The transitions start, extract context information, and accept location proof request are timed transitions and are represented as \textit{$T_{t}$} = \{ ``Start'', ``Extract Context Information'', ``Accept Location Proof Request''\}, respectively, whereas the transitions verify context information and verify location proof are immediate transitions and are represented as \textit{$T_{i}$} = \{ ``Verify Context Information'', ``Verify Location Proof''\}.

\vspace{-4mm}%Put here to reduce too much white space after your table 

\section{Incidence Marking} \label{IncidenceMarkingResults}

This appendix presents the results of the incidence markings of HLPN models, which comprise ECDSA* signature generation, ECDSA* signature verification, calculate location, generate location proof and verify location proof. The results of the incidence markings are determined using several matrices, including forwards, backwards, combined, and inhibition.

\begin{table}[!h]
\tiny
\centering
\caption{High Level Petri Net Incidences - ECDSA* Signature Generation}

%{\tiny\renewcommand{\arraystretch}{0.2}
%\resizebox{!}{0.15\paperheight}{%
\begin{adjustbox}{width=0.47\textwidth,center}
\begin{tabular}{|c|c|c|c|c|c|}
\hline
\multicolumn{6}{|l|}{\textbf{Forwards Incidence Matrix \textit{$I^{+}$}}} \\ \hline
   &  Start  & \makecell{Compute \\ Coordinates}   & \makecell{Compute \\ Hash}    & \makecell{Generate\\ Signature  Pair 1} & \makecell{Generate\\ Signature Pair 2}   \\ \hline
 $\varphi$ $(Inputs)$  &1    &0    & 0   &    0&0\\ \hline
  $\varphi$ $\makecell{(Coordinates \\ Store)}$  & 0   & 1   & 0   & 0 &0   \\ \hline
 $\varphi$ $\makecell{(Hash Integer\\ Store)}$  &    0& 0   &    1& 0 & 0  \\ \hline
  $\varphi$ $\makecell{(Signature \\ Store)}$ & 0   & 0   & 0   & 1  &1 \\ \hline

\multicolumn{6}{|l|}{\textbf{Backwards Incidence Matrix \textit{$I^{-}$}}} \\ \hline
 &  Start  & \makecell{Compute \\ Coordinates}   & \makecell{Compute \\ Hash}    & \makecell{Generate\\ Signature  Pair 1} & \makecell{Generate\\ Signature Pair 2}   \\ \hline
  $\varphi$ $(Inputs)$ &  0  &    1&  1  &0   &1 \\ \hline
  $\varphi$ $\makecell{(Coordinates \\ Store)}$   & 0   & 0   & 0   & 1 & 0  \\ \hline
   $\varphi$ $\makecell{(Hash Integer\\ Store)}$ & 0   & 0   & 0   & 0 & 1   \\ \hline
 $\varphi$ $\makecell{(Signature \\ Store)}$  & 0   &0    & 0   &    0&0\\ \hline

\multicolumn{6}{|l|}{\textbf{Combined Incidence Matrix \textit{I}}} \\ \hline
  &  Start  & \makecell{Compute \\ Coordinates}   & \makecell{Compute \\ Hash}    & \makecell{Generate\\ Signature  Pair 1} & \makecell{Generate\\ Signature Pair 2}   \\ \hline
 $\varphi$ $(Inputs)$  & 1   &    -1&  -1  & 0 & -1  \\ \hline
$\varphi$ $\makecell{(Coordinates \\ Store)}$   &    0& 1   &    0& -1 & 0   \\ \hline
  $\varphi$ $\makecell{(Hash Integer\\ Store)}$  & 0   & 0   &1    &0 & -1    \\ \hline
$\varphi$ $\makecell{(Signature \\ Store)}$    & 0   &    0&  0  &1 & 1   \\ \hline

\multicolumn{6}{|l|}{\textbf{Inhibition Matrix \textit{H}}} \\ \hline
&  Start  & \makecell{Compute \\ Coordinates}   & \makecell{Compute \\ Hash}    & \makecell{Generate\\ Signature  Pair 1} & \makecell{Generate\\ Signature Pair 2}  \\ \hline
 $\varphi$ $(Inputs)$ &   0 & 0   & 0   & 0 & 0   \\ \hline
$\varphi$ $\makecell{(Coordinates \\ Store)}$  &    0& 0   &    0& 0 & 0  \\ \hline
  $\varphi$ $\makecell{(Hash Integer\\ Store)}$  & 0   & 0   & 0   & 0 & 0  \\ \hline
 $\varphi$ $\makecell{(Signature \\ Store)}$   & 0   &    0&  0  &    0& 0\\ \hline
\end{tabular}
\end{adjustbox}
\label{table:petri2}
\vspace{-4mm}%Put here to reduce too much white space after your table 
\end{table}

Table \ref{table:petri2} presents the results of forwards, backwards, combined, and inhibition incidence matrices for an ECDSA* Signature Generation Petri net.

The forwards, backwards, combined, and inhibition matrices for the ECDSA* Signature Verification Petri Net are shown in Table \ref{table:petri3}.

\begin{table}[!h]
\tiny
\centering
\caption{High Level Petri Net Incidences - ECDSA* Signature Verification}
%{\tiny\renewcommand{\arraystretch}{0.2}
%\resizebox{!}{0.15\paperheight}{%
\begin{adjustbox}{width=0.47\textwidth,center}

\begin{tabular}{|c|c|c|c|c|c|c|}
\hline
\multicolumn{7}{|l|}{\textbf{Forwards Incidence Matrix \textit{$I^{+}$}}} \\ \hline
   &  Start  & \makecell{Get \\ Signature \\ Integers}   & \makecell{Compute \\ Hash}    & \makecell{Calculate\\ Point} & \makecell{Compute\\ Coordinates} & \makecell{Verify \\ Signatures} \\ \hline
   
 $\varphi$ $(Inputs)$  &1    &0    & 0   &    0&0&0\\ \hline
  $\varphi$ $\makecell{(Signatures \\ Store)}$  & 0   & 1   & 0   & 0 &0  &0 \\ \hline
 $\varphi$ $\makecell{(Hash \\ Integer\\ Store)}$  &    0& 0   &    1& 0 & 0 & 0  \\ \hline
  $\varphi$ $\makecell{(Point \\ Store)}$ & 0   & 0   & 0   & 1  &0 &0 \\ \hline
  $\varphi$ $\makecell{(Coordinates \\ Store)}$ & 0   & 0   & 0   & 0  &1 & 0 \\ \hline
  $\varphi$ $\makecell{(Accept \\ Reject)}$ & 0   & 0   & 0   & 0  &0&1 \\ \hline

\multicolumn{7}{|l|}{\textbf{Backwards Incidence Matrix \textit{$I^{-}$}}} \\ \hline
&  Start  & \makecell{Get \\ Signature \\ Integers}   & \makecell{Compute \\ Hash}    & \makecell{Calculate\\ Point} & \makecell{Compute\\ Coordinates} & \makecell{Verify \\ Signatures} \\ \hline

  $\varphi$ $(Inputs)$  &0    &1    & 1   &0    &0&1\\ \hline
  $\varphi$ $\makecell{(Signatures \\ Store)}$  & 0   & 0  & 0   & 1 &1&0   \\ \hline
 $\varphi$ $\makecell{(Hash \\ Integer\\ Store)}$  &    0& 0   &    0& 0 & 1 &0  \\ \hline
  $\varphi$ $\makecell{(Point \\ Store)}$ & 0   & 0   & 0   & 0  &1 &0\\ \hline
  $\varphi$ $\makecell{(Coordinates \\ Store)}$ & 0   & 0   & 0   & 0  &0 &1 \\ \hline
  $\varphi$ $\makecell{(Accept \\ Reject)}$ & 0   & 0   & 0   & 0 &0&0 \\ \hline

\multicolumn{7}{|l|}{\textbf{Combined Incidence Matrix \textit{I}}} \\ 
\hline
 &  Start  & \makecell{Get \\ Signature \\ Integers}   & \makecell{Compute \\ Hash}    & \makecell{Calculate\\ Point} & \makecell{Compute\\ Coordinates} & \makecell{Verify \\ Signatures} \\ \hline
 
 $\varphi$ $(Inputs)$  &1    &-1    & -1   &    0&0&-1\\ \hline
  $\varphi$ $\makecell{(Signatures \\ Store)}$  & 0   & 1   & 0   & -1 &-1&0   \\ \hline
 $\varphi$ $\makecell{(Hash \\ Integer\\ Store)}$  &    0& 0   &    1& 0 & -1&0  \\ \hline
  $\varphi$ $\makecell{(Point \\ Store)}$ & 0   & 0   & 0   & 1  &-1&0 \\ \hline
  $\varphi$ $\makecell{(Coordinates \\ Store)}$ & 0   & 0   & 0   & 0 &1&-1 \\ \hline
  $\varphi$ $\makecell{(Accept \\ Reject)}$ & 0   & 0   & 0   & 0 &0&1 \\ \hline

\multicolumn{7}{|l|}{\textbf{Inhibition Matrix \textit{H}}} \\ \hline
&  Start  & \makecell{Get \\ Signature \\ Integers}   & \makecell{Compute \\ Hash}    & \makecell{Calculate\\ Point} & \makecell{Compute\\ Coordinates} & \makecell{Verify \\ Signatures} \\ \hline

 $\varphi$ $(Inputs)$  &0&0    &0    & 0   &    0&0\\ \hline
  $\varphi$ $\makecell{(Signatures \\ Store)}$  & 0   & 0&0   & 0   & 0 &0   \\ \hline
 $\varphi$ $\makecell{(Hash \\ Integer\\ Store)}$  &    0& 0   &    &0& 0 & 0  \\ \hline
  $\varphi$ $\makecell{(Point \\ Store)}$ & 0   & 0   & 0   & 0  &0&0 \\ \hline
  $\varphi$ $\makecell{(Coordinates \\ Store)}$ & 0   & 0   & 0   & 0 &0&0 \\ \hline
  $\varphi$ $\makecell{(Accept \\ Reject)}$ & 0   & 0   & 0   & 0 &0&0 \\ \hline
\end{tabular}
\end{adjustbox}
\label{table:petri3}
\vspace{-2mm}%Put here to reduce too much white space after your table 
\end{table}

Table \ref{table:petri4} shows the results of forward, backward, combined and inhibition matrices for the scenario of Calculate Location in the proposed clone node detection mechanism.

\begin{table}[!h]

\centering
\caption{High Level Petri Net Incidences - Calculate Location}
%{\tiny\renewcommand{\arraystretch}{.8}
%\resizebox{!}{.09\paperheight}{%
\begin{adjustbox}{width=0.47\textwidth,center}
\begin{tabular}{|c|c|c|c|c|c|c|c|}
\hline
\multicolumn{4}{|l|}{\textbf{Forwards Incidence Matrix \textit{$I^{+}$}}} & \multicolumn{4}{l|}{\textbf{Backwards Incidence Matrix \textit{$I^{-}$}}} \\ \hline
       &    Start   &  \makecell{Determine \\2-D \\ Point\\ Space}     & \makecell{Calculate \\ Distance}    &  &    Start   &  \makecell{Determine \\2-D \\ Point\\ Space}     & \makecell{Calculate \\ Distance}  \\ \hline
       
   $\varphi$ $(Inputs)$   &     1  & 0               & 0 &  $\varphi$ $(Inputs)$  & 0     &1  &0 \\ \hline
   
      $\varphi$ $\makecell{(Points\\Store)}$ &    0  & 1   & 0   &   $\varphi$ $\makecell{(Points \\Store)}$    &    0   &  0 &    1\\ \hline

      $\varphi$ $\makecell{(Provers' \\ Distance \\ Store)}$ &   0   &    0& 1   &   $\varphi$ $\makecell{(Provers' \\ Distance\\ Store)}$    &    0   &  0 &   0 \\ \hline

\multicolumn{4}{|l|}{\textbf{Combined Incidence Matrix \textit{I}}} & \multicolumn{4}{l|}{\textbf{Inhibition Matrix \textit{H}}} \\ \hline

 &    Start   &  \makecell{Determine\\ 2-D \\ Point\\ Space}     & \makecell{Calculate \\ Distance}    &  &    Start   &  \makecell{Determine \\2-D \\ Point \\Space}     & \makecell{Calculate \\ Distance}  \\ \hline
       
   $\varphi$ $(Inputs)$   &    1   & -1               & 0 &  $\varphi$ $(Inputs)$  &0      & 0 & 0\\ \hline
   
      $\varphi$ $\makecell{(Points\\Store)}$ &    0  & 1   &-1    &   $\varphi$ $\makecell{(Points\\Store)}$    &    0   &  0 & 0   \\ \hline

     $\varphi$ $\makecell{(Provers' \\ Distance \\ Store)}$ &    0  &    0&  1  &   $\varphi$ $\makecell{(Provers' \\ Distance \\ Store)}$   &    0   &  0 & 0   \\ \hline
      
\end{tabular}
\end{adjustbox}
\label{table:petri4}
\vspace{-4mm}%Put here to reduce too much white space after your table 
\end{table}

\begin{table}[!h]
\tiny
\centering
\caption{High Level Petri Net Incidences - Generate Location Proof}
%{\tiny\renewcommand{\arraystretch}{.8}
%\resizebox{!}{.35\paperheight}{%
\begin{adjustbox}{width=0.47\textwidth,center}
\begin{tabular}{|c|c|c|c|c|c|}
\hline
\multicolumn{6}{|l|}{\textbf{Forwards Incidence Matrix \textit{$I^{+}$}}} \\ \hline
   &  Start  & \makecell{Sense \\ Context \\ Information}   & \makecell{Stored\\ Context \\ Information}& \makecell{Request\\ Location  \\Proof} & \makecell{Generate\\ Location Proof}   \\ \hline
 $\varphi$ $(Inputs)$  &1    &0    & 0   &    0&0\\ \hline
  $\varphi$ $\makecell{(Context \\ Information \\ Store)}$  & 0   & 1   & 0   & 0 &0   \\ \hline
 $\varphi$ $\makecell{(LBS\\ Information \\ Store)}$  &    0& 0   &    1& 0 & 0  \\ \hline
  $\varphi$ $\makecell{(Location \\ Proof \\Store)}$ & 0   & 0   & 0   & 1  &0 \\ \hline
  
  $\varphi$ $\makecell{(Signed \\ Location \\ Proofs \\ Store)}$ & 0   & 0   & 0   & 0  &1 \\ \hline

\multicolumn{6}{|l|}{\textbf{Backwards Incidence Matrix \textit{$I^{-}$}}} \\ \hline
 &  Start  & \makecell{Sense \\ Context \\ Information}   & \makecell{Stored\\ Context \\ Information}& \makecell{Request\\ Location\\ Proof} & \makecell{Generate\\ Location  Proof}   \\ \hline
  $\varphi$ $(Inputs)$ &  0  &    1&  0  &0   &1 \\ \hline
  $\varphi$ $\makecell{(Context \\ Information \\ Store)}$  & 0   & 0   & 1   & 0 & 1  \\ \hline
   $\varphi$ $\makecell{(LBS\\ Information \\ Store)}$ & 0   & 0   & 0   & 1 & 0   \\ \hline
  $\varphi$ $\makecell{(Location \\ Proof \\Store)}$  & 0   &0    & 0   &    0&1\\ \hline
  $\varphi$ $\makecell{(Signed \\ Location \\ Proofs \\Store)}$ & 0   & 0   & 0   & 0 &0 \\ \hline

\multicolumn{6}{|l|}{\textbf{Combined Incidence Matrix \textit{I}}} \\ \hline
  &  Start  & \makecell{Sense \\ Context \\ Information}   & \makecell{Stored\\ Context \\ Information}& \makecell{Request\\ Location \\Proof} & \makecell{Generate\\ Location Proof}   \\ \hline
 $\varphi$ $(Inputs)$  & 1   &    -1&  0 & 0 & -1  \\ \hline
$\varphi$ $\makecell{(Context \\ Information \\ Store)}$   &    0& 1   &    -1& 0 & -1   \\ \hline
  $\varphi$ $\makecell{(LBS\\ Information \\ Store)}$ & 0   &  0  &1    &-1 & 0    \\ \hline
 $\varphi$ $\makecell{(Location \\ Proof \\Store)}$    & 0   &    0&  0  &1 & -1   \\ \hline
 $\varphi$ $\makecell{(Signed \\ Location \\ Proofs \\Store)}$ & 0   & 0   & 0   & 0  &1 \\ \hline

\multicolumn{6}{|l|}{\textbf{Inhibition Matrix \textit{H}}} \\ \hline
&  Start  & \makecell{Sense \\ Context \\ Information}   & \makecell{Stored\\ Context \\ Information}& \makecell{Request\\ Location \\ Proof} & \makecell{Generate\\ Location Proof}   \\ \hline
 $\varphi$ $(Inputs)$ &   0 & 0   & 0   & 0 & 0   \\ \hline
$\varphi$ $\makecell{(Context \\ Information \\ Store)}$  &    0& 0   &    0& 0 & 0  \\ \hline
  $\varphi$ $\makecell{(LBS\\ Information \\ Store)}$  & 0   & 0   & 0   & 0 & 0  \\ \hline
  $\varphi$ $\makecell{(Location \\ Proof \\Store)}$   & 0   &    0&  0  &    0& 0\\ \hline
  $\varphi$ $\makecell{(Signed \\ Location \\ Proofs \\Store)}$ & 0   & 0   & 0   & 0  &0 \\ \hline
  
\end{tabular}
\end{adjustbox}
\label{table:petri5}
\vspace{-4mm}%Put here to reduce too much white space after your table 

\end{table}

Table \ref{table:petri5} presents the results of forwards, backwards, combined, and inhibition incidence matrices for the Generate Location Proof Petri net in the proposed clone node attack detection technique.

\begin{table}[!h]
\tiny
\caption{High Level Petri Net Incidences - Verify Location Proof}
\centering
%{\tiny\renewcommand{\arraystretch}{.8}
%\resizebox{!}{.35\paperheight}{%
\begin{adjustbox}{width=0.47\textwidth,center}
\begin{tabular}{|c|c|c|c|c|c|}
\hline
\multicolumn{6}{|l|}{\textbf{Forwards Incidence Matrix \textit{$I^{+}$}}} \\ \hline
   &  Start  & \makecell{Extract \\ Context \\ Information}   & \makecell{Accept\\ Location \\Proof \\ Request}& \makecell{Verify\\ Context \\ Information} & \makecell{Verify\\ Location \\ Proof}   \\ \hline
 $\varphi$ $(Inputs)$  &1    &0    & 0   &    0&0\\ \hline
 
  $\varphi$ $\makecell{(Extracted \\ Context  Information \\ Store)}$  & 0   & 1   & 0   & 0 &0   \\ \hline
 
 $\varphi$ $\makecell{(Location\\ Proofs \\ Store)}$  &    0& 0   &    1& 0 & 0  \\ \hline
  
  $\varphi$ $\makecell{(Verified \\ Information \\Store)}$ & 0   & 0   & 0   & 1  &0 \\ \hline
  
  $\varphi$ $\makecell{(Accept / \\ Reject \\ Location \\Proof)}$ & 0   & 0   & 0   & 0  &1 \\ \hline

\multicolumn{6}{|l|}{\textbf{Backwards Incidence Matrix \textit{$I^{-}$}}} \\ \hline
 &  Start  & \makecell{Extract \\ Context \\ Information}   & \makecell{Accept\\ Location \\Proof \\ Request}& \makecell{Verify\\ Context \\ Information} & \makecell{Verify\\ Location \\ Proof} \\ \hline
 
  $\varphi$ $(Inputs)$ &  0  &    1&  0  &0   &1 \\ \hline
  
  $\varphi$ $\makecell{(Extracted \\ Context  Information \\ Store)}$ & 0   & 0   & 1   & 1& 0  \\ \hline
  
   $\varphi$ $\makecell{(Location\\ Proofs \\ Store)}$ & 0   & 0   & 0   & 1 & 0   \\ \hline
   
  $\varphi$ $\makecell{(Verified \\ Information \\Store)}$  & 0   &0    & 0   &    0&1\\ \hline
  
  $\varphi$ $\makecell{(Accept / \\ Reject \\ Location \\Proof)}$ & 0   & 0   & 0   & 0  &0 \\ \hline

\multicolumn{6}{|l|}{\textbf{Combined Incidence Matrix \textit{I}}} \\ \hline
  &  Start  & \makecell{Extract \\ Context \\ Information}   & \makecell{Accept\\ Location \\Proof \\ Request}& \makecell{Verify\\ Context \\ Information} & \makecell{Verify\\ Location \\ Proof} \\ \hline
  
 $\varphi$ $(Inputs)$  & 1   &    -1&  0 & 0 & -1  \\ \hline
$\varphi$ $\makecell{(Extracted \\ Context  Information \\ Store)}$  &    0& 1   &    -1& -1 & 0   \\ \hline
 $\varphi$ $\makecell{(Location\\ Proofs \\ Store)}$ & 0   &  0  &1    &-1 & 0    \\ \hline
 $\varphi$ $\makecell{(Verified \\ Information \\Store)}$    & 0   &    0&  0  &1 & -1   \\ \hline
 
 $\varphi$ $\makecell{(Accept / \\ Reject \\ Location \\Proof)}$ & 0   & 0   & 0   & 0  &1 \\ \hline

\multicolumn{6}{|l|}{\textbf{Inhibition Matrix \textit{H}}} \\ \hline
&  Start  & \makecell{Extract \\ Context \\ Information}   & \makecell{Accept\\ Location \\Proof \\ Request}& \makecell{Verify\\ Context \\ Information} & \makecell{Verify\\ Location \\ Proof} \\ \hline

$\varphi$ $(Inputs)$  & 0  &    0&  0 & 0 & 0 \\ \hline

$\varphi$ $\makecell{(Extracted \\ Context  Information \\ Store)}$& 0   & 0   & 0   & 0 & 0 \\ \hline

 $\varphi$ $\makecell{(Location\\ Proofs \\ Store)}$  & 0   & 0   & 0   & 0 & 0  \\ \hline
 
  $\varphi$ $\makecell{(Verified \\ Information \\Store)}$ & 0   & 0   & 0   & 0  &0 \\ \hline
  
  $\varphi$ $\makecell{(Accept / \\ Reject \\ Location \\Proof)}$ & 0   & 0   & 0   & 0  &0 \\ \hline
  
\end{tabular}
\end{adjustbox}
\label{table:petri6}
\vspace{-4mm}%Put here to reduce too much white space after your table 

\end{table}

The forwards, backwards, combined, and inhibition matrices for the Verify Location Proof Petri net of the proposed clone node attack detection scheme are shown in Table \ref{table:petri6}.

\section{Confidence Intervals} \label{ConfidenceInterval}

This appendix contains the confidence intervals for several HLPNs in the average number of tokens with minimum and maximum threshold values. These HLPNs include ECDSA* signature generation, ECDSA* signature verification, calculate location, generate location proof, and verify location proof.

Table \ref{Token2} shows the results of ECDSA* signature generation HLPN in the form of the minimum and maximum threshold values obtained by throwing an average number of tokens at each place.

\begin{table}[!h]
\centering
\tiny
\caption{Average Number of Token with Minimum and Maximum Threshold Values - ECDSA* Signature Generation}
\begin{adjustbox}{width=0.47\textwidth,center}

\begin{tabular}{|c|c|c|c|c|}
\hline
\textbf{\makecell{Places}} &\textbf{ \makecell{Average Number\\ of Tokens}} & \textbf{\makecell{Minimum \\ Threshold \\ Values}} & \textbf{\makecell{Average Number \\ of Tokens}} & \textbf{\makecell{Maximum \\ Threshold\\  Values}}  \\ \hline

Inputs &      0.86139     &     0.356      &     1.02997      &   0.196        \\ \hline

\makecell{Coordinates\\ Store} &     0.26733     &   0.4366        &   0.2684      &     0.57542         \\ \hline
 
 \makecell{Hash \\ Integer \\Store}&    0.23762     &     0.0734      &  0.028           &       0.21279    \\ \hline
 
 \makecell{Signature \\Store}&     13.67327      &      0.5872     &   2.612       &        144.92008    \\ \hline

\end{tabular}
\end{adjustbox}
\label{Token2}
\vspace{-4mm}%Put here to reduce too much white space after your table 

\end{table}

Table \ref{Token3} shows the results of ECDSA* signature verification HLPN in the form of the minimum and maximum threshold values obtained by throwing an average number of tokens at each place.

\begin{table}[!h]
\centering
\tiny
\caption{Average Number of Token with Minimum and Maximum Threshold Values - ECDSA* Signature Verification}
\begin{adjustbox}{width=0.47\textwidth,center}

\begin{tabular}{|c|c|c|c|c|}
\hline
\textbf{\makecell{Places}} &\textbf{ \makecell{Average Number\\ of Tokens}} & \textbf{\makecell{Minimum \\ Threshold \\ Values}} & \textbf{\makecell{Average Number \\ of Tokens}} & \textbf{\makecell{Maximum \\ Threshold\\  Values}}  \\ \hline

Inputs &   1.0297        &    0.3697       &     1.17183      &    0.27562       \\ \hline

\makecell{Signature\\ Store} &     0.68317      &    0.35744       &  0.46254         &   0.13784        \\ \hline
 
 \makecell{Hash \\ Integer \\Store}&     1.9604      &    4.48262       &    45.97802       &    17.2552       \\ \hline
 
 \makecell{Point \\Store}&     0.45545      &     0.29094      &       0.32068    &    0.16362       \\ \hline
 
 \makecell{Coordinates \\Store}&   0.17822        &     0.0359      &   0.12987        &      0.01708     \\ \hline

 \makecell{Accept \\ Reject}&     4.08911      &      0.84402     &       40.39461    &    3.425       \\ \hline

\end{tabular}
\end{adjustbox}
\label{Token3}
\vspace{-4mm}%Put here to reduce too much white space after your table 

\end{table}

Table \ref{Token4} shows the results of calculate location HLPN in the form of the minimum and maximum threshold values obtained by throwing an average number of tokens at each place.

\begin{table}[!h]
\centering
\tiny
\caption{Average Number of Token with Minimum and Maximum Threshold Values - Calculate Location}
\begin{adjustbox}{width=0.47\textwidth,center}

\begin{tabular}{|c|c|c|c|c|}
\hline
\textbf{\makecell{Places}} &\textbf{ \makecell{Average Number\\ of Tokens}} & \textbf{\makecell{Minimum \\ Threshold \\ Values}} & \textbf{\makecell{Average Number \\ of Tokens}} & \textbf{\makecell{Maximum \\ Threshold\\  Values}}  \\ \hline
Inputs &     2.65347        &   5.153333      &       33.81019      &   15.71      \\ \hline
\makecell{Points \\ Store} &      1.60396     &        3.087666   &   9.436666        &     13.28272      \\ \hline
\makecell{Provers' \\ Distance \\ Store}&   14.71287        &      2.4348     &  146.54146         &     8.12      \\ \hline
\end{tabular}
\end{adjustbox}
\label{Token4}
\vspace{-4mm}%Put here to reduce too much white space after your table 

\end{table}

Table \ref{Token5} shows the results of generate location proof HLPN in the form of the minimum and maximum threshold values obtained by throwing an average number of tokens at each place.

\begin{table}[!h]
\centering
\tiny
\caption{Average Number of Token with Minimum and Maximum Threshold Values - Generate Location Proof}
\begin{adjustbox}{width=0.47\textwidth,center}

\begin{tabular}{|c|c|c|c|c|}
\hline
\textbf{\makecell{Places}} &\textbf{ \makecell{Average Number\\ of Tokens}} & \textbf{\makecell{Minimum \\ Threshold \\ Values}} & \textbf{\makecell{Average Number \\ of Tokens}} & \textbf{\makecell{Maximum \\ Threshold\\  Values}}  \\ \hline

Inputs &   1.33663        &       1.2186    &     2.27473      &    1.074       \\ \hline

\makecell{Context \\ Information \\ Store} &      0.71287     &     0.512      &       1.02897    &   0.8534        \\ \hline
 
 \makecell{LBS \\ Information \\Store}&   1.0396        &    0.3306       &          1.3037 &  0.3034         \\ \hline
 
 \makecell{Location \\ Proofs \\ Store}&     1.62376      &   0.617        &       0.86513    &     0.361      \\ \hline
 
 \makecell{Signed \\ Location  \\ Proofs \\ Store}&    4.70297       &   0.3852        &    61.03696       &   0.2952        \\ \hline

\end{tabular}
\end{adjustbox}
\label{Token5}
\vspace{-4mm}%Put here to reduce too much white space after your table 

\end{table}

Table \ref{Token6} shows the results of verify location proof HLPN in the form of the minimum and maximum threshold values obtained by throwing an average number of tokens at each place.

\begin{table}[!h]
\tiny
\centering
\tiny
\caption{Average Number of Token with Minimum and Maximum Threshold Values - Verify Location Proof}
\begin{adjustbox}{width=0.47\textwidth,center}

\begin{tabular}{|c|c|c|c|c|}
\hline
\textbf{\makecell{Places}} &\textbf{ \makecell{Average Number\\ of Tokens}} & \textbf{\makecell{Minimum \\ Threshold \\ Values}} & \textbf{\makecell{Average Number \\ of Tokens}} & \textbf{\makecell{Maximum \\ Threshold\\  Values}}  \\ \hline

Inputs &       1.11881    &       2.968    &     4.7033      &     1.456      \\ \hline

\makecell{Extracted \\ Context \\ Information \\ Store} &     0.78218      &      0.57056     &    0.91409       &     0.30618      \\ \hline
 
 \makecell{Location \\ Proofs \\Store}&   0.24752        &      0.0806     &     0.26174      &      0.033476     \\ \hline
 
 \makecell{Verified \\Information\\ Store}&    0.19802       &    0.082582       &         0.14785  &    0.02452       \\ \hline
 
 \makecell{Accept / Reject  \\ Location  \\ Proof \\ Store}&      5.67327     &    0.4008       &    61.47453       &       0.199048    \\ \hline

\end{tabular}
\end{adjustbox}
\label{Token6}
\vspace{-4mm}%Put here to reduce too much white space after your table 

\end{table}

\end{document}